\documentclass[useAMS,usenatbib]{mn2e}

\usepackage{natbib}
\usepackage{amsmath,amsfonts,amssymb}
\usepackage[dvips]{graphicx}
\usepackage{longtable}
\usepackage{float}
\usepackage{caption}
\usepackage{subfig}
\usepackage{txfonts}
\usepackage{multirow}
\usepackage[multiple]{footmisc}
\usepackage[dvips]{color}

\title[Deep 20-GHz survey of the Chandra Deep Field South and SDSS Stripe 82]
  {Deep 20-GHz survey of the Chandra Deep Field South and SDSS Stripe 82: source catalogue and spectral properties}
\author[Franzen et al.]
  {Thomas~M.~O.~Franzen,$^1$\thanks{Email: Thomas.Franzen@csiro.au}
  Elaine~M.~Sadler,$^2$
  Rajan~Chhetri,$^{1,3}$
  Ronald~D.~Ekers,$^{1}$  
  \newauthor  
  Elizabeth~K.~Mahony,$^{4}$
  Tara~Murphy,$^{2,5}$     
  Ray~P.~Norris,$^{1,6}$  
  Elizabeth~M.~Waldram,$^7$
  \newauthor  
  and Imogen~H.~Whittam$^7$\\
  $^1$Australia Telescope National Facility, CSIRO Astronomy \& Space Science, PO Box 76, Epping, NSW 1710, Australia \\
  $^2$Sydney Institute for Astronomy, School of Physics, The University of Sydney, NSW 2006, Australia \\
  $^3$Department of Astrophysics \& Optics, School of Physics, University of New South Wales, NSW 2052, Australia \\
  $^4$ASTRON, the Netherlands Institute for Radio Astronomy, Postbus 2, 7990 AA, Dwingeloo, The Netherlands \\
  $^5$School of Information Technologies, University of Sydney, NSW 2006, Australia \\
  $^6$ARC Centre of Excellence for All-sky Astrophysics (CAASTRO) \\
  $^7$Astrophysics Group, Cavendish Laboratory, 19 J.~J.~Thomson Avenue, Cambridge CB3 0HE}      
\date{Accepted ????. Received ????}
\pagerange{\pageref{firstpage}--\pageref{lastpage}}
\pubyear{2012}

\begin{document}
\maketitle
\label{firstpage}

\begin{abstract}

\noindent

We present a source catalogue and first results from a deep, blind radio survey carried out at 20~GHz with the Australia Telescope Compact Array, with follow-up observations at 5.5, 9 and 18~GHz. The Australia Telescope 20~GHz (AT20G) deep pilot survey covers a total area of 5~$\mathrm{deg}^{2}$ in the Chandra Deep Field South and in Stripe 82 of the Sloan Digital Sky Survey. We estimate the survey to be 90\% complete above 2.5~mJy. Of the 85 sources detected, 55\% have steep spectra ($\alpha_{1.4}^{20} < -0.5$) and 45\% have flat or inverted spectra ($\alpha_{1.4}^{20} \geq -0.5$).

The steep-spectrum sources tend to have single power-law spectra between 1.4 and 18~GHz, while the spectral indices of the flat- or inverted-spectrum sources tend to steepen with frequency. Among the 18 inverted-spectrum ($\alpha_{1.4}^{20} \geq 0.0$) sources, 10 have clearly defined peaks in their spectra with $\alpha_{1.4}^{5.5} > 0.15$ and $\alpha_{9}^{18} < -0.15$. On a 3-yr timescale, at least 10 sources varied by more than 15\% at 20~GHz, showing that variability is still common at the low flux densities probed by the AT20G-deep pilot survey. 

We find a strong and puzzling shift in the typical spectral index of the 15--20-GHz source population when combining data from the AT20G, Ninth Cambridge and Tenth Cambridge surveys: there is a shift towards a steeper-spectrum population when going from $\sim 1$~Jy to $\sim 5$~mJy, which is followed by a shift back towards a flatter-spectrum population below $\sim 5$~mJy. The 5-GHz source-count model by \cite{jackson1999}, which only includes contributions from FRI and FRII sources, and star-forming galaxies, does not reproduce the observed flattening of the flat-spectrum counts below $\sim 5$~mJy. It is therefore possible that another population of sources is contributing to this effect.

\end{abstract}


\section{Introduction}\label{Introduction}

Our knowledge of the sky at high radio frequency ($\gtrsim 10~\mathrm{GHz}$) is poor compared with that at 1.4~GHz because the small field-of-view of ground-based diffraction-limited telescopes makes high-radio-frequency surveys very time-consuming. However, they are extremely important both to complete some of the gaps in information obtained from lower-frequency surveys as well as to enable different classes of sources to be studied.

In a high-frequency survey, there is preferential detection of sources with flat spectra up to high frequencies or with peaks in their spectra at high frequencies. The emission from such sources mainly originates from the compact cores of radio galaxies (rather than from the extended jets and lobes). It follows that a survey at high frequency probes current and very recent AGN activity, whereas a survey at lower frequencies (e.g. at 1.4~GHz) probes AGN activity on longer timescales (as well as a large population of star-forming galaxies). In combination with a 1.4-GHz survey, it can therefore provide new insights into the lifetimes and duty cycles of radio-loud AGN.

Two examples of sources which are preferentially detected at high frequency are: blazars (the beamed products of FRI and FRII sources), which are prone to variability in flux density, and gigahertz-peaked-spectrum (GPS) sources. The latter are compact, powerful radio sources with spectral peaks at GHz frequencies and sizes $\lesssim 1$~kpc \citep[e.g.][]{odea1998}. They are thought to represent the earliest stages of radio-galaxy evolution and evolve into large-scale radio galaxies, although there are still large gaps in our knowledge of their evolutionary properties.

Our knowledge of the sky at 15--20~GHz has much improved over the past 10 years or so thanks to the completion of the Australia Telescope 20~GHz \citep[AT20G;][]{murphy2010} survey, the 15.2-GHz Ninth Cambridge \citep[9C;][]{waldram2003,waldram2010} survey and the 15.7-GHz Tenth Cambridge \citep[10C;][]{franzen2011,davies2011} survey. Correlation of the catalogues from these high-frequency surveys with the 1.4-GHz NRAO VLA Sky Survey \citep[NVSS;][]{condon1998} catalogue suggests strong changes in the 15--20-GHz source population with flux density. The AT20G survey, completed in 2008, covers the whole Southern sky down to a limiting flux density of 40~mJy. The AT20G source catalogue is dominated by compact, flat-spectrum QSOs and BL-Lac objects, but the fraction of sources which are unbeamed galaxies begins to rise at fainter flux densities \citep{mahony2011} and this shift is accompanied by a steepening in the median 1--20-GHz radio spectral index \citep{massardi2011}. A similar trend was found in the 9C survey which covers a smaller area but probes deeper flux-density levels (complete to 5.5~mJy).

In contrast, in the 10C survey, which covers $\approx 27$~deg$^2$ complete to 1.0~mJy and $\approx 12$~deg$^2$ complete to 0.5~mJy, a completely different trend was found in which the fraction of steep-spectrum sources {\it declined} with flux density for these fainter sources. A sub-sample of 10C sources in the Lockman Hole was studied in detail by \cite{whittam2013} at 610~MHz and 1.4~GHz. They think it is likely that the faint, flat-spectrum sources are a result of the cores of FRI sources becoming dominant at 15.7~GHz. 

The advent of the Compact Array Broadband Backend \citep[CABB;][]{wilson2011} on the Australia Telescope Compact Array (ATCA)\footnote{http://www.narrabri.atnf.csiro.au/} makes it possible to carry out a new large-area continuum survey going significantly deeper than the AT20G survey; the Evolutionary Map of the Universe \citep[EMU;][]{norris2011}, a wide-field radio continuum survey planned for the new Australian Square Kilometre Array Pathfinder telescope, would provide deep ($\approx 10 \mu \mathrm{Jy}$ rms) coverage at 1.4~GHz. We have carried out a pilot survey, in advance of a larger AT20G-deep survey, covering two regions which are well studied at other wavelengths: the Chandra Deep Field South (03~hr; $-27$~deg) and part of SDSS Stripe 82 (21~hr; $-0.5$~deg). Sources detected in the survey were followed up with ATCA at 5.5, 9 and 18~GHz, allowing a detailed study of their radio properties.

In this paper, we present some first results from the AT20G-deep pilot survey. We also attempt to provide an interpretation of the spectral index shifts in the 15--20-GHz source population described above. The layout of the paper is as follows. Section~\ref{Choice of fields} describes the survey fields. Section~\ref{Survey observations and data reduction} outlines our observing strategy, and imaging and source finding techniques. In Section~\ref{Correlation with NVSS, ATLAS and FIRST at 1.4 GHz}, we measure the spectral index distribution of our sample between 1.4 and 20~GHz. Section~\ref{Multi-frequency follow-up observations and data reduction} describes the reduction of the data from the follow-up observations. The results from the follow-up observations are presented in Sections~\ref{Flux density variability at 20 GHz} and~\ref{Spectral curvature between 1.4 and 18 GHz}; in Section~\ref{Flux density variability at 20 GHz}, we look at variability at 20~GHz over a 3-yr timescale and in Section~\ref{Spectral curvature between 1.4 and 18 GHz}, we investigate spectral curvature between 1.4 and 18~GHz. In Section~\ref{Exploring the relationship between spectral index and flux density in the high-radio-frequency source population}, we combine data from the AT20G, 9C and 10C surveys to explore the relationship between spectral index and flux density. Our results are summarized in Section~\ref{Conclusions}.

The sources in the AT20G-deep pilot survey will be studied further in an accompanying paper, hereafter Paper II, where we will study their angular sizes and polarisation properties, and look at their optical and infrared characteristics. 

\section{Choice of fields}\label{Choice of fields}

Two fields form part of the AT20G-deep pilot survey: the Chandra Deep Field South (CDFS), hereafter referred to as the `03hr' field, and the SDSS Stripe 82 region, hereafter referred to as the `21hr' field. Each field covers an area of $\approx 2.5~\mathrm{deg}^{2}$. The lines of RA and Dec. bounding the region complete to 2.5~mJy, for each of the fields, are given in Table~\ref{tab:survey_regions}.

\begin{table*}
\begin{center}
\caption{The survey areas}
\label{tab:survey_regions}
\begin{tabular}{@{} c c c c }
\hline
Field      & RA range                 & Dec. range                 & Area (deg$^{2}$) \\
\hline
03hr & 03:25:59.3 to 03:35:04.6 & $-$28:42:16 to $-$27:27:44 & 2.49 \\
21hr & 21:30:01.2 to 21:38:02.3 & $-$01:15:06 to $-$00:00:34 & 2.49 \\
\hline
\end{tabular}
\end{center}
\end{table*}

These fields were chosen as they are already well studied at other wavelengths: the 03hr field is covered by the NRAO VLA Sky Survey \citep[NVSS;][]{condon1998} and by the Australia Telescope Large Area Survey \citep[ATLAS;][]{norris2006} at 1.4~GHz. The field is also coincident with the Spitzer Wide-area Infrared Extragalactic \citep[SWIRE;][]{lonsdale2003} survey, the Spitzer Extragalactic Representative Volume Survey \citep[SERVS;][]{mauduit2012}, the VISTA Deep Extragalactic Observations \citep[VIDEO;][]{jarvis2013} survey, and has good spectroscopic coverage \citep{mao2012}. The 21hr field is covered by the NVSS and by the Faint Images of the Radio Sky at Twenty Centimetres \citep[FIRST;][]{becker1995} survey. The region also has good coverage in the 2dF-SDSS LRG and QSO (2SLAQ) Luminous Red Galaxy survey \citep{cannon2006} and 2SLAQ QSO survey \citep{croom2008}, as well as deep photometry from the Sloan Digital Sky Survey \citep[SDSS;][]{abazajian2009}.

\section{Survey observations and data reduction}\label{Survey observations and data reduction}

\subsection{Observations}\label{Observations}

The AT20G-deep pilot survey was carried out with ATCA in July 2009, shortly after the telescope was provided with a new wide-bandwidth correlator, the Compact Array Broadband Backend \citep[CABB;][]{wilson2011}. As a result of this upgrade to the telescope, the observing bandwidth was increased by a factor of 16, from $2 \times 128$ to $2 \times 2048$~MHz, in all bands (ranging from 1.1 to 105~GHz), greatly increasing the sensitivity of continuum observations. Our observations were made in continuum mode using two 2048-MHz CABB bands centred at 19 and 21~GHz, with each 2048-MHz band divided into 2048 1-MHz channels. All four Stokes parameters were measured.

Each 2.5-$\mathrm{deg}^{2}$ field was surveyed using 3332 pointing centres in a $68 \times 49$ hexagonal array, spaced by 1.8~arcmin. The full width at half-maximum (FWHM) of the primary beam at the upper end (22~GHz) of the higher-frequency band, where it is smallest, is 2.32~arcmin. A pointing spacing of 1.8~arcmin provides close to uniform sensitivity in the fields at 22~GHz: the variation in sensitivity intrinsic to the rastering set-up at this frequency is 8\%. Each pointing was observed twice (at different hour angles) for 10~s. The total observing time was $\approx 60$ hours. The observations were made at night when conditions are most stable. 

The H75 array configuration was used, which consists of six antennas on east-west and north-south arms. Five of the antennas lie in a compact configuration, with a range of baselines of 31 to 89~m. The sixth antenna lies at a distance of about 4.5~km to the west of the other antennas and was not used for any of the ATCA observations presented in this paper. This array configuration, excluding baselines to the $6^{\mathrm{th}}$ antenna, yields an angular resolution of $\approx 30$~arcsec at 20~GHz, providing good surface-brightness sensitivity. 

The primary flux density calibrator used was PKS B1934-638, which is the standard calibrator for ATCA observations \citep[S = 0.948~Jy at 20.000~GHz;][]{sault2003}. PKS B1921-293 was used to calibrate the bandpass. A secondary calibrator was observed every 30--40 minutes to calibrate the complex antenna gains. The pointing was calibrated on every second visit to the secondary calibrator. For the 03hr field, PKS B0327-241 was used as the secondary calibrator and for the 21hr field, PKS B2134+004. 

\subsection{Calibration and imaging}\label{Calibration and imaging}

The data were calibrated and imaged using the \textsc{miriad} \citep{sault1995} software. The `birdie' option was used in the \textsc{miriad} task \textsc{atlod} to remove channels affected by self-interference due to the 640-MHz clock harmonics and 100 channels at both edges of the band, where the response of the receiver is poor. In addition, the `opcorr' option was used to correct the data for atmospheric opacity. The data were flagged for antenna shadowing. The data were also inspected visually using the \textsc{pgflag} task and any RFI was manually excised; the task display allows bad channels and timestamps for each baseline to be readily identified. The lower end of the band centred at 21~GHz was found to be contaminated by RFI in observations of the 21hr field, probably from geostationary satellites. In total, about 14\% of the data were flagged for the 03hr field and about 23\% for the 21hr field. The bandpass and gains were calibrated by following the standard procedure described in the \textsc{miriad} calibration manual\footnote{http://www.atnf.csiro.au/computing/software/miriad}.

The so-called `joint approach' was used to image each field, where all pointings are deconvolved together: a dirty image of each pointing was made using the multi-frequency synthesis technique \citep{sault1994}. Natural weighting was used to maximize the signal-to-noise ratio (SNR). The resultant pointing images were combined in a linear mosaicking process. The mosaicked image was then deconvolved using the \textsc{mossdi} task, which performs a Steer CLEAN \citep{steer1984} on a mosaicked image. The number of CLEAN iterations was set to 10000, as the total CLEANed flux was found to reach a plateau after this number of iterations. 

\begin{figure*}
 \begin{center}
  \includegraphics[scale=0.86,angle=270]{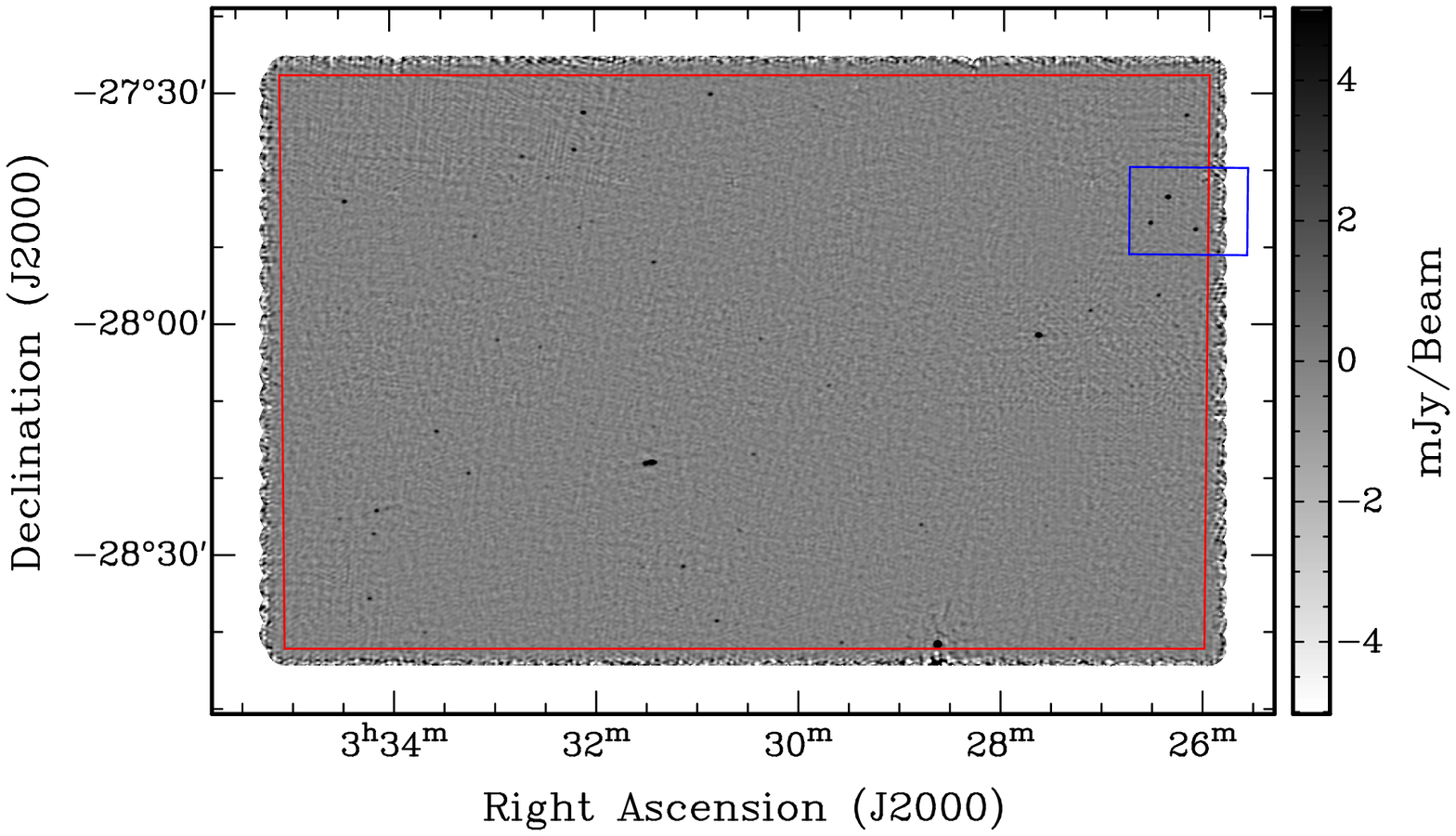}
  \includegraphics[scale=0.76,angle=270]{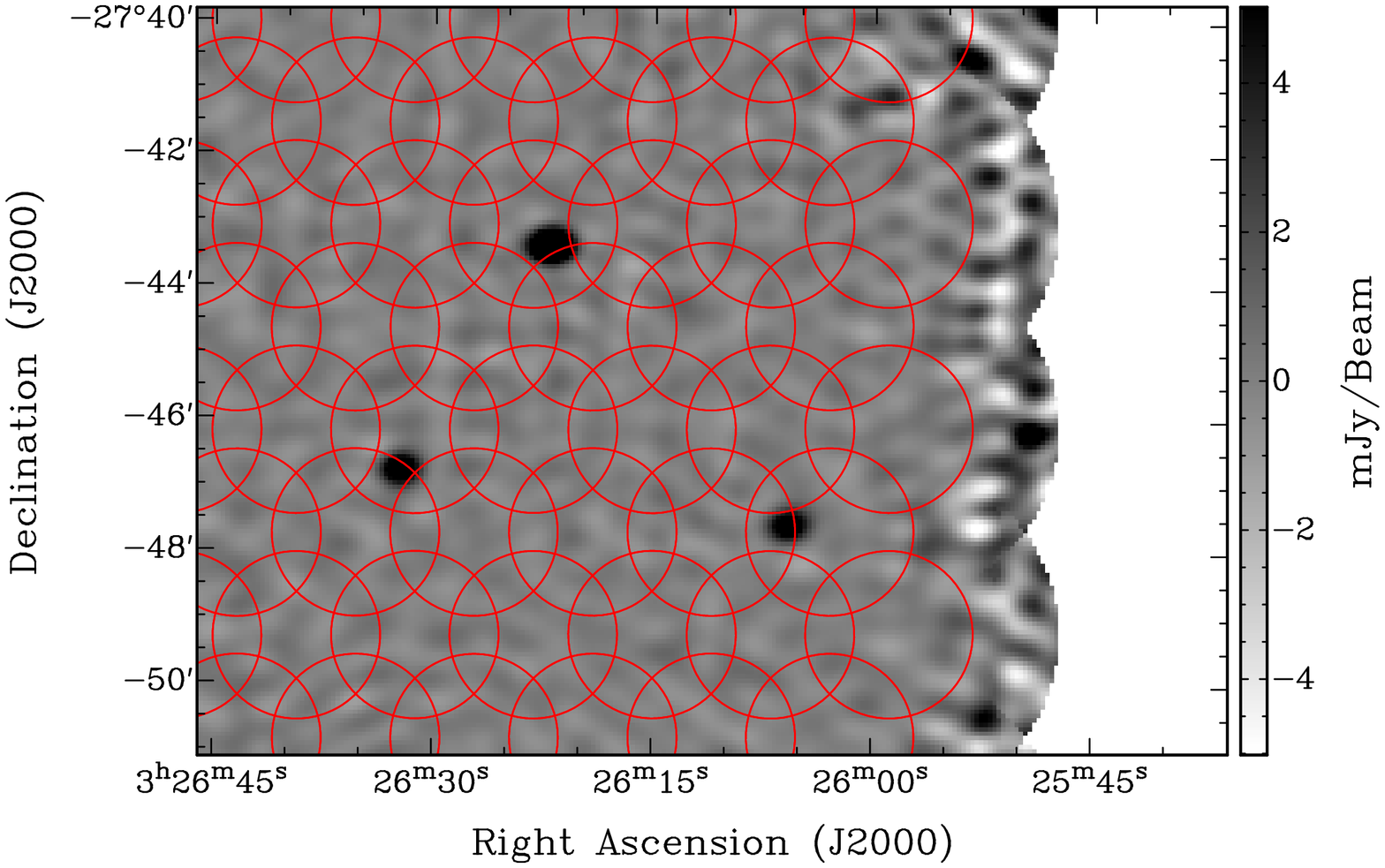}
  \caption{Top: Map of the 03hr field produced by combining approximately 3500 individual constituent maps. The survey area of the field defined in Table~\ref{tab:survey_regions} is shown by the red box. The blue box delineates the region of the field which is shown in the bottom panel. Bottom: Region of the 03hr field containing three point sources. The half-power beamwidth of the ATCA antennas at 20~GHz, centred on the pointing positions used in the mosaicked observations, are shown as circles.}
  \label{fig:field03}
 \end{center}
\end{figure*}

The map of the 03hr field was restored with a beam of 29.1 by 21.9 arcsec, the median synthesized beam of all pointings in the 03hr field; the map of the 21hr field was restored  with a beam of 29.0 by 25.2 arcsec, the median synthesized beam of all pointings in the 21hr field. Fig.~\ref{fig:field03} shows the final mosaic (top) of the 03hr field, with the area in which the survey is complete to 2.5~mJy shown in red, and a region (bottom) of the field containing three point sources. The ATCA primary beams at 20~GHz are shown as circles in the bottom image. 

\subsection{Source extraction}\label{Source extraction}

\subsubsection{Noise estimation}\label{Noise estimation} 

\begin{figure}
 \begin{center}
 \includegraphics[scale=0.43,angle=270]{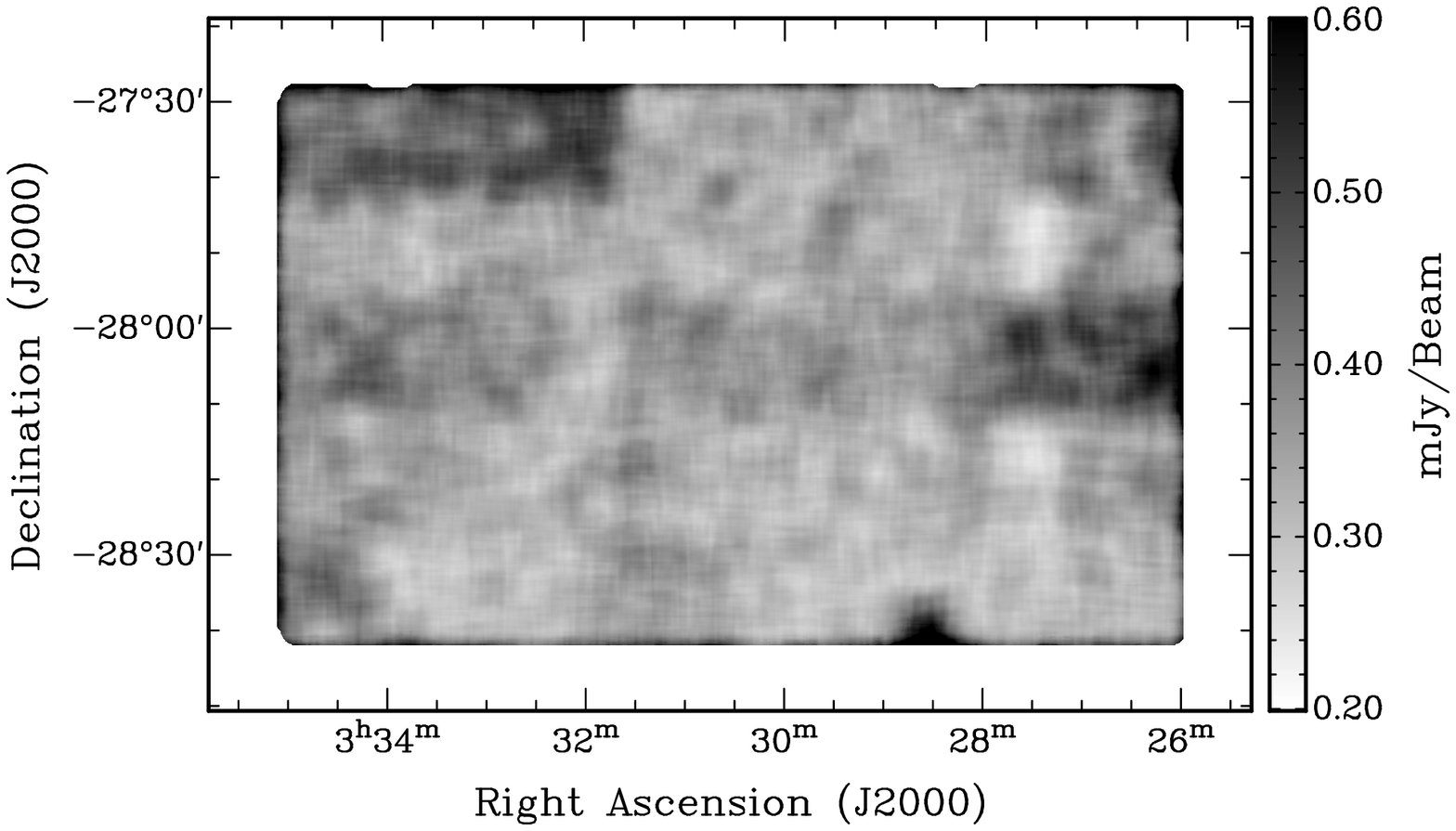}
  \includegraphics[scale=0.3,angle=270]{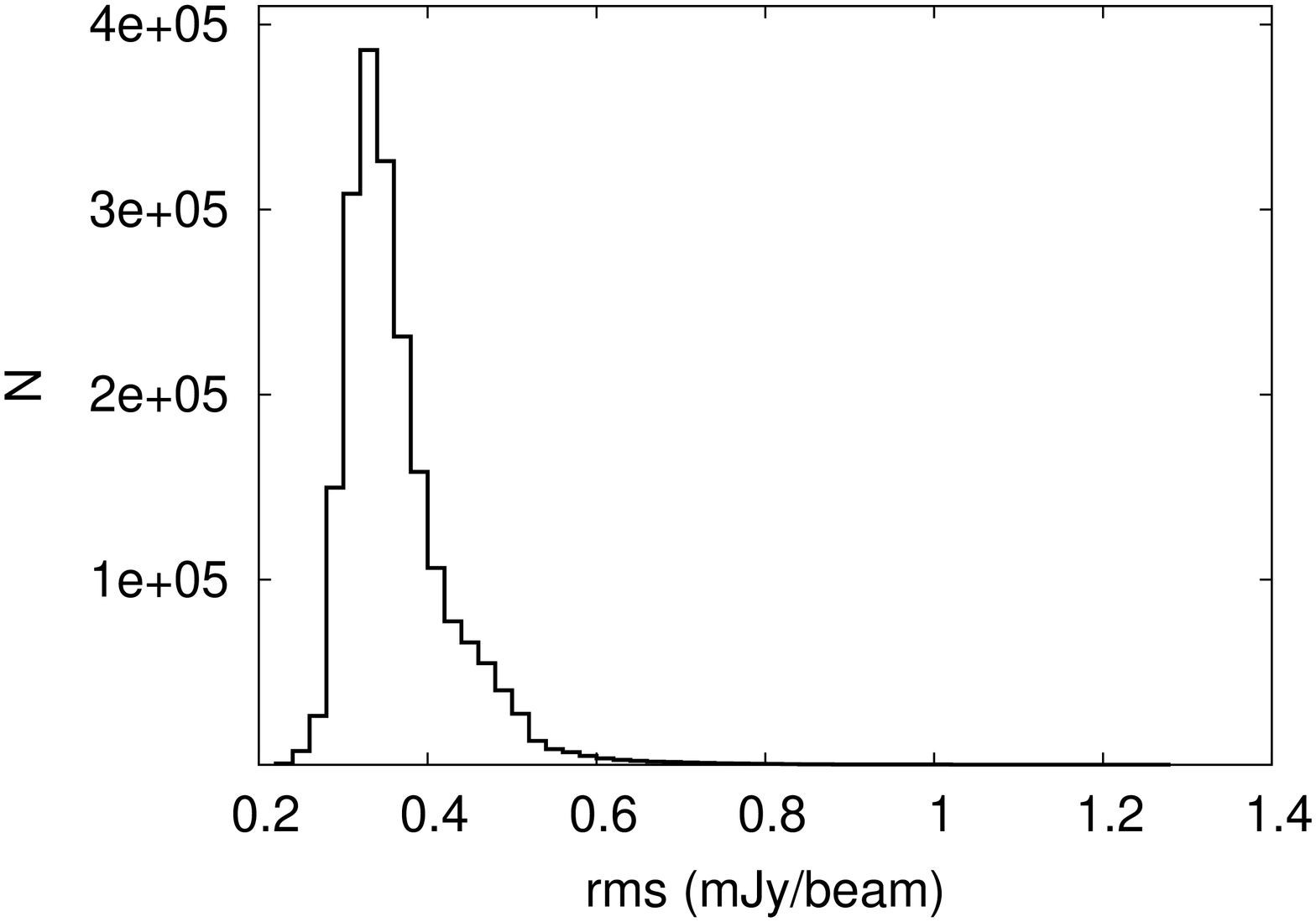}
  \caption{Top: The noise map of the 03hr field, produced by evaluating the noise power at each pixel position as described in Section~\ref{Noise estimation}. Bottom: The distribution of pixel values in the noise map. The distribution peaks at ~0.33~mJy.}
  \label{fig:field03_noise}
 \end{center}
\end{figure}

The first step in the source extraction process involved determination of the noise level. Noise maps were produced for the fields as follows: at each pixel position, the noise was taken as the rms inside a square centred on the pixel with a width of approximately 10 times the synthesized beam, but in order to avoid the noise estimate from being significantly affected by source emission, points were clipped iteratively until convergence at $\pm 3 \sigma$ was reached. We note that the noise could not be reliably measured close to the map edges. The area in which the noise was measured for each field (and in which source finding was subsequently carried out) is defined in Table~\ref{tab:survey_regions}. The typical noise level in both fields is 0.3--0.4~mJy. Fig.~\ref{fig:field03_noise} shows the noise map of the 03hr field together with the distribution of noise pixel values. The noise is somewhat higher around the very brightest sources as a result of calibration and deconvolution errors. 

\subsubsection{Source catalogue}\label{Source catalogue} 

Individual sources in the image were identified and characterised using a similar method to that employed by \cite{franzen2011} for the 10C survey. The noise maps were used to identify component `peaks' on the basis of their signal-to-noise ratios. Local maxima above $5 \sigma$ were identified as component peaks. Peak positions and flux densities were measured by interpolating between the pixels. An integration area, consisting of contiguous pixels down to a level of $2.5 \sigma$, was measured for each component. Using this information, the \textsc{aips}\footnote{\textsc{astronomical image processing system} -- www.aips.nrao.edu/} task \textsc{jmfit} was employed to fit a 2D elliptical Gaussian to each component in an automated fashion. Integrated flux densities and positions were derived from these Gaussian fits. In addition, an estimate of the true size of each component was made by deconvolving the point-source response from the fitted Gaussian (both represented by 2D elliptical Gaussians) and by taking the resulting 2D elliptical Gaussian to represent the sky brightness distribution. A component was classified as extended if $e_{\mathrm{maj}} \geq e_{\mathrm{crit}}$, where $e_{\mathrm{maj}}$ is the major axis of the component after deconvolution, 
\begin{eqnarray}
\label{eqn:ecrit}
e_{\mathrm{crit}} =
\left\{
\begin{array}{ll}
3.0 b_{\mathrm{maj}} \rho^{-1/2} & \mathrm{if~} 3.0 b_{\mathrm{maj}} \rho^{-1/2} > 10~\mathrm{arcsec,} \\
10~\mathrm{arcsec} & \mathrm{otherwise,}
\end{array}
\right.
\end{eqnarray}
$\rho$ is the signal-to-noise ratio and $b_{\mathrm{maj}}$ is the major axis of the restoring beam. The formula for $e_{\mathrm{crit}}$ is based on the theoretical expectation that the minimum component size that can be measured is proportional to the FWHM of the synthesized beam and inversely proportional to the square root of the SNR.

A total of 51 components were detected in the 03hr field and 36 in the 21hr field. Only three components were classified as extended. Contour plots from NVSS, ATLAS and FIRST were examined to help identify multiple-component sources. Two doubles (S28 and S71) were found, one in each field, resulting in a total of 50 sources in the 03hr field and 35 sources in the 21hr field. Both fields cover the same area of sky to approximately the same depth. Assuming a Poisson distribution of sources in the sky, the difference between the numbers of sources detected in the two fields is not statistically significant. The source catalogues for the 03hr and 21hr fields are given in Tables~\ref{tab:source_catalogue_03hr} and~\ref{tab:source_catalogue_21hr} respectively. The columns are as follows:

\indent Column (1) - Source number. \\
\indent Column (2) - Source IAU name. \\
\indent Columns (3) and (4) - Source position: RA (J2000), $\alpha$, and Dec. (J2000), $\delta$. This is the centroid position of all components included in the source. The peak position, interpolated between pixels, is taken to best represent the position of a component classified as point-like; the peak position from the Gaussing fitting is taken to best represent the position of a component classified as extended. \\
\indent Column (5) - Total flux density at 20~GHz (in mJy), $S_{20}$. This is the total flux density of all components included in the source. The peak flux density is taken to best represent the flux density of a component classified as point-like; the integrated flux density from the Gaussian fitting is taken to best represent the flux density of a component classified as extended. \\
\indent Column (6) - Uncertainty on the total flux density at 20~GHz (in mJy), $\delta S_{20}$. This is the sum of the uncertainties on the flux densities of all components included in the source. For a component classified as point-like, the uncertainty is 5\% of the peak flux density added in quadrature with the local noise. For a component classified as extended, the uncertainty is 5\% of the integrated flux density added in quadrature with the error on the integrated flux density reported by the Gaussian-fitting task (which accounts for the local noise). The 5\% calibration error is based on findings shown in Murphy et al. (2010). To provide errors on the 20-GHz flux density measurements in the AT20G catalogue, Murphy et al. (2010) estimated the calibration error for each observational epoch from the scatter in the visibility amplitudes of the calibrators in each observing run. The calibration error was found to be typically 4--5\% of the total flux density.\\
\indent Column (7) - Total flux density at 1.4~GHz taken from the NVSS catalogue or estimated from the NVSS image (in mJy), $S_{1.4~(\mathrm{NVSS})}$. Flux densities derived from the NVSS image are indicated in square brackets. Flux densities were derived from the NVSS image only for sources that did not appear in the NVSS catalogue (see Section~\ref{Correlation with NVSS, ATLAS and FIRST at 1.4 GHz} for details). \\
\indent Column (8) - Total flux density at 1.4~GHz taken from the ATLAS/FIRST catalogue or estimated from the ATLAS/FIRST image in the case of the 03hr/21hr field (in mJy), $S_{1.4~(\mathrm{ATLAS/FIRST})}$. Flux densities were derived from the ATLAS/FIRST image only for sources that did not appear in the ATLAS/FIRST catalogue (see Section~\ref{Correlation with NVSS, ATLAS and FIRST at 1.4 GHz} for details). \\
\indent Column (9) - `Best' 1.4-GHz flux density, $S_{1.4}$ (in mJy). This is the NVSS flux density if the source is present in the NVSS catalogue, otherwise it is the ATLAS/FIRST flux density in the 03hr/21hr field. \\
\indent Column (10) - Spectral index between 1.4 and 20~GHz obtained using $S_{1.4}$ and $S_{20}$, $\alpha_{1.4}^{20}$. \\
\indent Column (11) - Spectral class: steep (S), flat (F) or inverted (I), determined as described in Section~\ref{Correlation with NVSS, ATLAS and FIRST at 1.4 GHz}.\\ 

\begin{table*}
\caption{The 20-GHz source catalogue for the 03hr field.}
\label{tab:source_catalogue_03hr}
\begin{tabular}{@{} c c c c c r r r r r r } 
\hline
\multicolumn{1}{c}{(1)}
&\multicolumn{1}{c}{(2)}
&\multicolumn{1}{c}{(3)}
&\multicolumn{1}{c}{(4)}
&\multicolumn{1}{c}{(5)}
&\multicolumn{1}{c}{(6)}
&\multicolumn{1}{c}{(7)}
&\multicolumn{1}{c}{(8)}
&\multicolumn{1}{c}{(9)}
&\multicolumn{1}{c}{(10)}
&\multicolumn{1}{c}{(11)}\\
\multicolumn{1}{c}{Source}
&\multicolumn{1}{c}{Source name}
&\multicolumn{1}{c}{$\alpha$}
&\multicolumn{1}{c}{$\delta$}
&\multicolumn{1}{c}{$S_{20}$}
&\multicolumn{1}{c}{$\delta S_{20}$}
&\multicolumn{1}{c}{$S_{1.4~(\mathrm{NVSS})}$}
&\multicolumn{1}{c}{$S_{1.4~(\mathrm{ATLAS})}$}
&\multicolumn{1}{c}{$S_{1.4}$}
&\multicolumn{1}{c}{$\alpha_{1.4}^{20}$}
&\multicolumn{1}{c}{Spectral}\\
\multicolumn{1}{c}{number}
&
&
&
&\multicolumn{1}{c}{(mJy)}
&\multicolumn{1}{c}{(mJy)}
&\multicolumn{1}{c}{(mJy)}
&\multicolumn{1}{c}{(mJy)}
&\multicolumn{1}{c}{(mJy)}
&
&class\\
\hline
S01 &  AT20GDP J032605-274735 &  03:26:05.7 &   -27:47:35 &    8.30 &    0.58 &         81.00 &        74.70 &        81.00 &        -0.86 &        S \\           
S02 &  AT20GDP J032611-273243 &  03:26:11.4 &   -27:32:43 &    6.46 &    0.58 &         108.20 &       110.90 &       108.20 &       -1.06 &        S \\           
S03 &  AT20GDP J032616-280013 &  03:26:16.2 &   -28:00:13 &    2.59 &    0.48 &         [$<1.35$] &    1.70 &         1.70 &         0.16 &         I \\           
S04 &  AT20GDP J032622-274323 &  03:26:22.0 &   -27:43:23 &    19.55 &   1.06 &         26.50 &        27.80 &        26.50 &        -0.11 &        F \\           
S05 &  AT20GDP J032627-275612 &  03:26:27.0 &   -27:56:12 &    5.22 &    0.50 &         3.80 &         4.10 &         3.80 &         0.12 &         I \\           
S06 &  AT20GDP J032632-274646 &  03:26:32.2 &   -27:46:46 &    9.45 &    0.60 &         25.00 &        24.30 &        25.00 &        -0.37 &        F \\           
S07 &  AT20GDP J032642-280804 &  03:26:42.4 &   -28:08:04 &    2.69 &    0.47 &         45.50 &        42.30 &        45.50 &        -1.06 &        S \\           
S08 &  AT20GDP J032707-275816 &  03:27:07.1 &   -27:58:16 &    5.00 &    0.56 &         [$<1.35$] &    1.40 &         1.40 &         0.48 &         I \\           
S09 &  AT20GDP J032715-275430 &  03:27:15.9 &   -27:54:30 &    1.98 &    0.35 &         2.70 &         2.20 &         2.70 &         -0.12 &        F \\           
S10 &  AT20GDP J032717-284103 &  03:27:17.2 &   -28:41:03 &    2.17 &    0.38 &         [$<1.35$] &    0.50 &         0.50 &         0.55 &         I \\           
S11 &  AT20GDP J032732-282622 &  03:27:32.3 &   -28:26:22 &    1.86 &    0.29 &         7.40 &         8.60 &         7.40 &         -0.52 &        S \\           
S12 &  AT20GDP J032737-280130 &  03:27:37.7 &   -28:01:30 &    39.60 &   2.05 &         100.20 &       96.40 &        100.20 &       -0.35 &        F \\           
S13 &  AT20GDP J032836-284149 &  03:28:36.5 &   -28:41:49 &    165.56 &  8.50 &         1449.30 &      1357.80 &      1449.30 &      -0.82 &        S \\           
S14 &  AT20GDP J032843-282714 &  03:28:43.5 &   -28:27:14 &    1.82 &    0.34 &         [$<1.35$] &    [$<0.23$] &    [$<0.23$] &    $>0.78$ &      I \\           
S15 &  AT20GDP J032846-282618 &  03:28:46.6 &   -28:26:18 &    4.20 &    0.39 &         61.30 &        58.00 &        61.30 &        -1.01 &        S \\           
S16 &  AT20GDP J032933-284142 &  03:29:33.6 &   -28:41:42 &    3.98 &    0.42 &         12.70 &        14.90 &        12.70 &        -0.44 &        F \\           
S17 &  AT20GDP J032941-280811 &  03:29:41.4 &   -28:08:11 &    3.16 &    0.44 &         21.90 &        23.00 &        21.90 &        -0.73 &        S \\           
S18 &  AT20GDP J032945-273216 &  03:29:45.4 &   -27:32:16 &    1.88 &    0.37 &         - &            11.10 &        11.10 &        -0.67 &        S \\           
S19 &  AT20GDP J032949-273150 &  03:29:49.3 &   -27:31:50 &    2.04 &    0.36 &         - &            18.90 &        18.90 &        -0.84 &        S \\           
S20 &  AT20GDP J033021-280207 &  03:30:21.8 &   -28:02:07 &    3.05 &    0.39 &         16.80 &        15.10 &        16.80 &        -0.64 &        S \\           
S21 &  AT20GDP J033026-281710 &  03:30:26.0 &   -28:17:10 &    3.44 &    0.35 &         3.00 &         3.60 &         3.00 &         0.05 &         I \\           
S22 &  AT20GDP J033034-282705 &  03:30:34.5 &   -28:27:05 &    2.42 &    0.34 &         17.80 &        18.60 &        17.80 &        -0.75 &        S \\           
S23 &  AT20GDP J033047-283853 &  03:30:47.8 &   -28:38:53 &    5.74 &    0.44 &         53.20 &        59.10 &        53.20 &        -0.84 &        S \\           
S24 &  AT20GDP J033051-273014 &  03:30:51.4 &   -27:30:14 &    6.99 &    0.48 &         49.10 &        43.80 &        49.10 &        -0.73 &        S \\           
S25 &  AT20GDP J033107-283147 &  03:31:07.6 &   -28:31:47 &    4.90 &    0.45 &         2.30 &         2.00 &         2.30 &         0.28 &         I \\           
S26 &  AT20GDP J033124-275208 &  03:31:24.8 &   -27:52:08 &    5.28 &    0.43 &         40.90 &        35.50 &        40.90 &        -0.77 &        S \\           
S27 &  AT20GDP J033125-281334 &  03:31:25.0 &   -28:13:34 &    2.19 &    0.36 &         3.40 &         5.70 &         3.40 &         -0.17 &        F \\        
S28 &  AT20GDP J033128-281816 &  03:31:28.0 &   -28:18:16 &    56.43 &   3.15 &         399.30 &       372.40 &       399.30 &       -0.74 &        S \\           
S29 &  AT20GDP J033130-273817 &  03:31:30.7 &   -27:38:17 &    1.86 &    0.37 &         13.80 &        10.50 &        13.80 &        -0.75 &        S \\           
S30 &  AT20GDP J033131-283153 &  03:31:31.6 &   -28:31:53 &    1.98 &    0.40 &         82.10 &        79.50 &        82.10 &        -1.40 &        S \\           
S31 &  AT20GDP J033201-281305 &  03:32:01.3 &   -28:13:05 &    1.67 &    0.32 &         11.60 &        9.60 &         11.60 &        -0.73 &        S \\           
S32 &  AT20GDP J033201-274647 &  03:32:01.5 &   -27:46:47 &    2.14 &    0.38 &         55.90 &        49.10 &        55.90 &        -1.23 &        S \\           
S33 &  AT20GDP J033206-273236 &  03:32:06.0 &   -27:32:36 &    12.30 &   0.79 &         8.10 &         9.10 &         8.10 &         0.16 &         I \\           
S34 &  AT20GDP J033208-274734 &  03:32:08.8 &   -27:47:34 &    3.08 &    0.36 &         [1.35] &       1.70 &         1.70 &         0.22 &         I \\           
S35 &  AT20GDP J033211-273725 &  03:32:11.7 &   -27:37:25 &    7.07 &    0.62 &         4.50 &         3.60 &         4.50 &         0.17 &         I \\           
S36 &  AT20GDP J033219-275408 &  03:32:19.3 &   -27:54:08 &    1.72 &    0.34 &         10.90 &        8.80 &         10.90 &        -0.69 &        S \\           
S37 &  AT20GDP J033227-274106 &  03:32:27.1 &   -27:41:06 &    2.81 &    0.46 &         22.50 &        16.60 &        22.50 &        -0.78 &        S \\           
S38 &  AT20GDP J033232-280308 &  03:32:32.0 &   -28:03:08 &    3.12 &    0.42 &         29.60 &        23.90 &        29.60 &        -0.85 &        S \\           
S39 &  AT20GDP J033242-273818 &  03:32:42.2 &   -27:38:18 &    4.64 &    0.52 &         93.30 &        72.30 &        93.30 &        -1.13 &        S \\           
S40 &  AT20GDP J033257-280212 &  03:32:57.2 &   -28:02:12 &    3.65 &    0.41 &         25.10 &        23.10 &        25.10 &        -0.73 &        S \\           
S41 &  AT20GDP J033310-274840 &  03:33:10.2 &   -27:48:40 &    3.46 &    0.38 &         24.20 &        19.30 &        24.20 &        -0.73 &        S \\           
S42 &  AT20GDP J033314-281933 &  03:33:14.8 &   -28:19:33 &    4.72 &    0.42 &         12.20 &        9.40 &         12.20 &        -0.36 &        F \\           
S43 &  AT20GDP J033333-281404 &  03:33:33.5 &   -28:14:04 &    5.56 &    0.45 &         [2.08] &       2.30 &         2.30 &         0.33 &         I \\           
S44 &  AT20GDP J033341-284016 &  03:33:41.1 &   -28:40:16 &    2.79 &    0.38 &         8.70 &         10.00 &        8.70 &         -0.43 &        F \\           
S45 &  AT20GDP J033403-282404 &  03:34:03.1 &   -28:24:04 &    2.19 &    0.41 &         18.70 &        17.80 &        18.70 &        -0.81 &        S \\           
S46 &  AT20GDP J033409-282420 &  03:34:09.2 &   -28:24:20 &    7.03 &    0.54 &         80.00 &        79.40 &        80.00 &        -0.91 &        S \\           
S47 &  AT20GDP J033411-282723 &  03:34:11.0 &   -28:27:23 &    4.50 &    0.42 &         [$<1.35$] &    1.30 &         1.30 &         0.47 &         I \\           
S48 &  AT20GDP J033413-283547 &  03:34:13.8 &   -28:35:47 &    5.25 &    0.48 &         15.00 &        13.20 &        15.00 &        -0.39 &        F \\           
S49 &  AT20GDP J033427-274402 &  03:34:27.0 &   -27:44:02 &    8.74 &    0.57 &         58.80 &        51.30 &        58.80 &        -0.72 &        S \\           
S50 &  AT20GDP J033431-282524 &  03:34:31.1 &   -28:25:24 &    2.80 &    0.38 &         42.90 &        39.20 &        42.90 &        -1.03 &        S \\  
\hline
\end{tabular}
\end{table*}

\begin{table*}								    								    	
\caption{The 20-GHz source catalogue for the 21hr field.}
\label{tab:source_catalogue_21hr}
\begin{tabular}{@{} c c c c c r r r r r r }
\hline
\multicolumn{1}{c}{(1)}
&\multicolumn{1}{c}{(2)}
&\multicolumn{1}{c}{(3)}
&\multicolumn{1}{c}{(4)}
&\multicolumn{1}{c}{(5)}
&\multicolumn{1}{c}{(6)}
&\multicolumn{1}{c}{(7)}
&\multicolumn{1}{c}{(8)}
&\multicolumn{1}{c}{(9)}
&\multicolumn{1}{c}{(10)}
&\multicolumn{1}{c}{(11)}\\
\multicolumn{1}{c}{Source}
&\multicolumn{1}{c}{Source name}
&\multicolumn{1}{c}{$\alpha$}
&\multicolumn{1}{c}{$\delta$}
&\multicolumn{1}{c}{$S_{20}$}
&\multicolumn{1}{c}{$\delta S_{20}$}
&\multicolumn{1}{c}{$S_{1.4~(\mathrm{NVSS})}$}
&\multicolumn{1}{c}{$S_{1.4~(\mathrm{ATLAS})}$}
&\multicolumn{1}{c}{$S_{1.4}$}
&\multicolumn{1}{c}{$\alpha_{1.4}^{20}$}
&\multicolumn{1}{c}{Spectral}\\
\multicolumn{1}{c}{number}
&
&
&
&\multicolumn{1}{c}{(mJy)}
&\multicolumn{1}{c}{(mJy)}
&\multicolumn{1}{c}{(mJy)}
&\multicolumn{1}{c}{(mJy)}
&\multicolumn{1}{c}{(mJy)}
&
&class\\
\hline 
S51 &  AT20GDP J213004-010239 &  21:30:04.7 &   -01:02:39 &    39.14 &   2.30 &         388.80 &       372.43 &       388.80 &       -0.86 &        S \\           
S52 &  AT20GDP J213023-000308 &  21:30:23.3 &   -00:03:08 &    2.23 &    0.41 &         38.60 &        4.43 &         38.60 &        -1.07 &        S \\           
S53 &  AT20GDP J213039-002703 &  21:30:39.5 &   -00:27:03 &    5.72 &    0.55 &         4.50 &         2.40 &         4.50 &         0.09 &         I \\           
S54 &  AT20GDP J213054-000904 &  21:30:54.6 &   -00:09:04 &    2.20 &    0.42 &         3.00 &         3.49 &         3.00 &         -0.12 &        F \\           
S55 &  AT20GDP J213055-005547 &  21:30:55.6 &   -00:55:47 &    2.25 &    0.41 &         19.30 &        14.39 &        19.30 &        -0.81 &        S \\           
S56 &  AT20GDP J213112-003929 &  21:31:12.3 &   -00:39:29 &    6.08 &    0.60 &         34.80 &        32.28 &        34.80 &        -0.66 &        S \\           
S57 &  AT20GDP J213144-004700 &  21:31:44.9 &   -00:47:00 &    1.94 &    0.38 &         12.00 &        12.22 &        12.00 &        -0.69 &        S \\           
S58 &  AT20GDP J213202-003155 &  21:32:02.5 &   -00:31:55 &    2.25 &    0.43 &         17.90 &        15.84 &        17.90 &        -0.78 &        S \\           
S59 &  AT20GDP J213250-005355 &  21:32:50.2 &   -00:53:55 &    1.70 &    0.35 &         2.30 &         3.31 &         2.30 &         -0.11 &        F \\           
S60 &  AT20GDP J213253-010815 &  21:32:53.7 &   -01:08:15 &    2.35 &    0.36 &         20.10 &        17.27 &        20.10 &        -0.81 &        S \\           
S61 &  AT20GDP J213259-004008 &  21:32:59.2 &   -00:40:08 &    6.97 &    0.54 &         28.10 &        26.16 &        28.10 &        -0.52 &        S \\           
S62 &  AT20GDP J213308-011128 &  21:33:08.1 &   -01:11:28 &    3.49 &    0.38 &         5.00 &         4.69 &         5.00 &         -0.14 &        F \\           
S63 &  AT20GDP J213336-005111 &  21:33:36.3 &   -00:51:11 &    8.18 &    0.54 &         198.90 &       199.60 &       198.90 &       -1.20 &        S \\           
S64 &  AT20GDP J213348-004152 &  21:33:48.6 &   -00:41:52 &    9.20 &    0.61 &         34.60 &        10.27 &        34.60 &        -0.50 &        F \\           
S65 &  AT20GDP J213353-004405 &  21:33:53.3 &   -00:44:05 &    5.27 &    0.50 &         4.20 &         1.64 &         4.20 &         0.09 &         I \\           
S66 &  AT20GDP J213425-003427 &  21:34:25.5 &   -00:34:27 &    2.44 &    0.38 &         [$<1.35$] &    [0.73] &       [0.73] &       0.09 &         I \\           
S67 &  AT20GDP J213453-010639 &  21:34:53.7 &   -01:06:39 &    8.24 &    0.53 &         130.50 &       120.32 &       130.50 &       -1.04 &        S \\           
S68 &  AT20GDP J213459-010849 &  21:34:59.3 &   -01:08:49 &    5.82 &    0.45 &         4.30 &         6.94 &         4.30 &         0.11 &         I \\           
S69 &  AT20GDP J213504-001312 &  21:35:04.6 &   -00:13:12 &    4.20 &    0.41 &         9.30 &         6.24 &         9.30 &         -0.30 &        F \\           
S70 &  AT20GDP J213508-002243 &  21:35:08.2 &   -00:22:43 &    2.41 &    0.40 &         18.70 &        13.48 &        18.70 &        -0.77 &        S \\     
S71 &  AT20GDP J213513-005244 &  21:35:13.3 &   -00:52:44 &    22.57 &   1.32 &         346.40 &       336.26 &       346.40 &       -1.03 &        S \\             
S72 &  AT20GDP J213514-001120 &  21:35:14.4 &   -00:11:20 &    4.87 &    0.38 &         91.60 &        79.94 &        91.60 &        -1.10 &        S \\           
S73 &  AT20GDP J213530-004211 &  21:35:30.6 &   -00:42:11 &    2.93 &    0.41 &         [$<1.35$] &    1.50 &         1.50 &         0.25 &         I \\           
S74 &  AT20GDP J213540-001428 &  21:35:40.6 &   -00:14:28 &    5.18 &    0.44 &         72.90 &        74.75 &        72.90 &        -0.99 &        S \\           
S75 &  AT20GDP J213550-011249 &  21:35:50.7 &   -01:12:49 &    65.22 &   3.30 &         156.20 &       140.51 &       156.20 &       -0.33 &        F \\           
S76 &  AT20GDP J213616-003713 &  21:36:16.5 &   -00:37:13 &    2.69 &    0.42 &         8.00 &         7.70 &         8.00 &         -0.41 &        F \\           
S77 &  AT20GDP J213616-010840 &  21:36:16.8 &   -01:08:40 &    7.62 &    0.54 &         17.10 &        16.74 &        17.10 &        -0.30 &        F \\           
S78 &  AT20GDP J213637-002652 &  21:36:37.7 &   -00:26:52 &    1.85 &    0.37 &         34.90 &        32.26 &        34.90 &        -1.10 &        S \\           
S79 &  AT20GDP J213640-004520 &  21:36:40.7 &   -00:45:20 &    1.96 &    0.39 &         3.10 &         4.19 &         3.10 &         -0.17 &        F \\           
S80 &  AT20GDP J213640-004601 &  21:36:40.8 &   -00:46:01 &    9.36 &    0.59 &         84.00 &        71.45 &        84.00 &        -0.83 &        S \\           
S81 &  AT20GDP J213644-002752 &  21:36:44.6 &   -00:27:52 &    4.92 &    0.46 &         [6.62] &       5.24 &         5.24 &         -0.02 &        F \\           
S82 &  AT20GDP J213733-005727 &  21:37:33.6 &   -00:57:27 &    9.85 &    0.58 &         124.40 &       121.48 &       124.40 &       -0.95 &        S \\           
S83 &  AT20GDP J213743-000329 &  21:37:43.8 &   -00:03:29 &    7.54 &    0.52 &         106.50 &       109.31 &       106.50 &       -1.00 &        S \\           
S84 &  AT20GDP J213746-010101 &  21:37:46.1 &   -01:01:01 &    3.63 &    0.38 &         4.00 &         6.98 &         4.00 &         -0.04 &        F \\           
S85 &  AT20GDP J213752-011124 &  21:37:52.7 &   -01:11:24 &    4.67 &    0.45 &         [2.08] &       2.48 &         2.48 &         0.24 &         I \\     
 \hline
 \end{tabular}
\end{table*}

\subsection{Estimating the completeness}\label{Estimating the completeness} 

\begin{figure}
 \includegraphics[width=0.35\textwidth,angle=270]{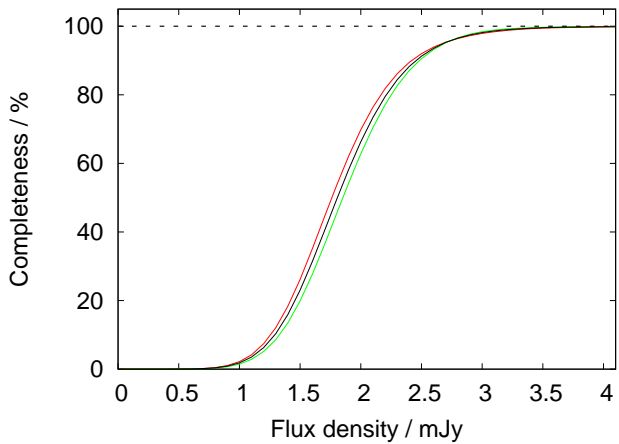}
 \caption{Completeness as a function of flux density for the 03hr field (red), 21hr field (green) and both fields (black).}
 \label{fig:completeness}
\end{figure}

Assuming that the noise has a Gaussian distribution, which is the case away from bright sources, the probability of detecting a source of peak flux density $\hat{S}$, above $5\sigma$, is given by
\begin{eqnarray}
\label{eqn:completeness}
P(\hat{S} \geq 5 \sigma) = \int^{\infty}_{5\sigma} \frac{1}{\sqrt{2\pi\sigma^{2}}}\exp{-\frac{\left(x - \hat{S}\right)^{2}}{2\sigma^{2}}}~\mathrm{d}x.
\end{eqnarray}
In practice, the noise varies over the map. This can be taken into account by averaging the probabilities of detecting the source at each pixel position given the noise map. Completeness curves were calculated in this way for regions covered by the 03hr field, 21hr field, and both the 03hr and 21hr fields. These are shown in Fig.~\ref{fig:completeness}. 

\begin{figure}
\includegraphics[width=0.35\textwidth,angle=270]{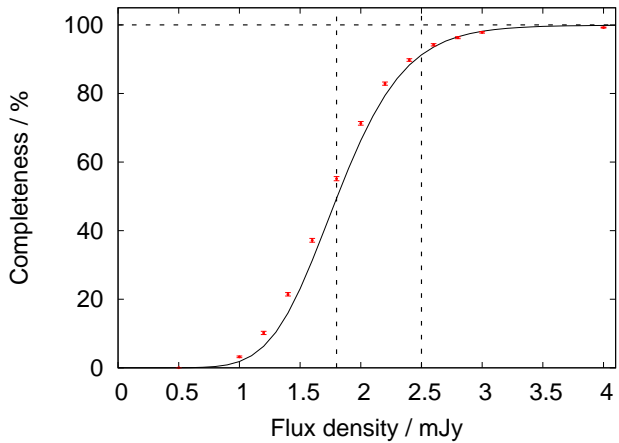}
\caption{Results of a simulation to investigate the completeness of the survey. The filled circles show the proportion of the simulated sources recovered as a function of flux density. The solid line shows the completeness predicted based on the noise-map pixel values, assuming Gaussian statistics for the noise. The completeness is estimated to be 50\% at 1.8~mJy and 90\% at 2.5~mJy, as indicated by the dashed vertical lines.}
\label{fig:completeness_vs_sim}
\end{figure}

Simulations were carried out in order to verify that, assuming Gaussian statistics, the noise maps can be used to provide reasonable estimates of the survey completeness. The completeness of the survey was investigated by inserting 100 point sources with flux densities ranging from 0.5 to 4.0~mJy at random positions into the 03hr and 21hr fields and by measuring the proportion of simulated sources recovered as a function of flux density. Sources were simulated in the image plane using the \textsc{miriad} task \textsc{imgen}. The simulation was repeated 50 times to build up enough statistics. Fig.~\ref{fig:completeness_vs_sim} shows the mean proportion of simulated sources recovered as a function of flux density. The error bars are standard errors of the means. The predicted completeness curve for the two fields is also shown for comparison.

The results obtained from the Monte Carlo analysis are broadly in agreement with the expected detection rate, although for sources with $S \lesssim 2.5~\mathrm{mJy}$, the detection rate is somewhat higher than expected. This probably arises, in large part, from source confusion. For example, two faint sources below the detection threshold may lie sufficiently close to one another to be detected as a single source above the detection threshold. The completeness is estimated to be 50\% at 1.8~mJy and 90\% at 2.5~mJy.

The simulations were also used to gauge the positional accuracy of the sources at low SNR. For a source with a SNR of 5, the positional errors in RA and Dec. are 2.4 and 1.7~arcsec, respectively.

The number of sources with $S > 2.5$~mJy in the 03hr and 21hr fields, which cover a total area of $4.98~\mathrm{deg}^{2}$, is 60. Given that the survey is 90\% complete above 2.5~mJy, the number of sources per square degree with $S_{20} > 2.5$~mJy is $\frac{60}{0.9 \times 4.98} = 13.4$. The Poissonian error on this figure is $\frac{\sqrt{60}}{0.9 \times 4.98} = 1.7$.

\section{Correlation with NVSS, ATLAS and FIRST at 1.4~GH\lowercase{z}}\label{Correlation with NVSS, ATLAS and FIRST at 1.4 GHz}

We matched the source catalogues for both fields with the NVSS catalogue. In addition, we matched the source catalogue for the 03hr field with the ATLAS catalogue from the first data release \citep{norris2006} and for the 21hr field with the FIRST catalogue \citep[version 03Apr11;][]{becker2003}. The resulting spectral indices are listed in Tables~\ref{tab:source_catalogue_03hr} and~\ref{tab:source_catalogue_21hr}.

We used a search radius of 40~arcsec when matching our 20-GHz sources with the NVSS catalogue. We inspected NVSS images of the sources to verify that the automated matching procedure had worked well. In a few cases where the matching was not straightforward, the sources were matched manually -- higher-resolution images from FIRST and ATLAS were used to assist in the matching. If it was evident that a source had more than one counterpart in NVSS, the flux densities of the NVSS sources were summed for comparison with the 20-GHz flux density.  

We used the same search radius when matching our source catalogues with the FIRST and ATLAS catalogues. Images from FIRST and ATLAS were inspected to verify the source matching and identify cases where a 20-GHz source had more than one low-frequency counterpart.

Two 20-GHz sources (S18 and S19) are confused in NVSS so it was not possible to obtain spectral indices for these sources using NVSS. No matches were found in the NVSS catalogue for 11 of our sources. We have measured the flux densities of these sources by extracting the pixel values in the NVSS image at the positions at which the sources were detected at 20~GHz. The flux density of one source (S81) was measured to be as high as 6.6~mJy, despite the fact that it is not in the NVSS catalogue. This is because the source lies too close to a brighter source to be detected separately; the FIRST image suggests that the two sources are unrelated. Seven sources lie below 1.35~mJy, which is three times the typical noise in the NVSS image; we have quoted an upper limit on the flux densities of these sources equal to 1.35 mJy. 

One source (S14) in the 03hr field is not present in the ATLAS catalogue, nor does it appear in the ATLAS image. We measured the noise close to the source to be 75~$\mu \mathrm{Jy}$. We have quoted an upper limit on the flux density of the source equal to three times the noise (0.23 mJy).

Only one source (S66) in the 21hr field is not present in the FIRST catalogue. The flux density of this source was extracted from the FIRST image. The source is just below five times the noise in the FIRST image.

The resolution (45~arcsec) of NVSS is well matched to that ($\approx 30$~arcsec) of our 20-GHz survey. When measuring spectral indices between 1.4 and 20~GHz, we therefore chose to use NVSS flux densities for all sources present in the NVSS catalogue. For the remaining sources in the 03hr/21hr field, ATLAS/FIRST flux densities were used. FIRST has a resolution of 5~arcsec and ATLAS a resolution of 11~arcsec. Spectral indices obtained using our `best' estimates of the 1.4-GHz flux densities are listed in Tables~\ref{tab:source_catalogue_03hr} and~\ref{tab:source_catalogue_21hr}. 

\begin{figure}
\includegraphics[width=0.35\textwidth,angle=270]{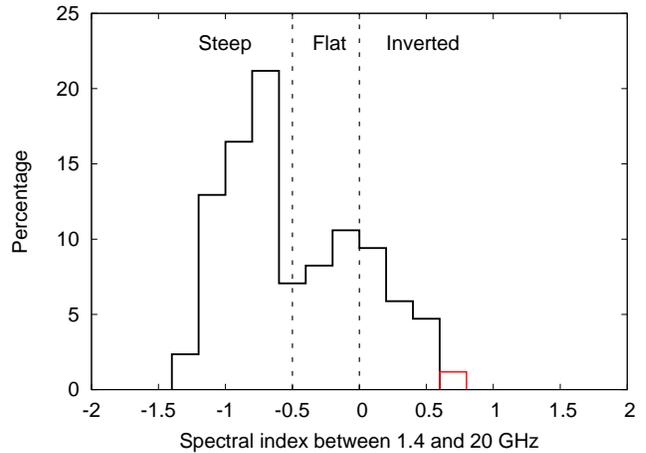}
\caption{Histogram of the spectral index distribution $\alpha_{1.4}^{20}$ for the 85 sources in our sample. One source has a lower limit on its spectral index; this source is shown in red.}
\label{fig:alpha_dist}
\end{figure}

The spectral index distribution of our source sample is shown in Fig.~\ref{fig:alpha_dist}. We used the spectral indices between 1.4 and 20~GHz to classify the sources as follows:
\begin{itemize}
\item Steep (S): $\alpha_{1.4}^{20} < -0.5$
\item Flat (F): $-0.5 \leq \alpha_{1.4}^{20} < 0$
\item Inverted (I): $\alpha_{1.4}^{20} \geq 0$
\end{itemize}
We find that 55\% of the sources have steep spectra, 24\% have flat spectra and 21\% have inverted spectra.

Some sources in our sample, particularly those that have flat or inverted spectra, could be blazars (the beamed products of FRI and FRII sources) with synchrotron self-absorbed spectra; the self-absorbed components of such sources often display flux density variations. A number of years separate the measurements at 1.4~GHz with those at 20~GHz. The spectral indices may therefore be affected by source variability. We note, however, that variability is not expected to play a significant role at 1.4~GHz given that the emission will mainly originate from the stable, extended, steep-spectrum component of the source. An assessment of the 20-GHz variability of our source sample, on a 3-yr timescale, is made in Section~\ref{Flux density variability at 20 GHz}.

\section{Multi-frequency follow-up observations and data reduction}\label{Multi-frequency follow-up observations and data reduction}

\subsection{Observations}\label{Observations2}

Each of the 85 sources detected in the 20-GHz survey was followed up at three frequencies (5.5, 9 and 18~GHz) with ATCA in September 2012. The sources were observed simultaneously at 5.5 and 9~GHz, with one 2-GHz CABB band centred at 5.5~GHz and the other at 9~GHz. The 18-GHz observations were carried out with two 2-GHz CABB bands centred at 17 and 19~GHz. 

We used the H214 array to obtain both good snapshot imaging data at 18~GHz and reasonable angular resolution at 5.5~GHz. This array configuration is somewhat larger than the H75 array, which was used for the survey observations, with baselines ranging between 82 and 247~m (excluding baselines to antenna 6).

Each source was observed twice (at different hour angles) at each frequency to improve the \textit{uv} coverage. The integration time per source was set so as to reach a SNR per baseline $\gtrsim 10$ at 5.5, 9 and 18~GHz in each \textit{uv} cut (the flux densities of the sources at the three frequencies were estimated using their spectra between 1.4 and 20~GHz as measured in Section~\ref{Correlation with NVSS, ATLAS and FIRST at 1.4 GHz}). The integration time per source per \textit{uv} cut at 5.5 and 9~GHz ranges between 1 and 17~min and at 18~GHz between 1 and 9~min.

At 5.5 and 9~GHz, both PKS B1934-638 (S = 4.965~Jy at 5.500~GHz; S = 2.701~Jy at 9.000~GHz) and PKS B0823-500 (S = 2.622~Jy at 5.500~GHz; S = 1.360~Jy at 9.000~GHz) were used as primary and bandpass calibrators. A secondary calibrator was observed every $\approx 30$ minutes. For the 03hr field, PKS B0346-279 and PKS B0327-241 were used as secondary calibrators and for the 21hr field, PKS B2134+004. At 18~GHz, PKS B1934-638 (S = 1.101~Jy at 18.000~GHz) was used as primary calibrator and PKS B1921-293 as bandpass calibrator. A secondary calibrator was observed every $\approx 20$ minutes and the pointing was calibrated on every third visit to the secondary calibrator. For the 03hr field, PKS B0346-279 was used as secondary calibrator and for the 21hr field, PKS B2134+004. 

\subsection{Data reduction}\label{Data reduction}

The data were flagged and calibrated following the same procedure as outlined for the survey data in Section~\ref{Calibration and imaging}. An image of each source at each frequency was made in \textsc{miriad}. Natural weighting was used to maximize the SNR. Each map was CLEANed using the \textsc{mfclean} task which models the spectral variation of the source emission using a two-term Taylor-Polynomial. One round of phase self-calibration was applied if the SNR was sufficiently high -- typically sources $> 4$~mJy -- to improve the dynamic range. A 2D elliptical Gaussian fitted to the central region of the dirty beam was used as the restoring beam. The angular resolution is $\approx 40$~arcsec at 5.5~GHz, $\approx 25$~arcsec at 9~GHz and $\approx 10$~arcsec at 18~GHz.

Integrated flux densities were measured by fitting elliptical 2D Gaussians to the sources. All 85 sources (including S14 which was not detected in ATLAS DR1) detected in the 20-GHz survey were confirmed to be real, demonstrating that despite the relatively poor \textit{uv} coverage of the survey data, our imaging and source finding techniques can provide us with a source catalogue which is close to 100\% reliable above 5$\sigma$. Source S14, which has a highly unusual spectrum, is discussed further in Section~\ref{Spectral curvature between 1.4 and 18 GHz}.

The flux densities of the sources at 5.5, 9 and 18~GHz for the 03hr and 21hr fields are listed in Tables~\ref{tab:source_catalogue_followup_03hr} and~\ref{tab:source_catalogue_followup_21hr} respectively. We were not able to measure $S_{5.5}$ and $S_{9}$ for S79 and S80 because the two sources are blended together in the two lower-frequency images. The columns are as follows:

\indent Column (1) - Source number.\\
\indent Column (2) - Flux density taken from the ATLAS 2.3-GHz source catalogue \citep{zinn2012} of the CDFS field, $S_{2.3}$ (in mJy).\\
\indent Column(3) - Uncertainty on the 2.3-GHz flux density, $\delta S_{2.3}$ (in mJy).\\
\indent Columns (4), (6) and (8) - Integrated flux densities at 5.5~GHz ($S_{5.5}$), 9~GHz ($S_{9}$) and 18~GHz ($S_{18}$), in mJy. In the case of a multi-component source, the sum of the integrated flux densities of the components is quoted.\\
\indent Columns (5), (7) and (9) - Uncertainties on the integrated flux densities at 5.5~GHz ($\delta S_{5.5}$), 9~GHz ($\delta S_{9}$) and 18~GHz ($\delta S_{18}$), in mJy. This is 5\% of the integrated flux density added in quadrature with the error on the integrated flux density reported by the Gaussian-fitting task (which accounts for the local noise). For a multi-component source, the sum of the uncertainties on the integrated flux densities of all components included in the source is quoted.\\
\indent Column (10) - The source class, variable (VAR) or non-variable (NVAR), determined as described in Section~\ref{Flux density variability at 20 GHz}.\\
\indent Column (11) - The variability index, $V$ (in \%), measured as described in Section~\ref{Flux density variability at 20 GHz}.\\
\indent Column (12) - Peak frequency, $\nu_{\mathrm{p}}$ (in GHz), measured as described in Section~\ref{Spectral curvature between 1.4 and 18 GHz}.\\

\begin{table*}
\caption{Data table for the 03hr field.}
\label{tab:source_catalogue_followup_03hr}
\begin{tabular}{@{} r r r r r r r r r r r r } 
\hline
\multicolumn{1}{c}{(1)}
&\multicolumn{1}{c}{(2)}
&\multicolumn{1}{c}{(3)}
&\multicolumn{1}{c}{(4)}
&\multicolumn{1}{c}{(5)}
&\multicolumn{1}{c}{(6)}
&\multicolumn{1}{c}{(7)}
&\multicolumn{1}{c}{(8)}
&\multicolumn{1}{c}{(9)}
&\multicolumn{1}{c}{(10)}
&\multicolumn{1}{c}{(11)}
&\multicolumn{1}{c}{(12)}\\
\multicolumn{1}{c}{Source}
&\multicolumn{1}{c}{$S_{2.3}$}
&\multicolumn{1}{c}{$\delta S_{2.3}$}
&\multicolumn{1}{c}{$S_{5.5}$}
&\multicolumn{1}{c}{$\delta S_{5.5}$}
&\multicolumn{1}{c}{$S_{9}$}
&\multicolumn{1}{c}{$\delta S_{9}$}
&\multicolumn{1}{c}{$S_{18}$}
&\multicolumn{1}{c}{$\delta S_{18}$}
&\multicolumn{1}{c}{Class}
&\multicolumn{1}{c}{$V$}
&\multicolumn{1}{c}{$\nu_{\mathrm{p}}$}\\
\multicolumn{1}{c}{number}
&\multicolumn{1}{c}{(mJy)}
&\multicolumn{1}{c}{(mJy)}
&\multicolumn{1}{c}{(mJy)}
&\multicolumn{1}{c}{(mJy)}
&\multicolumn{1}{c}{(mJy)}
&\multicolumn{1}{c}{(mJy)}
&\multicolumn{1}{c}{(mJy)}
&\multicolumn{1}{c}{(mJy)}
&
&\multicolumn{1}{c}{(\%)}
&\multicolumn{1}{c}{(GHz)}\\
\hline
S01 & - & - & 28.63 & 1.44 & 16.30 & 0.84 & 8.26 & 0.45 & NVAR & 2.5 & - \\
S02 & - & - & 31.29 & 1.57 & 16.73 & 0.87 & 8.54 & 0.48 & NVAR & 6.9 & - \\
S03 & 1.67 & 0.11 & 1.74 & 0.16 & 2.66 & 0.18 & 2.32 & 0.15 & NVAR & -8.3 & - \\
S04 & - & - & 37.04 & 1.87 & 34.16 & 1.72 & 28.45 & 1.43 & VAR & 16.7 & 4.3 \\
S05 & - & - & 23.23 & 1.17 & 21.23 & 1.08 & 12.76 & 0.65 & VAR & 38.4 & 6.0 \\
S06 & - & - & 13.59 & 0.72 & 12.20 & 0.65 & 9.40 & 0.52 & NVAR & -3.9 & - \\
S07 & - & - & 14.02 & 0.99 & 8.76 & 0.78 & 3.74 & 0.65 & NVAR & -9.2 & - \\
S08 & 1.57 & 0.11 & 2.78 & 0.23 & 2.52 & 0.25 & 2.23 & 0.26 & VAR & 38.0 & 9.2 \\
S09 & - & - & 2.41 & 0.17 & 2.01 & 0.16 & 2.13 & 0.14 & NVAR & -8.5 & - \\
S10 & 0.34 & 0.07 & 1.32 & 0.20 & 1.51 & 0.13 & 1.67 & 0.12 & NVAR & 6.1 & - \\
S11 & 6.74 & 0.34 & 4.55 & 0.25 & 3.38 & 0.20 & 2.07 & 0.12 & NVAR & -8.3 & - \\
S12 & - & - & 87.63 & 4.38 & 72.06 & 3.62 & 58.99 & 2.95 & VAR & 17.7 & - \\
S13 & - & - & 481.45 & 24.09 & 285.94 & 14.43 & 159.93 & 8.02 & NVAR & 4.8 & - \\
S14 & - & - & 1.01 & 0.15 & 1.55 & 0.21 & 1.85 & 0.13 & NVAR & -9.8 & - \\
S15 & 35.94 & 1.80 & 16.85 & 0.87 & 8.96 & 0.50 & 4.51 & 0.31 & NVAR & -5.9 & - \\
S16 & 10.07 & 0.51 & 5.41 & 0.32 & 4.29 & 0.29 & 2.93 & 0.23 & NVAR & 16.4 & - \\
S17 & 16.72 & 0.84 & 8.70 & 0.46 & 5.20 & 0.33 & 3.06 & 0.21 & NVAR & -6.0 & - \\
S18 & 7.62 & 0.54 & 4.09 & 0.25 & 3.03 & 0.27 & 1.76 & 0.15 & NVAR & -8.7 & - \\
S19 & - & - & 6.80 & 0.37 & 2.48 & 0.29 & 1.67 & 0.24 & NVAR & 4.4 & - \\
S20 & 12.47 & 0.63 & 7.44 & 0.39 & 4.82 & 0.29 & 2.80 & 0.19 & NVAR & 3.2 & - \\
S21 & 3.33 & 0.19 & 3.89 & 0.23 & 3.49 & 0.22 & 2.81 & 0.20 & NVAR & 9.6 & 5.4 \\
S22 & - & - & 6.33 & 0.34 & 3.83 & 0.25 & 2.18 & 0.16 & NVAR & 4.0 & - \\
S23 & - & - & 19.78 & 1.00 & 10.90 & 0.61 & 6.14 & 0.42 & NVAR & -5.4 & - \\
S24 & 35.80 & 1.79 & 20.61 & 1.04 & 12.39 & 0.64 & 7.30 & 0.40 & NVAR & -4.3 & - \\
S25 & 3.28 & 0.18 & 7.73 & 0.41 & 8.50 & 0.45 & 6.60 & 0.36 & NVAR & 11.7 & 9.8 \\
S26 & - & - & 12.06 & 0.62 & 7.08 & 0.40 & 4.73 & 0.29 & NVAR & 6.5 & - \\
S27 & - & - & 4.08 & 0.27 & 3.72 & 0.29 & 2.34 & 0.15 & NVAR & -9.1 & - \\
S28 & - & - & 137.94 & 6.96 & 71.16 & 3.78 & 30.71 & 2.98 & VAR & 34.8 & - \\
S29 & - & - & 4.37 & 0.26 & 2.54 & 0.20 & 1.12 & 0.10 & NVAR & 27.6 & - \\
S30 & - & - & 17.33 & 0.94 & 7.54 & 0.61 & 3.19 & 0.32 & NVAR & 13.2 & - \\
S31 & - & - & 3.90 & 0.25 & 2.18 & 0.18 & 1.11 & 0.13 & NVAR & 21.2 & - \\
S32 & - & - & 11.02 & 0.58 & 5.87 & 0.36 & 2.28 & 0.14 & NVAR & -9.0 & - \\
S33 & 11.28 & 0.57 & 13.12 & 0.68 & 13.20 & 0.69 & 14.36 & 0.73 & NVAR & 7.1 & - \\
S34 & - & - & 3.74 & 0.22 & 5.51 & 0.34 & 4.49 & 0.27 & VAR & 16.0 & 11.5 \\
S35 & 3.86 & 0.21 & 12.56 & 0.66 & 17.86 & 0.94 & 20.22 & 1.02 & VAR & 48.7 & - \\
S36 & - & - & 3.83 & 0.22 & 2.21 & 0.16 & 1.25 & 0.11 & NVAR & 15.8 & - \\
S37 & - & - & 7.60 & 0.42 & 5.17 & 0.33 & 3.39 & 0.20 & NVAR & -5.9 & - \\
S38 & - & - & 10.59 & 0.55 & 6.59 & 0.41 & 3.83 & 0.27 & NVAR & -4.6 & - \\
S39 & - & - & 23.67 & 1.20 & 12.01 & 0.65 & 6.00 & 0.60 & NVAR & -2.9 & - \\
S40 & - & - & 10.66 & 0.55 & 6.68 & 0.38 & 4.00 & 0.28 & NVAR & -6.9 & - \\
S41 & - & - & 9.65 & 0.51 & 6.18 & 0.35 & 3.29 & 0.21 & NVAR & 2.6 & - \\
S42 & - & - & 8.89 & 0.47 & 6.10 & 0.36 & 3.84 & 0.27 & NVAR & 12.3 & - \\
S43 & 3.75 & 0.20 & 10.24 & 0.53 & 9.41 & 0.49 & 7.91 & 0.44 & VAR & 15.4 & 9.3 \\
S44 & - & - & 4.72 & 0.27 & 3.78 & 0.27 & 2.62 & 0.20 & NVAR & -5.8 & - \\
S45 & - & - & 3.50 & 0.24 & 2.18 & 0.22 & 1.82 & 0.27 & NVAR & -7.3 & - \\
S46 & 52.29 & 2.62 & 25.98 & 1.32 & 14.92 & 0.78 & 7.11 & 0.41 & NVAR & -0.6 & - \\
S47 & - & - & 10.10 & 0.53 & 6.95 & 0.38 & 3.69 & 0.28 & NVAR & 13.1 & 4.4 \\
S48 & 9.44 & 0.48 & 6.90 & 0.39 & 5.00 & 0.32 & 4.30 & 0.29 & NVAR & 9.2 & - \\
S49 & 41.80 & 2.09 & 24.23 & 1.22 & 15.09 & 0.78 & 9.70 & 0.53 & NVAR & -4.2 & - \\
S50 & - & - & 11.12 & 0.59 & 6.44 & 0.43 & 2.77 & 0.20 & NVAR & -4.3 & - \\
\hline
\end{tabular}
\end{table*}

\begin{table*}
\caption{Data table for the 21hr field.}
\label{tab:source_catalogue_followup_21hr}
\begin{tabular}{@{} r r r r r r r r r r r r } 
\hline
\multicolumn{1}{c}{(1)}
&\multicolumn{1}{c}{(2)}
&\multicolumn{1}{c}{(3)}
&\multicolumn{1}{c}{(4)}
&\multicolumn{1}{c}{(5)}
&\multicolumn{1}{c}{(6)}
&\multicolumn{1}{c}{(7)}
&\multicolumn{1}{c}{(8)}
&\multicolumn{1}{c}{(9)}
&\multicolumn{1}{c}{(10)}
&\multicolumn{1}{c}{(11)}
&\multicolumn{1}{c}{(12)}\\
\multicolumn{1}{c}{Source}
&\multicolumn{1}{c}{$S_{2.3}$}
&\multicolumn{1}{c}{$\delta S_{2.3}$}
&\multicolumn{1}{c}{$S_{5.5}$}
&\multicolumn{1}{c}{$\delta S_{5.5}$}
&\multicolumn{1}{c}{$S_{9}$}
&\multicolumn{1}{c}{$\delta S_{9}$}
&\multicolumn{1}{c}{$S_{18}$}
&\multicolumn{1}{c}{$\delta S_{18}$}
&\multicolumn{1}{c}{Class}
&\multicolumn{1}{c}{$V$}
&\multicolumn{1}{c}{$\nu_{\mathrm{p}}$}\\
\multicolumn{1}{c}{number}
&\multicolumn{1}{c}{(mJy)}
&\multicolumn{1}{c}{(mJy)}
&\multicolumn{1}{c}{(mJy)}
&\multicolumn{1}{c}{(mJy)}
&\multicolumn{1}{c}{(mJy)}
&\multicolumn{1}{c}{(mJy)}
&\multicolumn{1}{c}{(mJy)}
&\multicolumn{1}{c}{(mJy)}
&
&\multicolumn{1}{c}{(\%)}
&\multicolumn{1}{c}{(GHz)}\\
\hline
S51 & - & - & 123.74 & 6.27 & 56.12 & 2.99 & 33.93 & 1.99 & NVAR & 10.0 & - \\
S52 & - & - & 7.49 & 0.42 & 3.48 & 0.42 & 1.96 & 0.38 & NVAR & -9.3 & - \\
S53 & - & - & 4.52 & 0.35 & 4.13 & 0.34 & 3.84 & 0.31 & VAR & 19.0 & - \\
S54 & - & - & 4.08 & 0.23 & 2.82 & 0.18 & 1.54 & 0.14 & NVAR & 18.1 & 3.1 \\
S55 & - & - & 6.28 & 0.41 & 3.35 & 0.48 & 1.52 & 0.27 & NVAR & 20.9 & - \\
S56 & - & - & 15.41 & 0.79 & 10.40 & 0.58 & 6.37 & 0.46 & NVAR & -6.3 & - \\
S57 & - & - & 5.13 & 0.28 & 3.41 & 0.25 & 2.00 & 0.18 & NVAR & -11.0 & - \\
S58 & - & - & 7.45 & 0.39 & 4.68 & 0.31 & 2.61 & 0.24 & NVAR & -10.4 & - \\
S59 & - & - & 2.56 & 0.17 & 2.44 & 0.15 & 1.72 & 0.19 & NVAR & -12.1 & - \\
S60 & - & - & 6.95 & 0.41 & 4.75 & 0.35 & 2.79 & 0.50 & NVAR & -12.2 & - \\
S61 & - & - & 12.46 & 0.66 & 9.06 & 0.58 & 6.50 & 0.40 & NVAR & 2.7 & - \\
S62 & - & - & 4.42 & 0.30 & 3.75 & 0.28 & 3.06 & 0.22 & NVAR & 4.2 & - \\
S63 & - & - & 44.46 & 2.23 & 23.90 & 1.24 & 10.04 & 0.54 & NVAR & -2.7 & - \\
S64 & - & - & 13.02 & 0.80 & 8.22 & 0.63 & 7.89 & 0.48 & NVAR & 6.4 & - \\
S65 & - & - & 6.45 & 0.41 & 5.50 & 0.41 & 4.68 & 0.34 & NVAR & 3.0 & 3.5 \\
S66 & - & - & 4.98 & 0.27 & 3.67 & 0.21 & 2.43 & 0.17 & NVAR & -8.3 & 4.4 \\
S67 & - & - & 38.90 & 1.98 & 21.27 & 1.22 & 9.92 & 0.86 & NVAR & -5.0 & - \\
S68 & - & - & 6.78 & 0.45 & 7.08 & 0.60 & 9.14 & 0.52 & VAR & 23.5 & - \\
S69 & - & - & 10.43 & 0.65 & 7.74 & 0.55 & 3.73 & 0.24 & NVAR & 9.5 & - \\
S70 & - & - & 6.45 & 0.36 & 3.80 & 0.27 & 2.38 & 0.17 & NVAR & -8.6 & - \\
S71 & - & - & 93.68 & 4.75 & 52.12 & 2.74 & 24.17 & 1.43 & NVAR & -3.8 & - \\
S72 & - & - & 22.80 & 1.18 & 11.35 & 0.70 & 5.50 & 0.38 & NVAR & -5.6 & - \\
S73 & - & - & 3.29 & 0.20 & 2.89 & 0.20 & 2.16 & 0.19 & NVAR & 14.7 & 4.7 \\
S74 & - & - & 20.42 & 1.05 & 12.49 & 0.68 & 5.74 & 0.39 & NVAR & -5.8 & - \\
S75 & - & - & 121.85 & 6.10 & 94.65 & 4.74 & 60.34 & 3.02 & NVAR & 6.2 & - \\
S76 & - & - & 4.37 & 0.26 & 3.66 & 0.28 & 2.97 & 0.19 & NVAR & -7.8 & - \\
S77 & - & - & 8.14 & 0.47 & 8.17 & 0.50 & 7.56 & 0.47 & NVAR & -5.0 & - \\
S78 & - & - & 8.95 & 0.53 & 3.00 & 0.37 & 1.83 & 0.24 & NVAR & -11.9 & - \\
S79 & - & - & - & - & - & - & 1.20 & 0.20 & NVAR & 19.8 & - \\
S80 & - & - & - & - & 16.56 & 0.89 & 10.97 & 0.68 & NVAR & -0.5 & - \\
S81 & - & - & 9.47 & 0.55 & 8.14 & 0.54 & 6.17 & 0.41 & NVAR & 7.0 & 4.1 \\
S82 & - & - & 37.74 & 1.89 & 22.77 & 1.17 & 11.22 & 0.60 & NVAR & -4.1 & - \\
S83 & - & - & 30.50 & 1.54 & 18.26 & 0.95 & 8.93 & 0.50 & NVAR & -3.7 & - \\
S84 & - & - & 2.83 & 0.19 & 2.50 & 0.22 & 2.82 & 0.22 & NVAR & 9.3 & - \\
S85 & - & - & 4.89 & 0.32 & 4.89 & 0.30 & 4.72 & 0.32 & NVAR & -6.2 & - \\
\hline
\end{tabular}
\end{table*}

\subsubsection{Measuring spectral indices across the individual ATCA bands}\label{Measuring spectral indices across the individual ATCA bands}

The model produced by \textsc{mfclean} contains two planes: the intensity ($I$) and intensity times spectral index ($I \times \alpha$) planes. The model can therefore be used to measure spectral indices across the ATCA bands. The `mfs' option in the \textsc{restor} task was used to write a second plane in the output image containing the ($I \times \alpha$) model convolved with the Gaussian beam. Spectral indices across the bands were then extracted from the $I$ and ($I \times \alpha$) images following the method described in \cite{whiting2012}: each component has its spectral index measured by taking the component fitted to the $I$ image, keeping all parameters fixed except the peak flux density, and fitting it to the $I \times \alpha$ image. The ratio of the integrated flux densities of this and the original component provides the spectral index for the component. For sources consisting of more than one component, the total flux densities of the components in the $I$ and $I \times \alpha$ planes were evaluated before perfoming the division.

Spectral indices were measured in this way for all sources, at each frequency. In most cases, however, the SNR was not high enough to obtain reliable measurements. Given the 2-GHz bandwidth available at each frequency, the error on the spectral index due to the thermal noise is expected to be $\sim 0.2$ at SNR = 50 at 5.5~GHz, and at SNR = 100 at 9 and 18~GHz. The measurements therefore only provide useful additional spectral information for the brightest sources in our sample.

\section{Flux density variability at 20~GH\lowercase{z}}\label{Flux density variability at 20 GHz}

\subsection{Previous work}\label{Previous work}

There have been some systematic studies of variability in high-frequency-selected samples, particularly at high flux densities. \cite{chen2013} analysed the variability of extragalactic radio sources selected from the \textit{Planck} Early Release Compact Source Catalogue over a period of 10 years at frequencies between 33 and 94~GHz; they used measurements from both the \textit{Wilkinson} Microwave Anisotropy Probe and \textit{Planck}. 50\% of sources with good quality measurements were found to be variable (at $> 99\%$ confidence level) at 33~GHz; the median fractional variation of the variable sources was 27\%. The general level of variability was found not to change significantly with frequency. Almost all the sources that showed variability were blazars. They compared the variability of the two classes of blazars, flat-spectrum radio quasars (FSRQs) and BL-Lac objects, but found no evidence that one class was more variable than the other.

\cite{franzen2009} measured variability at 16 and 33~GHz in a complete sample of 97 sources with $S_{33~\mathrm{GHz}}\gtrsim 1$~Jy. They found that, over timescales of about 1.5 yr, 15\% of the sources varied by more than 20\% at 16~GHz and 20\% varied by more than 20\% at 33~GHz. The proportion of flat-spectrum sources in the full sample was very high: 93\% had $\alpha_{14}^{34} > -0.5$. The median spectral index ($0.06 \pm 0.05$) of the variable sources was marginally flatter than that ($-0.13 \pm 0.04$) of the non-variable sources. 

\cite{sadler2006} studied the variability at 20~GHz of 108 sources from a complete sample with $S_{20~\mathrm{GHz}} > 100$~mJy. Over time-scales of 1--2 years, 42\% of the sources varied by more than 10\%. Most of the sources had significant spectral curvature over the range 0.8--20~GHz. They found no significant correlation between variability and spectral index. 

At lower flux densities, \cite{bolton2006} measured the variability of 51 sources from a complete sample with $S_{15~\mathrm{GHz}} > 25$~mJy over a 3-yr period. They divided their sample into four spectral classes: (1) gigahertz-peaked spectrum (GPS) sources with convex spectra peaking below 5~GHz; (2) high-frequency peaked (HFP) sources with convex spectra peaking above 5~GHz; (3) flat-spectrum sources with spectral indices at 10~GHz, $\alpha_{10}$, greater than -0.5 and (4) steep-spectrum sources with $\alpha_{10} \leq -0.5$. None of the GPS and steep-spectrum sources were found to be have varied above the flux density calibration errors of $\sim 6$\%; half of the flat-spectrum sources and the majority of the HFP sources were found to have varied significantly. The median fractional variation of the variable sources was 14\%. 

\subsection{Our work}\label{Our work}

Here, we look at variability, on a timescale of $\approx 3$ yr, in an even fainter sample than that studied by \citeauthor{bolton2006}, complete to 2.5~mJy at 20~GHz. The 18-GHz flux densities from the follow-up observations were extrapolated to 20~GHz, using $\alpha_{9}^{18}$, and then compared with the 20-GHz survey flux densities. The survey observations were carried out in July 2009 and the follow-up observations in September 2012.

\begin{figure}
\includegraphics[width=0.43\textwidth,angle=270, trim=0cm 2.5cm 0cm 0cm]{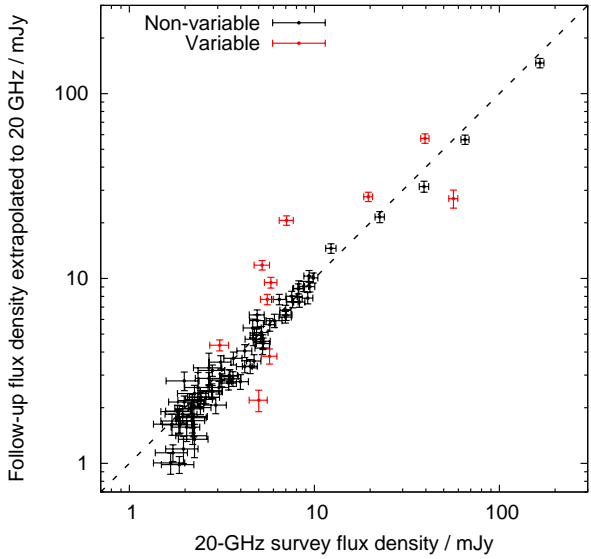}
\caption{Comparison of follow-up flux densities extrapolated to 20~GHz with 20-GHz survey flux densities. The dashed line indicates equal flux-density values. Sources classified as variable are shown in red and non-variable in black.}
\label{fig:variability_SS}
\end{figure}

Fig.~\ref{fig:variability_SS} shows how the 20-GHz flux densities of all sources compare at the two epochs. There is generally good agreement between the flux densities. At the faint end, the survey flux densities tend to be systematically higher than the follow-up flux densities. This is probably due to the Eddington bias \citep{eddington1913} causing the measured flux densities in the survey to be biased high close to the detection limit of the survey.

We have defined each source as variable (VAR) or non-variable (NVAR) using a $\chi^{2}$-test. Our null hypothesis is that the flux density of the source is the same at the two epochs of observation. The value of $\chi^{2}$ is given by
\begin{eqnarray}
\label{equation:chi_squared}
\chi^{2} =  \frac{(S_{1}-S_{2})^{2}}{\sigma^2_{1}+\sigma^2_{2}}
\mathrm{,}
\end{eqnarray}
where $S_{1}$ and $S_{2}$ are the flux densities at the two epochs, and $\sigma_{1}$ and $\sigma_{2}$ their respective uncertainties. When the probability that the data support the null hypothesis is less than 1\% the source is classified as variable (VAR), otherwise the source is classified as non-variable (NVAR). Tables~\ref{tab:source_catalogue_followup_03hr} and~\ref{tab:source_catalogue_followup_21hr} include a column specifying the classification for each source. Of the 85 sources, 10 (12 per cent) are classified as variable, seven of which have higher flux densities in the follow-up observations; these sources are highlighted in red in Fig.~\ref{fig:variability_SS}.

We have also quantified the variability for each source while applying a correction for the errors in the flux density measurements. Following~\cite{franzen2009}, we define the unbiased variability index as
\begin{eqnarray}
\label{equation:var_index}
V = \frac{1}{2 \bar{S}} 
\sqrt{(S_{1} - S_{2})^{2} 
- (\sigma^{2}_{1} + \sigma^{2}_{2})} 
\mathrm{,}
\end{eqnarray}
where $\bar{S}$ is the mean flux density. When the value inside the square root in Eqn.~\ref{equation:var_index} becomes negative, we set the unbiased variability index to be negative; $V$ is then given by
\begin{eqnarray}
\label{equation:var_index2}
V = - \frac{1}{2 \bar{S}} 
\sqrt{\vert{(S_{1} - S_{2})^{2} 
- (\sigma^{2}_{1} + \sigma^{2}_{2})}\vert} 
\mathrm{.}
\end{eqnarray}
We note that the minimum variability index required for a source to be classified as variable according to the $\chi^{2}$-test will depend upon the fractional errors on the flux densities for the source, which explains why a source classified as variable in Tables~\ref{tab:source_catalogue_followup_03hr} and~\ref{tab:source_catalogue_followup_21hr} may have a lower variability index than another source classified as non-variable. 

\begin{figure}
\includegraphics[width=0.43\textwidth,angle=270, trim=0cm 2.5cm 0cm 0cm]{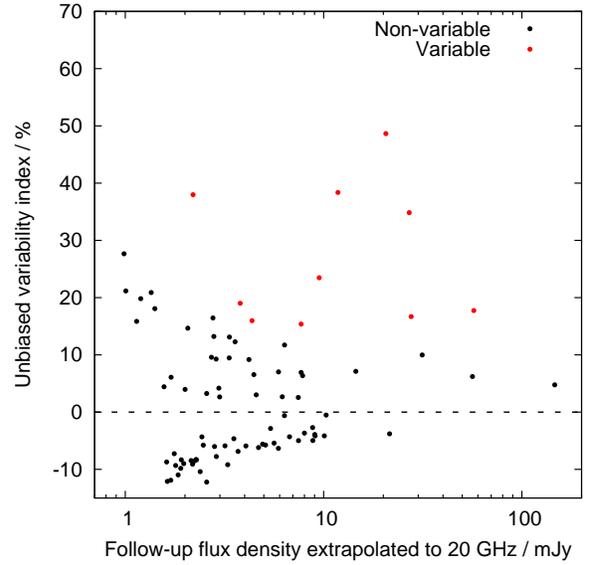}
\caption{Unbiased variabilty index versus follow-up flux density extrapolated to 20~GHz. Sources classified as variable are shown in red and non-variable in black.}
\label{fig:variability_SV}
\end{figure}

Fig. ~\ref{fig:variability_SV} shows how $V$ compares with the follow-up flux density extrapolated to 20~GHz. Sources classified as variable tend to be the brighter ones in our sample. This is likely to be, in large part, due to the larger fractional errors on the survey flux densities for the fainter sources. For the variable sources, $V$ ranges between 15.4 and 48.7\%; the median value is 23.5\%. The median 20-GHz flux density of the variable sources at the time of the follow-up observations is 11.8~mJy. The most variable source is S35, the source having increased in flux density by about a factor of 3 between the time of the survey and follow-up observations.

Table~\ref{tab:var_vs_alpha} shows the number of variable and non-variable sources in each spectral class. As expected, there is a strong tendency for the variable sources to have flat or inverted spectra.

\begin{table}
 \begin{center}
 \caption{Relationship between variability and spectral class.}
 \label{tab:var_vs_alpha}
 \begin{tabular}{@{} c c c c c }
 \hline
 & $N_{\mathrm{VAR}}$ & $N_{\mathrm{NVAR}}$ & Total & $N_{\mathrm{NVAR}}/\mathrm{Total}$ (\%) \\
 \hline
 Steep  &      1  &     46  &     47  &  2.1   \\
 Flat  &      2  &     18  &     20  & 10.0 \\
 Inverted  &      7  &     11  &     18 & 38.9  \\
 \hline
 \end{tabular}
 \end{center}
\end{table}

\section{Spectral curvature between 1.4 and 18~GH\lowercase{z}}\label{Spectral curvature between 1.4 and 18 GHz}

We used the measurements at 5.5, 9 and 18~GHz from the follow-up observations as well as low-frequency data at 1.4 and 2.3~GHz to investigate spectral curvature between 1.4 and 18~GHz. The 1.4-GHz flux densities were obtained from NVSS, ATLAS and FIRST, as described in Section~\ref{Correlation with NVSS, ATLAS and FIRST at 1.4 GHz}. 

Flux densities at 2.3~GHz were obtained for 18 sources by matching the 20-GHz source catalogue for the 03hr field with the ATLAS 2.3-GHz source catalogue from \cite{zinn2012}. The 2.3-GHz catalogue comes from a low-resolution (57 by 23~arcsec) ATCA survey of the CDFS field down to a rms noise of $80~\mu \mathrm{Jy}$. It should be noted that, in some cases, sources which were clearly separated in the ATLAS DR2 image \citep{hales2013} of the field at 1.4~GHz, which has higher resolution, were blended together in the 2.3-GHz image. Such blended sources were excluded from the 2.3-GHz catalogue. Consequently, a source which does not appear in the 2.3-GHz catalogue does not necessarily lie below the detection threshold of the catalogue.

\begin{figure*}
  \includegraphics[angle=270,scale=0.20]{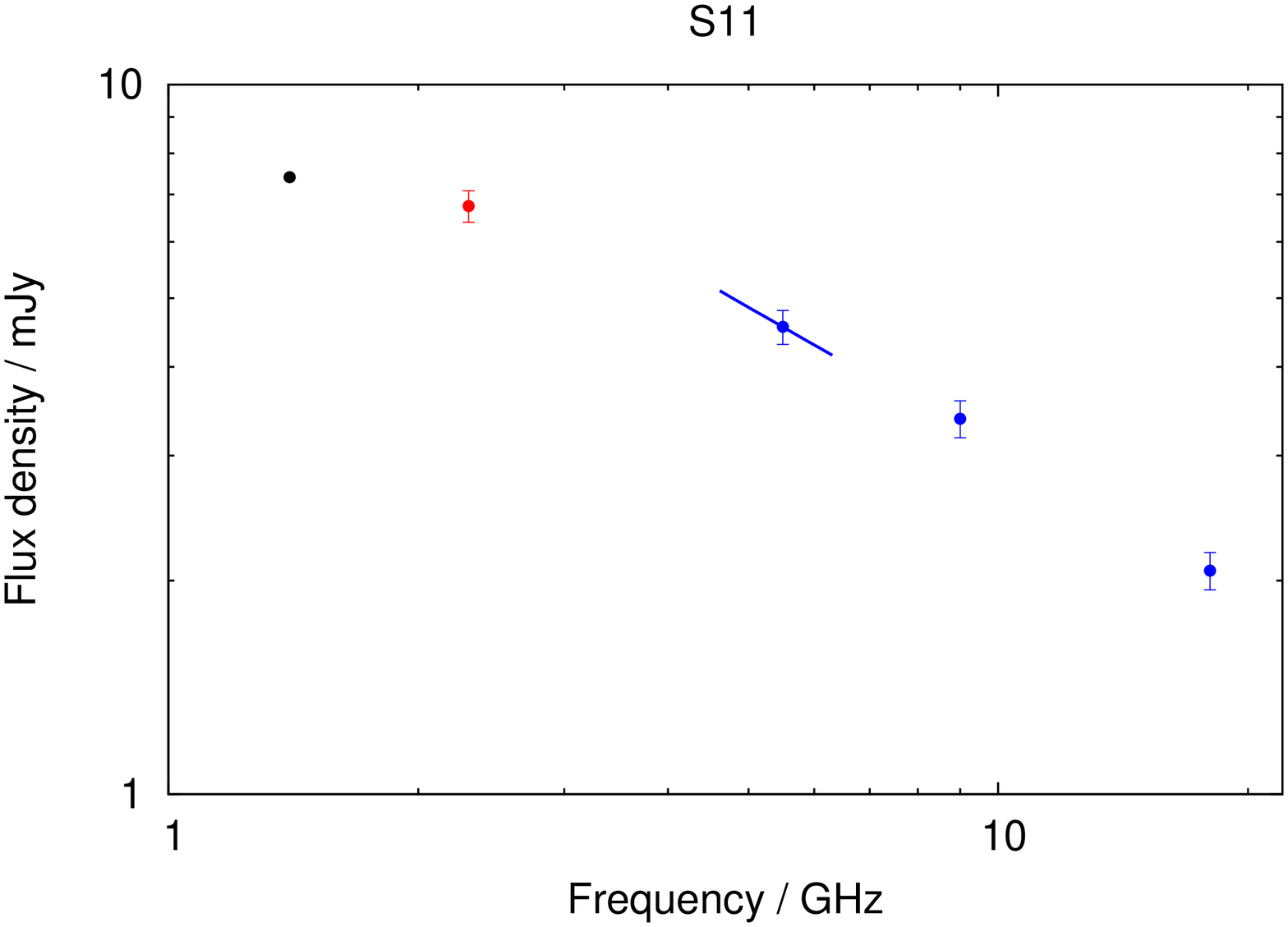}
  \includegraphics[angle=270,scale=0.20]{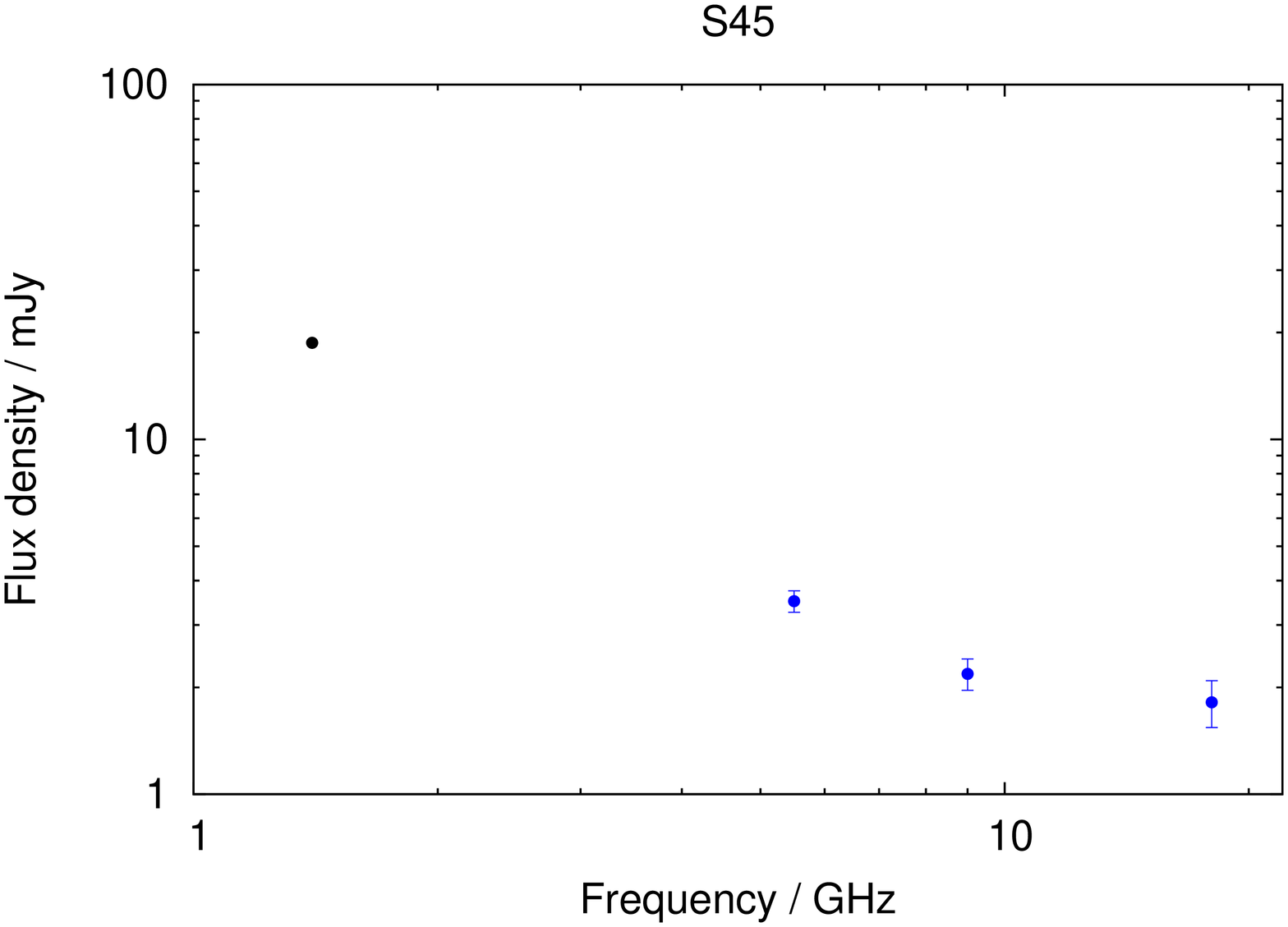}
  \includegraphics[angle=270,scale=0.20]{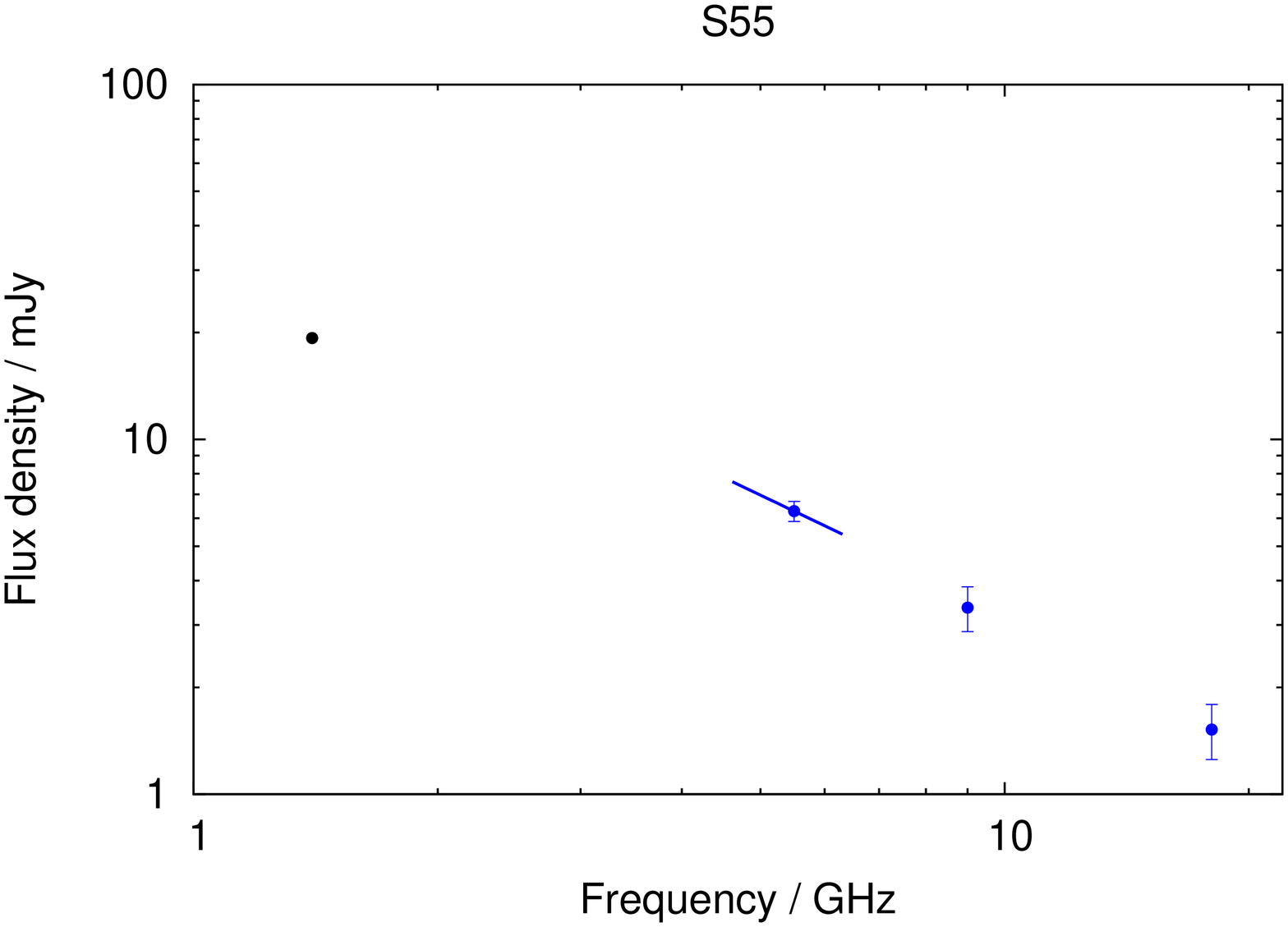} \\ 
  \includegraphics[angle=270,scale=0.20]{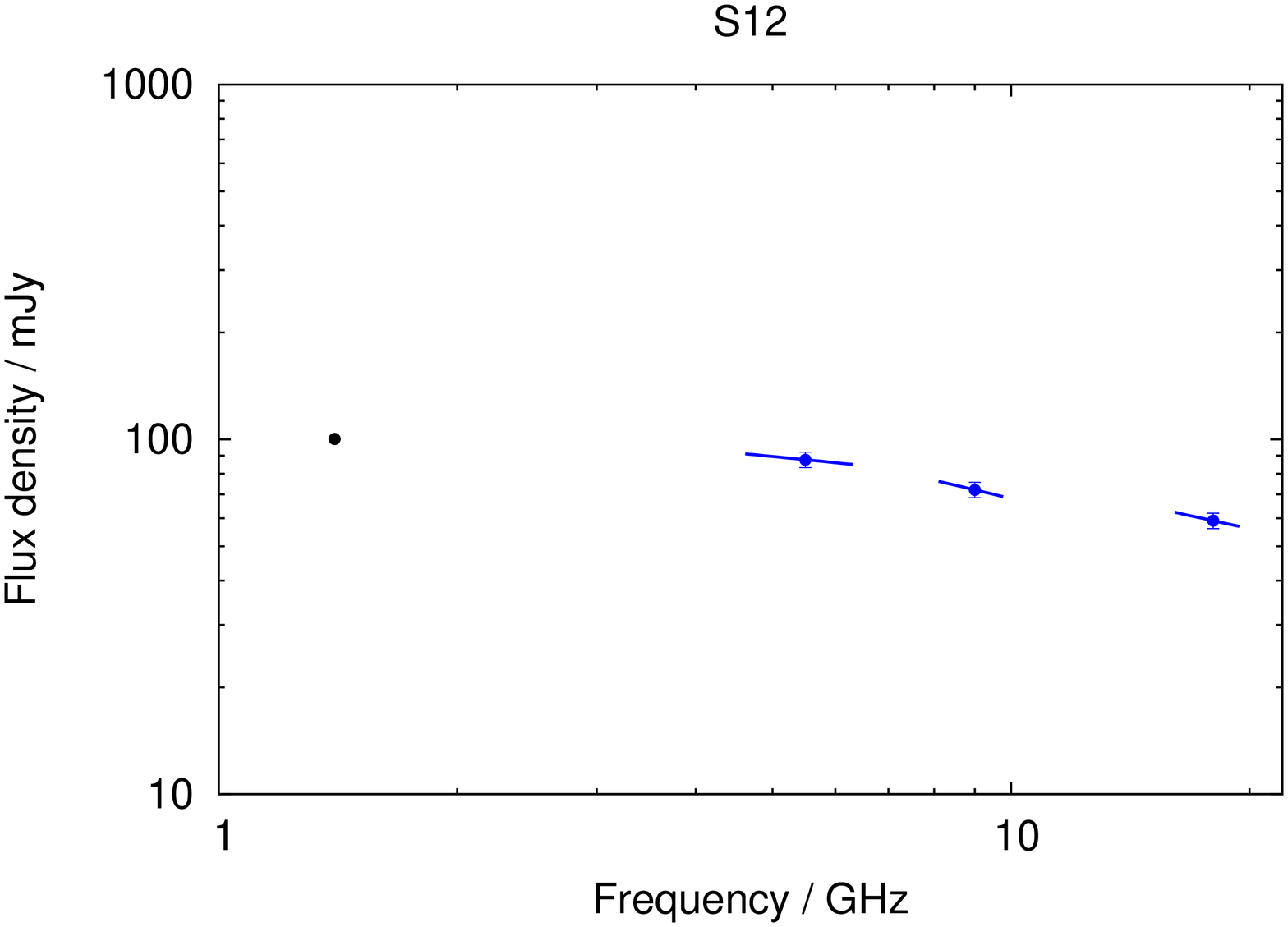}
  \includegraphics[angle=270,scale=0.20]{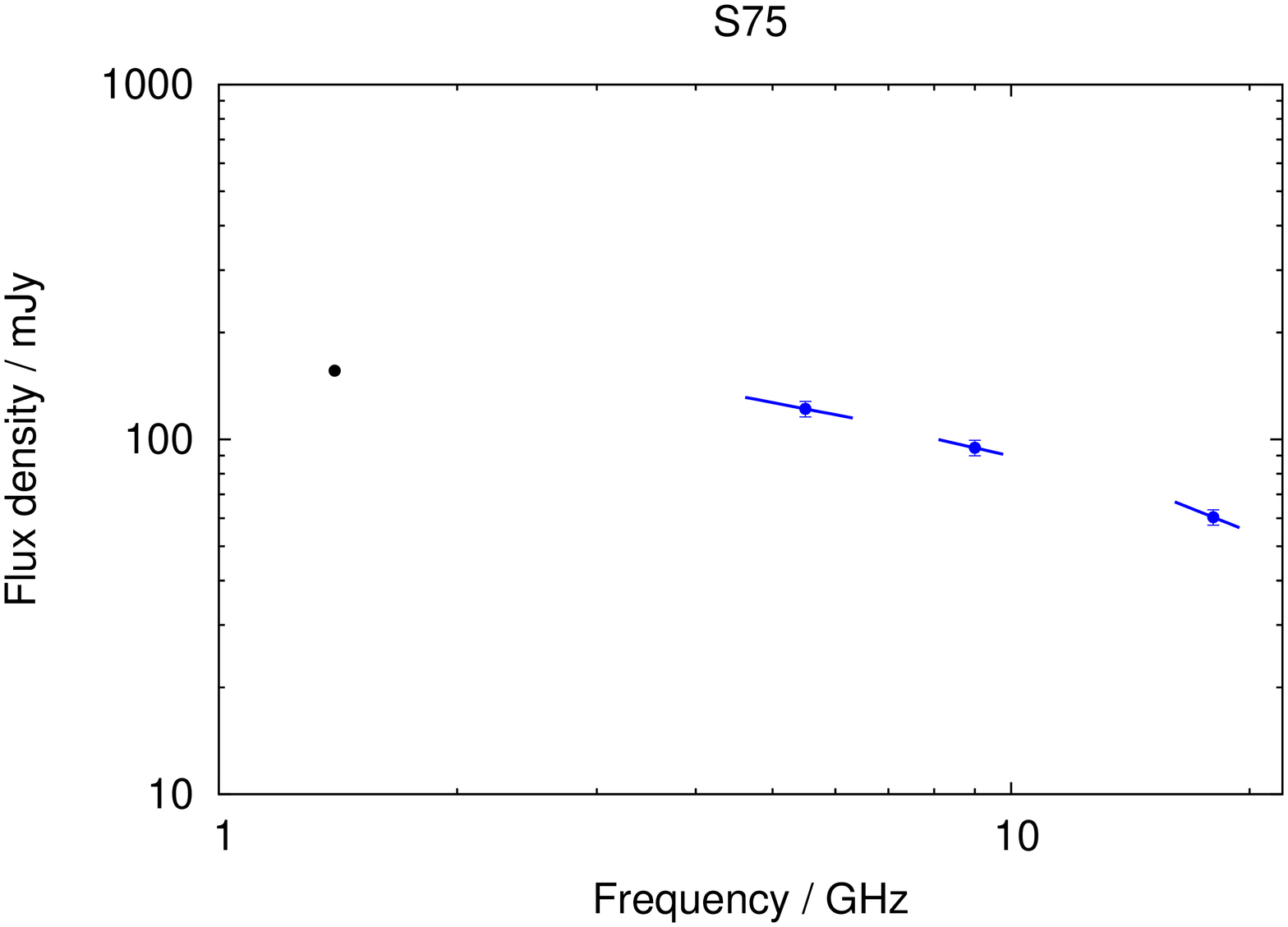}
  \includegraphics[angle=270,scale=0.20]{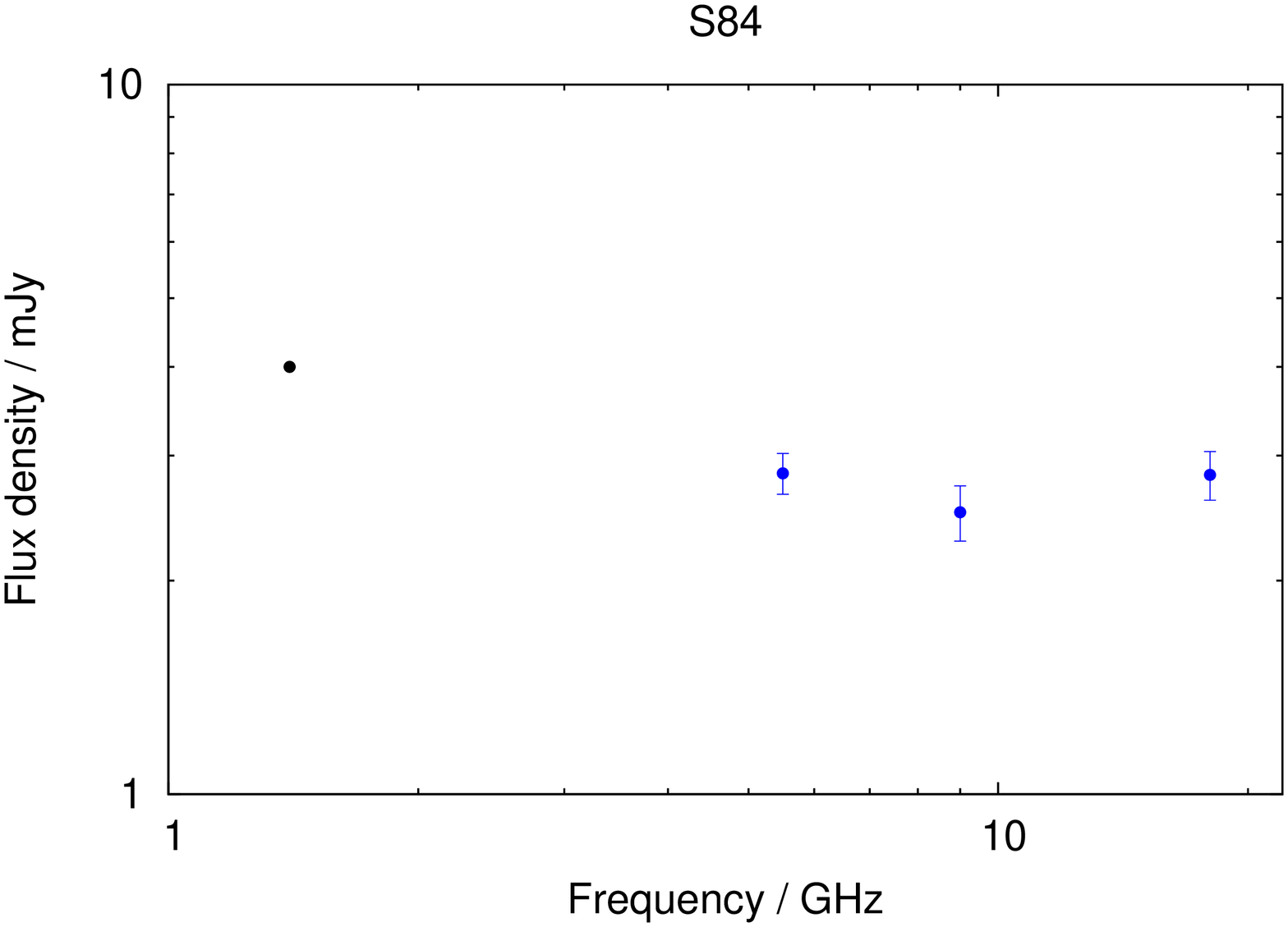} \\  
  \includegraphics[angle=270,scale=0.20]{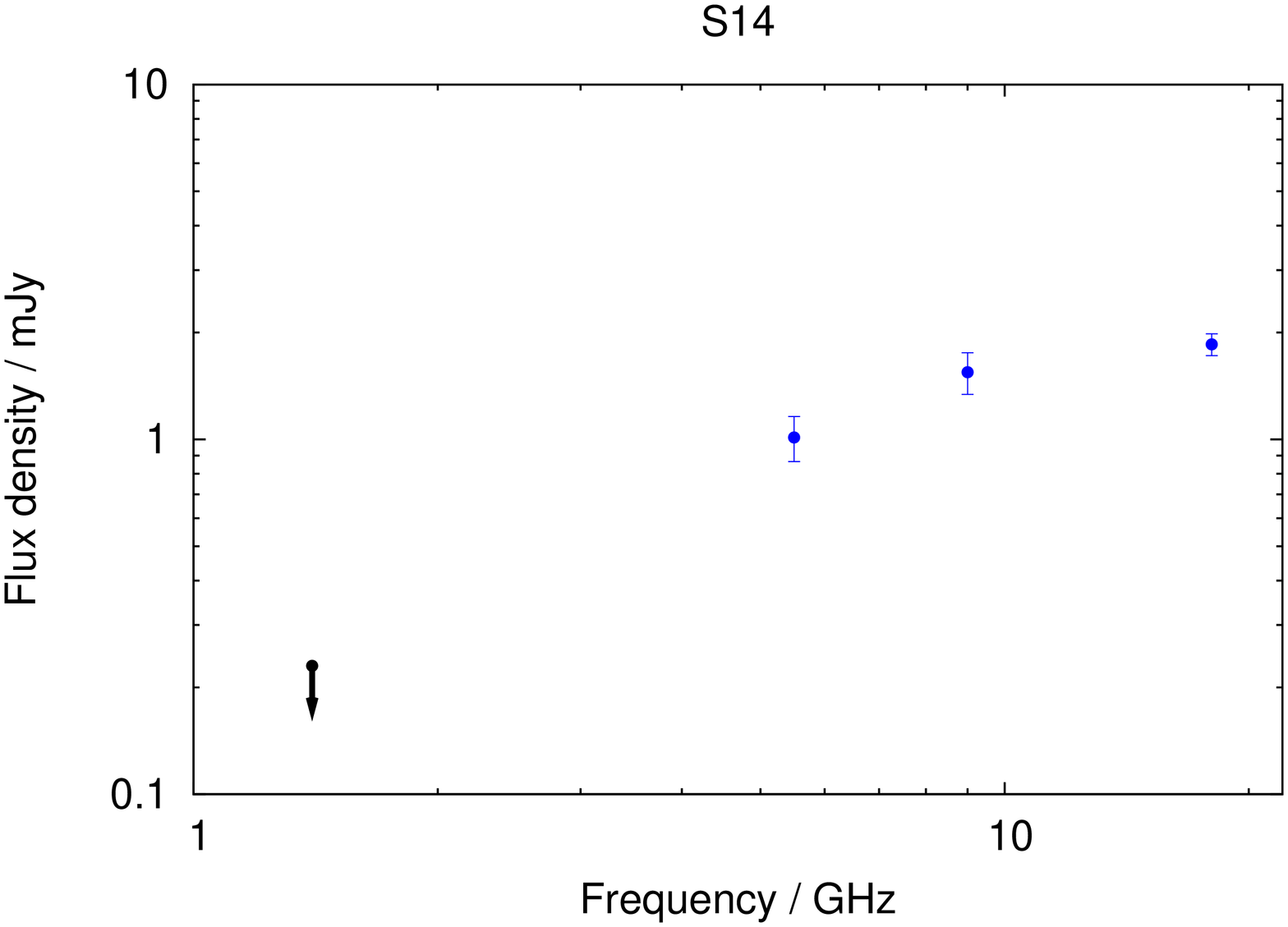}	      
  \includegraphics[angle=270,scale=0.20]{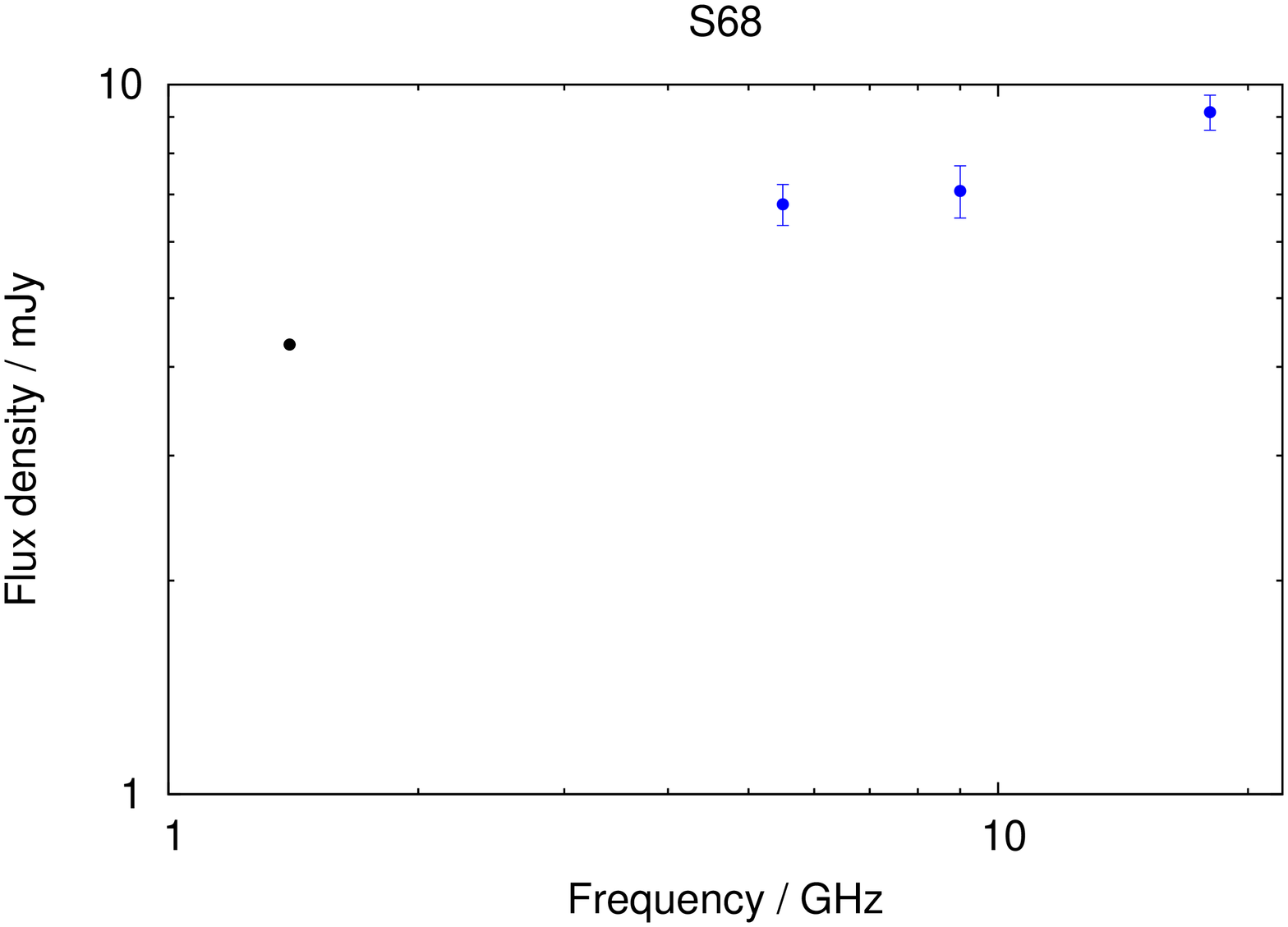}  
  \includegraphics[angle=270,scale=0.20]{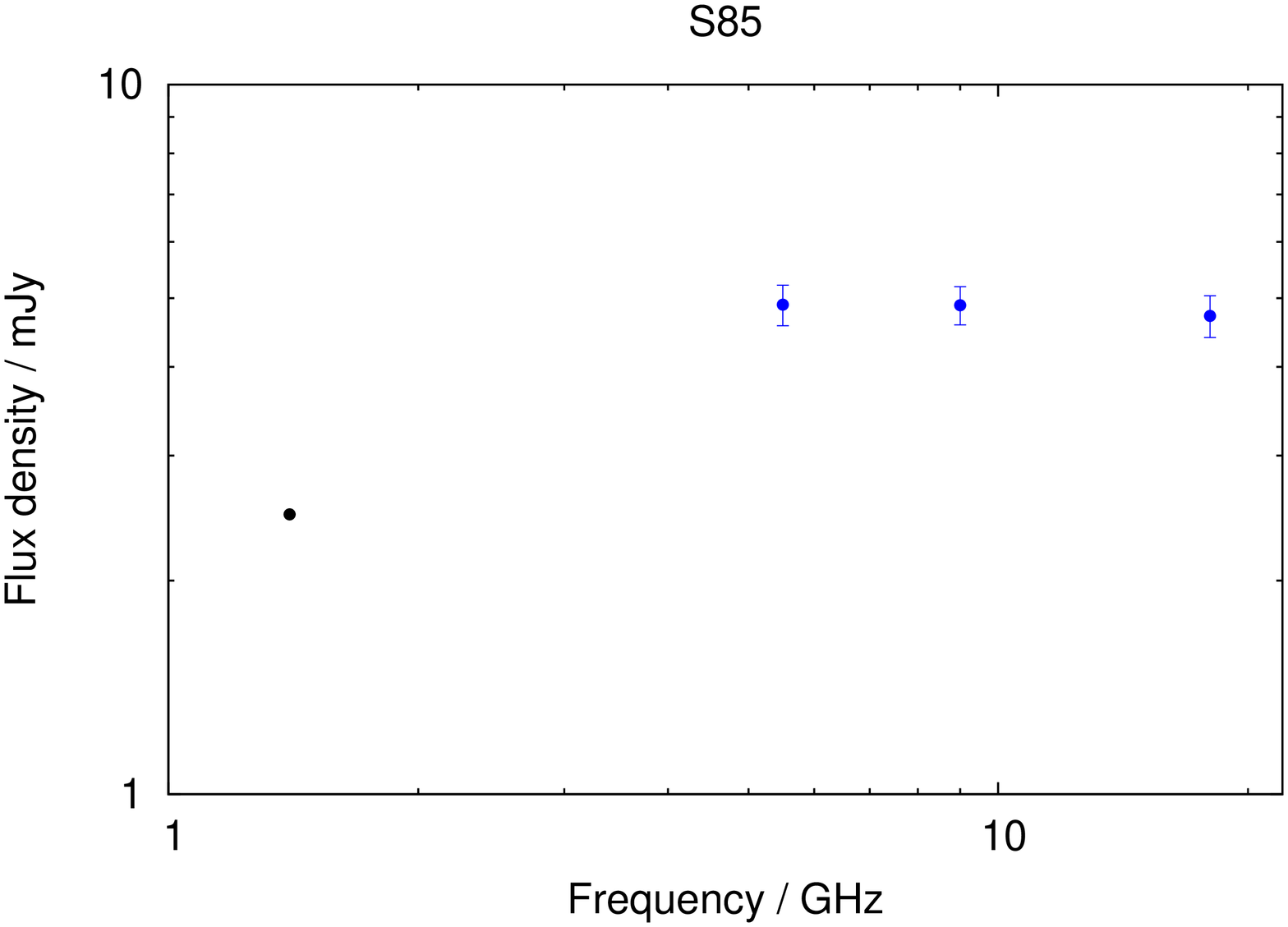} \\	   
  \includegraphics[angle=270,scale=0.20]{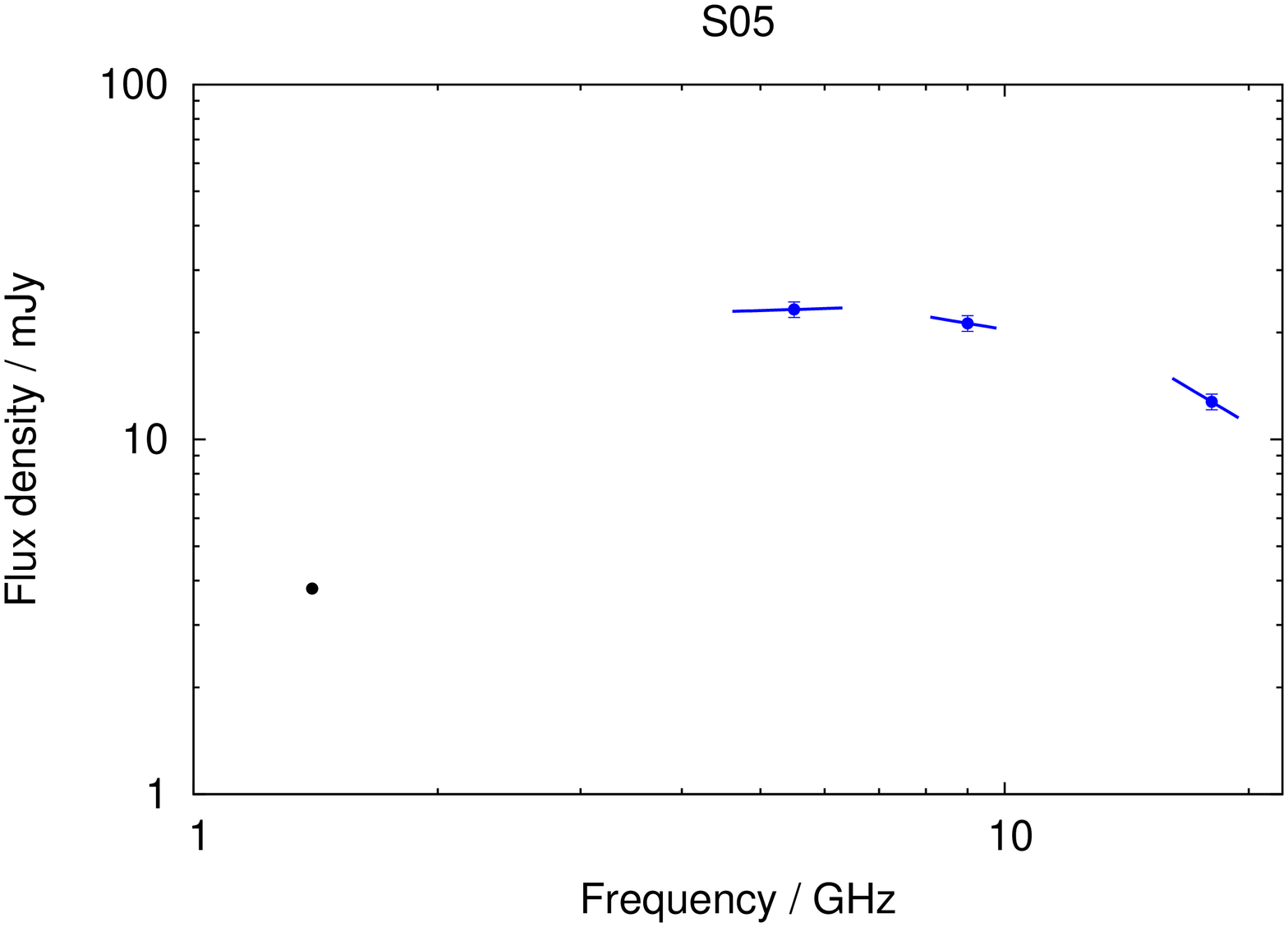}
  \includegraphics[angle=270,scale=0.20]{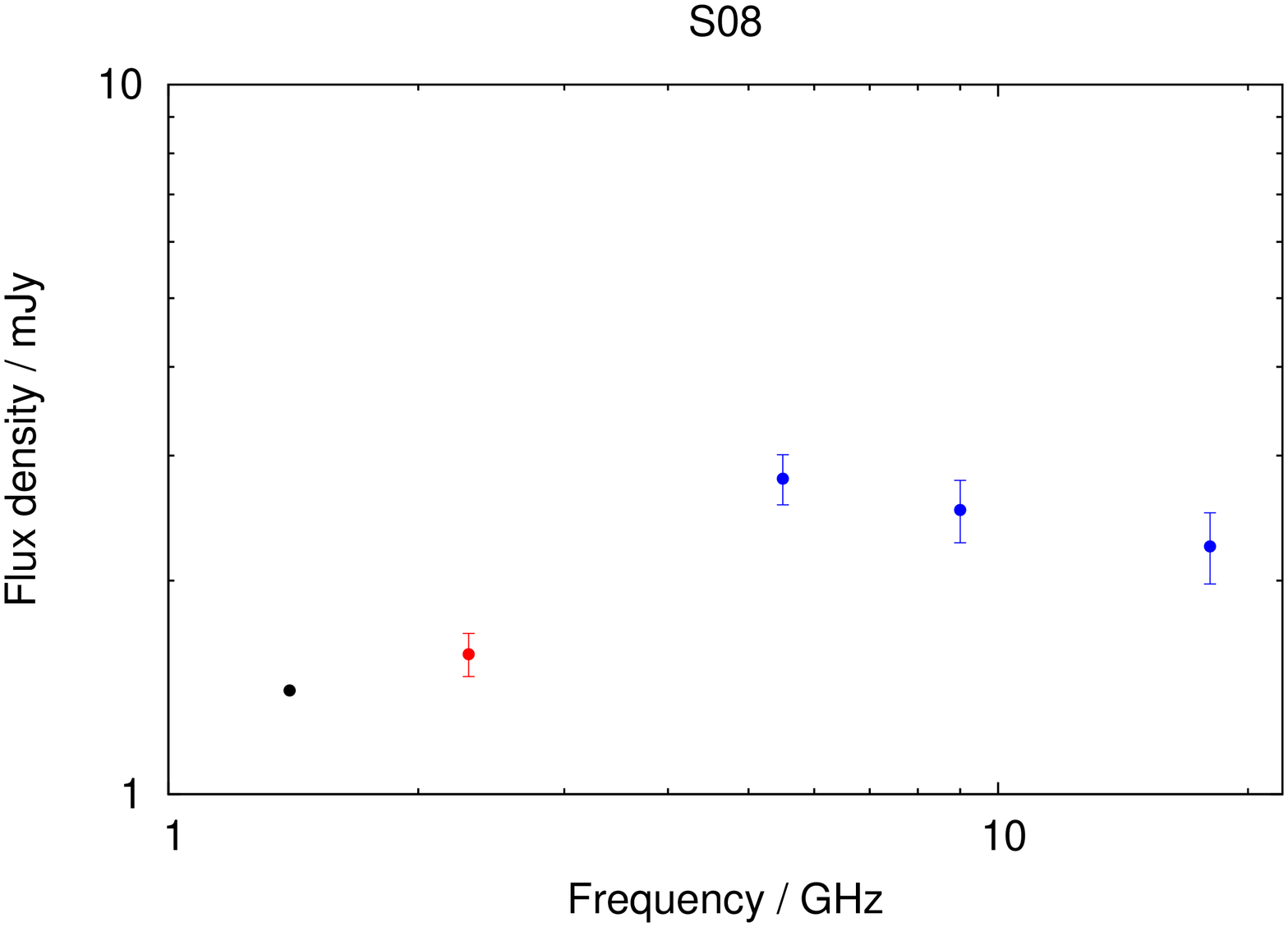}  
  \includegraphics[angle=270,scale=0.20]{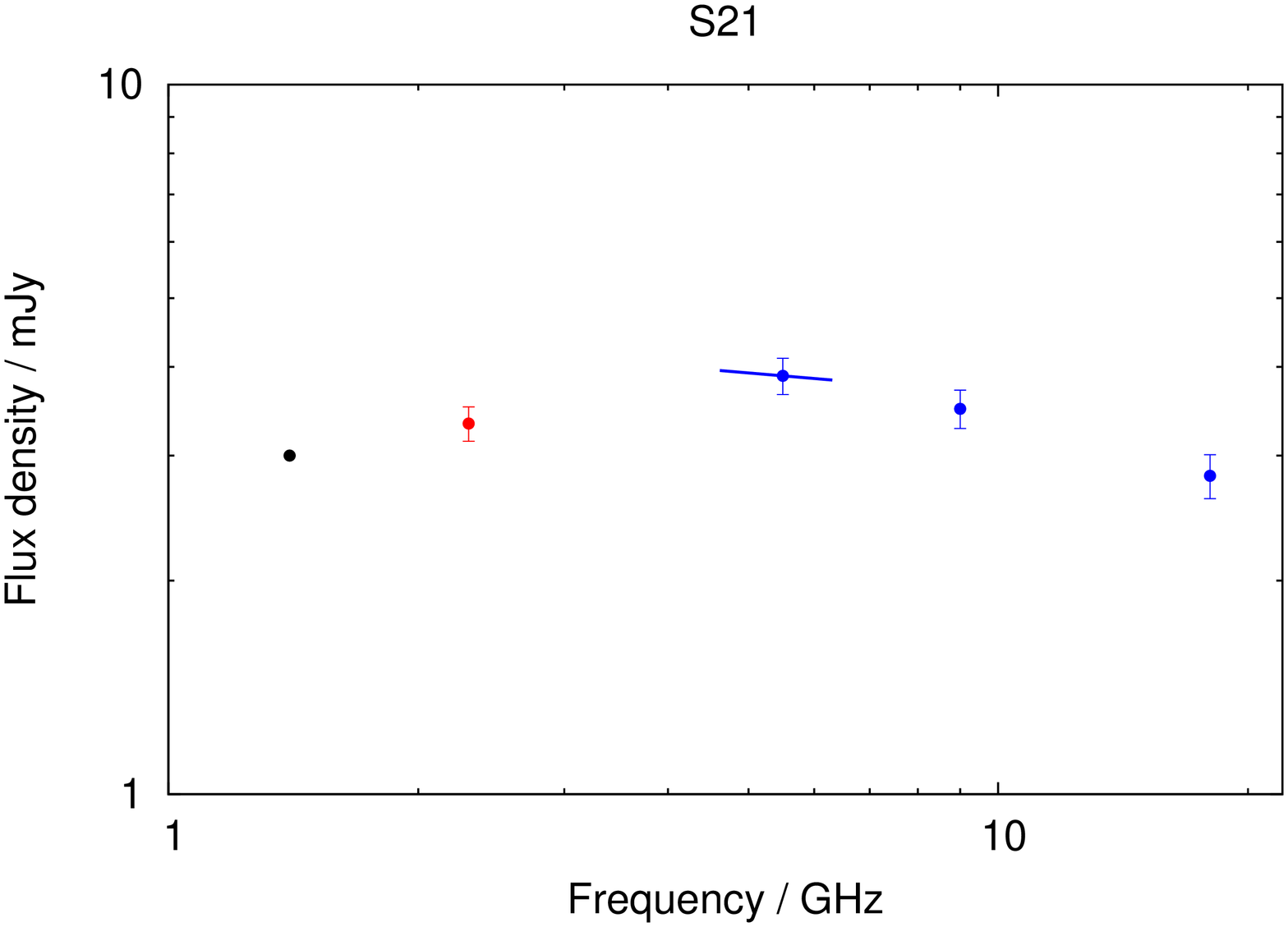} \\	 
  \includegraphics[angle=270,scale=0.20]{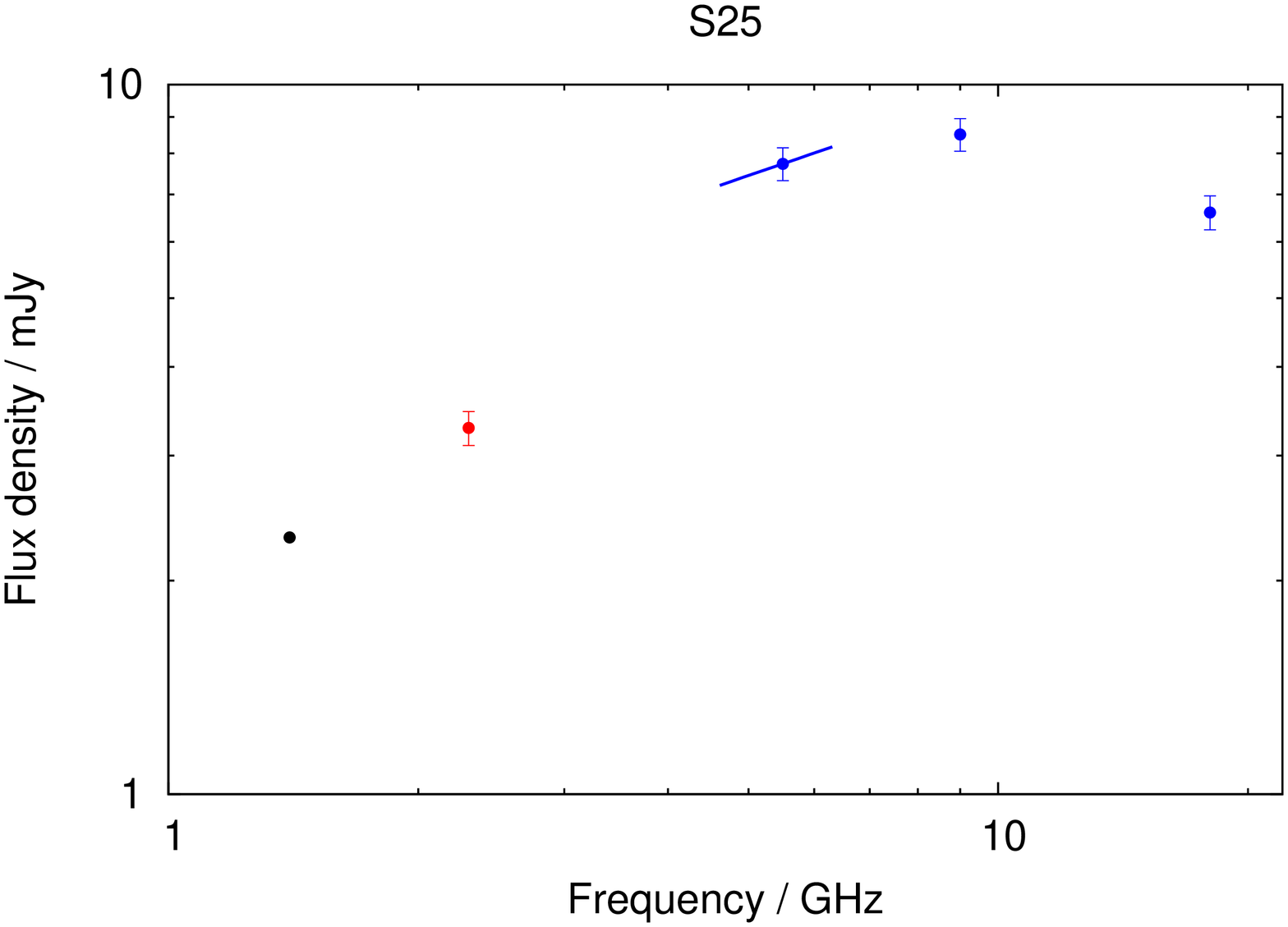}
  \includegraphics[angle=270,scale=0.20]{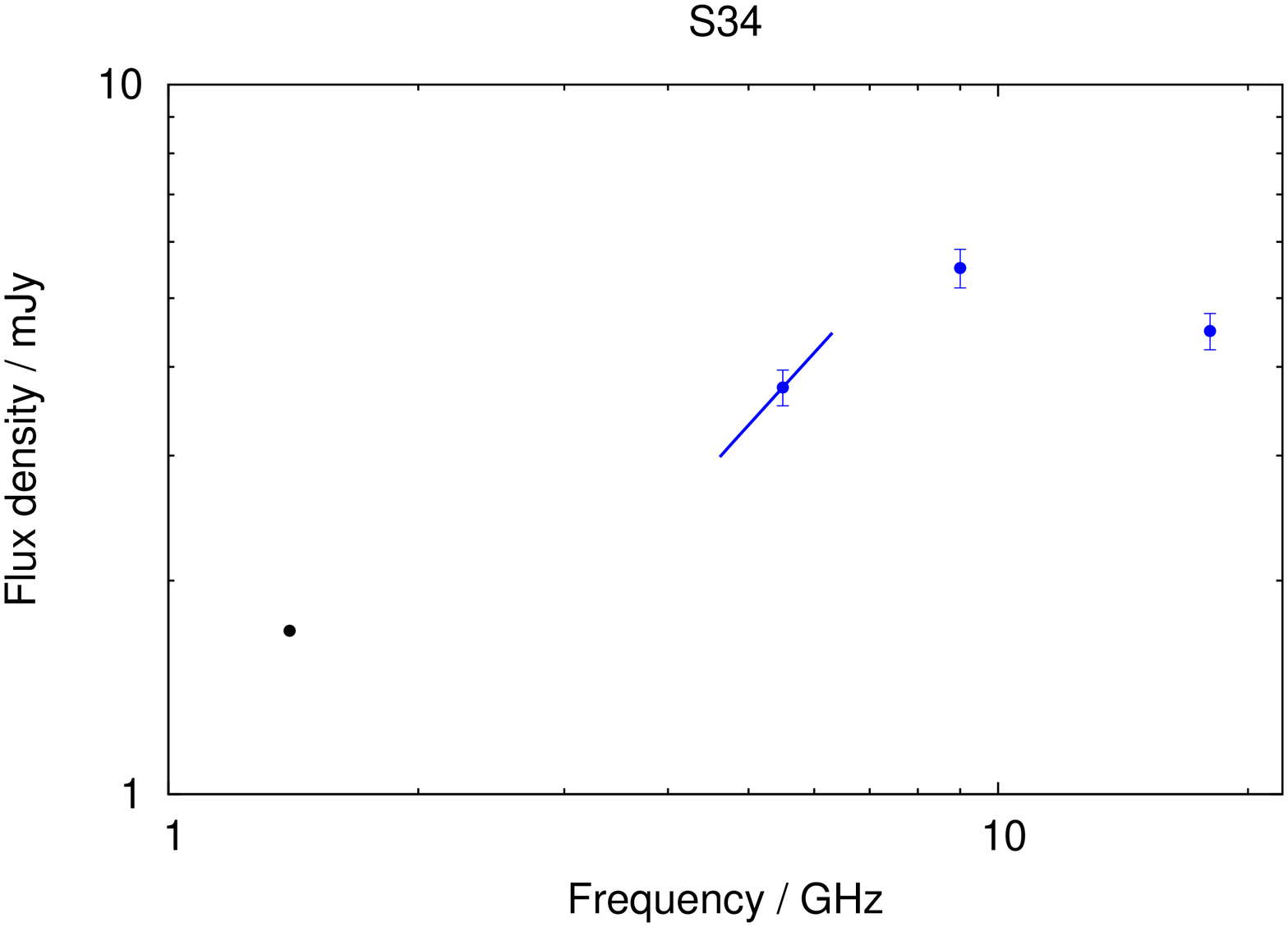}  
  \includegraphics[angle=270,scale=0.20]{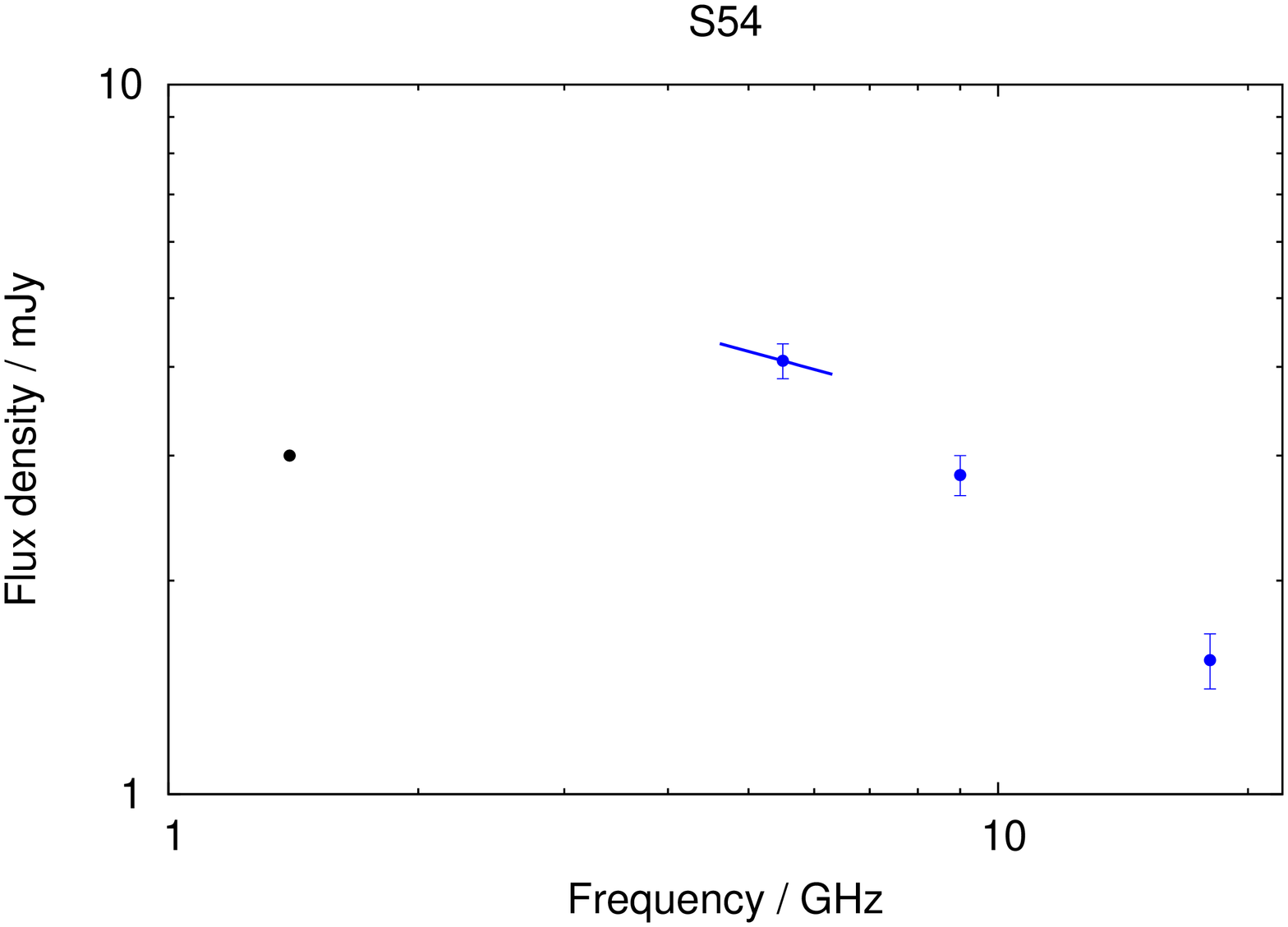} \\	       
  \caption{Examples of spectra between 1.4 and 18~GHz. The black points at 1.4~GHz are from NVSS, FIRST or ATLAS, the red points at 2.3~GHz from \citet{zinn2012}  and the blue points at 5.5, 9 and 18~GHz from the ATCA data presented in this paper. Spectra across the ATCA bands (4.5--6.5, 8--10 and 16--20~GHz) are shown as solid lines. Steep-spectrum sources are shown in the first row, flat-spectrum sources in the second row, inverted-spectrum sources in the third row, and sources with peaked spectra in the last two rows.}
  \label{fig:example_spectra}   
\end{figure*}

The spectra of a few sources showing different spectral characteristics are shown in Fig.~\ref{fig:example_spectra}. Spectra across the ATCA bands are indicated by solid lines. These are only shown if the SNR is greater than 50 at 5.5~GHz, and 100 at 9 and 18~GHz, to prevent the spectra from being significantly affected by the thermal noise. 

\begin{figure}
\includegraphics[width=0.43\textwidth,angle=270, trim=0cm 2.5cm 0cm 0cm]{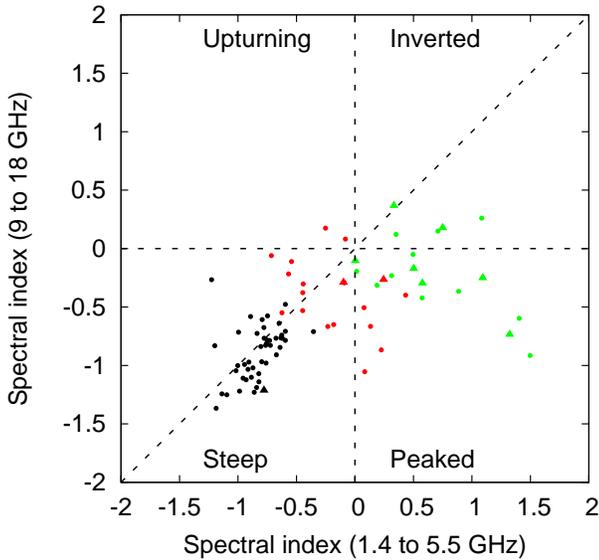}
\caption{Radio colour-colour plot for 83 sources with measured spectral indices $\alpha_{1.4}^{5.5}$ and $\alpha_{9}^{18}$. Sources classified as steep-spectrum are plotted in black, flat-spectrum in red and inverted-spectrum in green. Triangles refer to variable sources and circles to non-variable sources.}
\label{fig:two_colour_plot}
\end{figure}

We have measurements at 1.4, 5.5, 9 and 18~GHz for 83 out of the 85 sources in our sample. Fig.~\ref{fig:two_colour_plot} shows a radio colour-colour plot, comparing $\alpha_{1.4}^{5.5}$ with $\alpha_{9}^{18}$, for these 83 sources. We note that the axes in this diagram are independent. It can be seen that the steep-spectrum sources tend to have single power-law spectra as they lie close to the dashed diagonal line. One source which does not follow this trend is S45 which has $\alpha_{1.4}^{5.5} = -1.23$ and $\alpha_{9}^{18} = -0.27$  (see spectrum in Fig.~\ref{fig:example_spectra}); the most likely explanation for the pronounced spectral flattening is that the radio emission switches from being lobe-dominated at low frequency to core-dominated at high frequency. The flat-spectrum sources show a steepening in their spectra with frequency, consistent with the effects observed by \cite{chhetri2012} for compact sources, and hence, tend to lie below the diagonal line. Most of the inverted-spectrum sources lie in the lower-right quadrant of the plot, indicating that they have peaked spectra. Among the 18 inverted-spectrum sources, only five have $\alpha_{9}^{18} > 0$ and one has $\alpha_{9}^{18} > \alpha_{1.4}^{5.5}$.

We have identified sources with clearly defined peaks in their spectra between 1.4 and 18~GHz; these are defined as having $\alpha_{1.4}^{5.5} > 0.15$ and $\alpha_{9}^{18} < -0.15$. A total of 13 sources fit into this category, of which 10 are classified as inverted-spectrum and 3 as flat-spectrum. We have estimated the peak frequencies of these sources by fitting polynomial functions to their spectra; they lie between 3.1 and 11.5~GHz and are listed in Tables~\ref{tab:source_catalogue_followup_03hr} and~\ref{tab:source_catalogue_followup_21hr}. 

\subsection{Source not detected at 1.4~GHz}\label{Source not detected at 1.4 GHz}

Only one source (S14) is not detected in any of the three 1.4-GHz catalogues. The source has a flux density of ($1.85 \pm 0.13$)~mJy at 18~GHz. The rms noise close to the source in the ATLAS DR1 image is $75~\mu \mathrm{Jy}$. The source is also not present in the image from the third ATLAS data release \citep{franzen2013}, which is more sensitive. The rms noise close to the source in this image is $18~\mu \mathrm{Jy}$. Placing a 3-$\sigma$ upper limit to the flux density of the source at 1.4~GHz gives a limit on the spectral index of $\alpha_{1.4}^{18} > 1.38$. We note that the source happens to lie 1~arcmin from a 56-mJy source at 1.4~GHz. We checked the NASA Extragalactic Database but no identification for the source was found.

\cite{murphy2010} found that 27 sources in the AT20G survey, which has a detection limit of 40~mJy, did not have low-frequency counterparts in NVSS at 1.4~GHz, and the Sydney University Molonglo Sky Survey \citep[SUMSS;][]{mauch2007} and the 2nd-epoch Molonglo Galactic Plane Survey \citep[MGPS-2;][]{murphy2007} at 0.843~GHz, which together cover the whole AT20G survey area. All these sources had (non-simultaneous) spectral indices between $\sim 1$ and 20~GHz greater than 0.7. In general, these extreme spectral indices were not thought to be the result of source variability between the NVSS/SUMSS/MGPS-2 and AT20G observing epochs because the great majority of the sources had similarly extreme (simultaneous) spectral indices between 5 and 20~GHz.

Unlike the sample of ultra-inverted spectrum sources presented in \cite{murphy2010}, S14 has a spectrum which rapidly flattens with frequency between 1 and 20~GHz, as shown in Fig.~\ref{fig:example_spectra}. The spectrum of the source appears to turn over at $\sim 20$~GHz; higher-frequency measurements are required to provide a reliable estimate of the turnover frequency. The flux density of the source at 20~GHz did not vary significantly between July 2009 and September 2012, but monitoring of the source over longer timescales is required to rule out its blazar nature. Blazars typically only show small flux density variations on timescales of 1--2 years, with larger flares occuring on average once every 6 years \citep{hovatta2007}. Also, the error on the flux density measured in July 2009 from the survey observations is large given that the source was only detected at a level of 5.6$\sigma$.

If S14 is not a blazar, then there is the possibility that it is a truly young HFP source. In contrast to blazars, genuinely young sources are expected to maintain their convex spectra and to show little or no variability, discounting the spectral evolution associated with the source growth, where the peak frequency shifts towards lower frequencies as a consequence of adiabatic expansion \citep{odea1997}. Multi-frequency, multi-epoch observations would shed more light on the nature of this source. 

\section{Exploring the relationship between spectral index and flux density in the high-radio-frequency population}\label{Exploring the relationship between spectral index and flux density in the high-radio-frequency source population}

Strong variations in the typical spectral index of the 15--20-GHz source population, between 0.5~mJy and $\sim 1$~Jy, have been reported in the literature, the underlying cause of which is not well understood. These spectral index shifts suggest rapid changes in source population with flux density. There are therefore implications for theoretical modeling of the high-radio-frequency population and its evolution. 

In this section, we attempt to identify the source populations responsible for these spectral index shifts. We first review the relationship between spectral index and flux density for sources in the AT20G, 9C and 10C surveys, as reported in the literature. We then combine the data from these three surveys to get a more complete picture of the changes in spectral properties with flux density. We measure the fraction of sources with steep spectra ($\alpha_{1.4}^{15.7} < -0.5$) as a function of 15.7-GHz flux density between 0.5~mJy and $\sim 1$~Jy. We also measure the Euclidean normalized differential counts of flat- ($\alpha_{1.4}^{15.7} \geq -0.5$) and steep-spectrum sources to track how the two source classes change with flux density. Finally, we compare, qualitatively, the observed counts of the two source classes at 15.7~GHz with the 5-GHz source-count model by \cite{jackson1999}, who break down the source population into different types of FRI and FRII sources, and star-forming galaxies, and with the latest version of the 15-GHz source-count model by \cite{dezotti2005} for classical radio sources.

\subsection{Previous work}\label{Previous work2}

\cite{massardi2011} investigated the change in spectral properties with flux density of sources in the AT20G survey. The fraction of flat-spectrum ($\alpha_{8.6}^{20} > -0.5$) sources was found to decrease from 81\% for $S_{20} > 500$~mJy to 60\% for $S_{20} < 100$~mJy. This trend towards a steeper-spectrum population was found to continue at lower flux densities: in the 9C survey, the median value of $\alpha_{1.4}^{15.2}$ was found to decrease from --0.23 for sources with $S_{15.2} > 100$~mJy to --0.79 for sources with $S_{15.2} < 25$~mJy \citep{waldram2010}.

A completely different trend was found in \cite{davies2011} when going to even fainter flux densities. \citeauthor{davies2011} matched a complete and unbiased sample of sources from the deep areas of the 10C survey, complete to 0.5~mJy, with catalogues from the NVSS and FIRST survey. They measured the fractions of sources with $\alpha^{15.7}_{1.4} < -0.8$ in various 15.7-GHz flux density bins between 0.5 and 25~mJy. They found that the typical spectral index between 1.4 and 15.7~GHz becomes flatter for sources with decreasing flux densities between 25 and 0.5~mJy: when matching to NVSS, 49\% of sources with $5.0 \leq S < 25.0$~mJy were found to have $\alpha^{15.7}_{1.4} < -0.8$. For sources with $0.5 \leq S < 0.6$~mJy, the corresponding figure is 18\%. Many sources are too faint to be detected in NVSS; in the lowest flux density bin, the percentage of unmatched sources is as high as 60\%. The spectral index cut-off of --0.8 was chosen to ensure that in all flux density bins, any source with spectral index less than this value ought to be detected in NVSS, given its completeness limit of 3.4~mJy.

\cite{whittam2013} subsequently studied the radio properties of a sub-sample of 296 10C sources in the Lockman Hole using much deeper surveys at 1.4~GHz and 610~MHz. They found that the median spectral index between 15.7~GHz and 610~MHz changes from --0.75 for $S_{15.7} > 1.5$~mJy to --0.08 for $S_{15.7} < 0.8$~mJy. They found that the \textit{SKADS Simulated Skies} \citep[$S^{3}$;][]{wilman2008} simulation does not accurately reproduce the observed population at 15.7~GHz. They concluded that either there is a new population of faint, flat-spectrum sources which is not being accounted for in the model or the behaviour at 15.7~GHz of a known population is not being modeled correctly. 

\subsection{Combining data from the AT20G, 9C and 10C surveys}\label{Combining data from the AT20G, 9C and 10C surveys}

We have extended the analysis of \cite{davies2011} described in Section~\ref{Previous work2} to higher flux densities by using data from the shallower regions of the 10C survey, and from the 9C and AT20G surveys. The general properties of these surveys are listed in Table~\ref{tab:survey_properties}. Together, these surveys sample flux densities ranging from 0.5~mJy to several Jy.

\begin{table*}
\begin{minipage}{10cm}
\begin{center}
\caption{Coverage, resolution and completeness limits of the AT20G, 9C and 10C surveys. The numbers in brackets in the last column are upper completeness limits. We note that in the case of the AT20G survey, the detection limit (40~mJy) is more than a factor of two lower than the completeness limit.}
\label{tab:survey_properties}
\begin{tabular}{@{} c c c c c}
\hline
Survey & \multicolumn{1}{c}{Frequency} & \multicolumn{1}{c}{Coverage} & \multicolumn{1}{c}{Resolution} & \multicolumn{1}{c}{Completness limit} \\
& \multicolumn{1}{c}{(GHz)} & \multicolumn{1}{c}{(deg$^{2}$)} & \multicolumn{1}{c}{(arcsec)} & \multicolumn{1}{c}{(mJy)} \\
\hline
AT20G & 20 & 20086\footnote{\cite{massardi2011}} &10 & 100 \\
\hline
\multirow{3}{*}{9C} & \multirow{3}{*}{15.2} & 520\footnote{\cite{waldram2003}} & \multirow{3}{*}{25} & 25 \\
& & 115\footnote{\cite{waldram2010}} & & 10 (100) \\
& & $29^{\textstyle \textit{c}}$ & & 5.5 (100) \\
\hline
\multirow{2}{*}{10C\footnote{\cite{davies2011}}} & \multirow{2}{*}{15.7} & 27 & \multirow{2}{*}{30} & 1.0 (25) \\
& & 12 & & 0.5 (25) \\
\hline
\end{tabular}
\end{center}
\end{minipage}
\end{table*}

Sources in the shallower region of the 10C survey were matched to NVSS. For sources in the 9C survey, $S_{1.4}$ was obtained from \citet{waldram2003,waldram2010} who correlated the sources with NVSS. Only a very small fraction of sources were too faint to appear in NVSS in the deeper regions of the 9C survey. There is a small difference in the observing frequencies of the 9C (15.2~GHz) and 10C (15.7~GHz) surveys. Where $S_{1.4}$ was available, $S_{15.7}$ was estimated using $\alpha^{15.2}_{1.4}$, otherwise no correction was applied for the small difference in frequency.

All 5890 AT20G sources have flux-density measurements at 20~GHz, and 3538 of these also have near-simultaneous flux-density measurements at 4.8 and 8.6~GHz, from follow-up observations. In addition, all but 44 sources have flux density measurements at 1.4 and/or 0.843~GHz. The 1.4-GHz flux densities were obtained from NVSS and the 0.843-GHz flux densities from SUMSS and MGPS-2. The 44 uncatalogued sources at 1.4 and 0.843~GHz were removed from the analysis; these sources are either listed in \cite{murphy2010} as not having a low-frequency counterpart or they are nearby, large galaxies that were too complicated to be catalogued in MGPS-2. For the remaining sources, $S_{15.7}$ was estimated from $S_{20}$ and the flux density available at the closest frequency (8.6, 1.4 or 0.843~GHz) below 15.7~GHz, assuming a power-law dependence on frequency. If no measurement of $S_{1.4}$ was available from NVSS, $S_{1.4}$ was estimated from $S_{0.843}$ and the flux density available at the closest frequency (4.8 or 20~GHz) above 1.4~GHz.

We divided the flux density range 0.5~mJy --10~Jy into 21 bins. In any bin, we only used sources from survey regions which are close to 100\% complete. This is important because sources with higher flux densities are preferentially detected in incomplete bins. We adjusted the completeness limit of the AT20G catalogue to account for the frequency difference. As indicated in Table~\ref{tab:survey_properties}, the AT20G catalogue is estimated to be 100\% complete at 20~GHz above 100~mJy. We set the completeness limit at 15.7~GHz to 143.8~mJy in order to ensure that a source with a falling spectral index as extreme as $\alpha_{15.7}^{20}=-1.5$ still lies above 100~mJy at 20~GHz.

\begin{figure}
 \begin{center}
 \includegraphics[width=0.35\textwidth,angle=270]{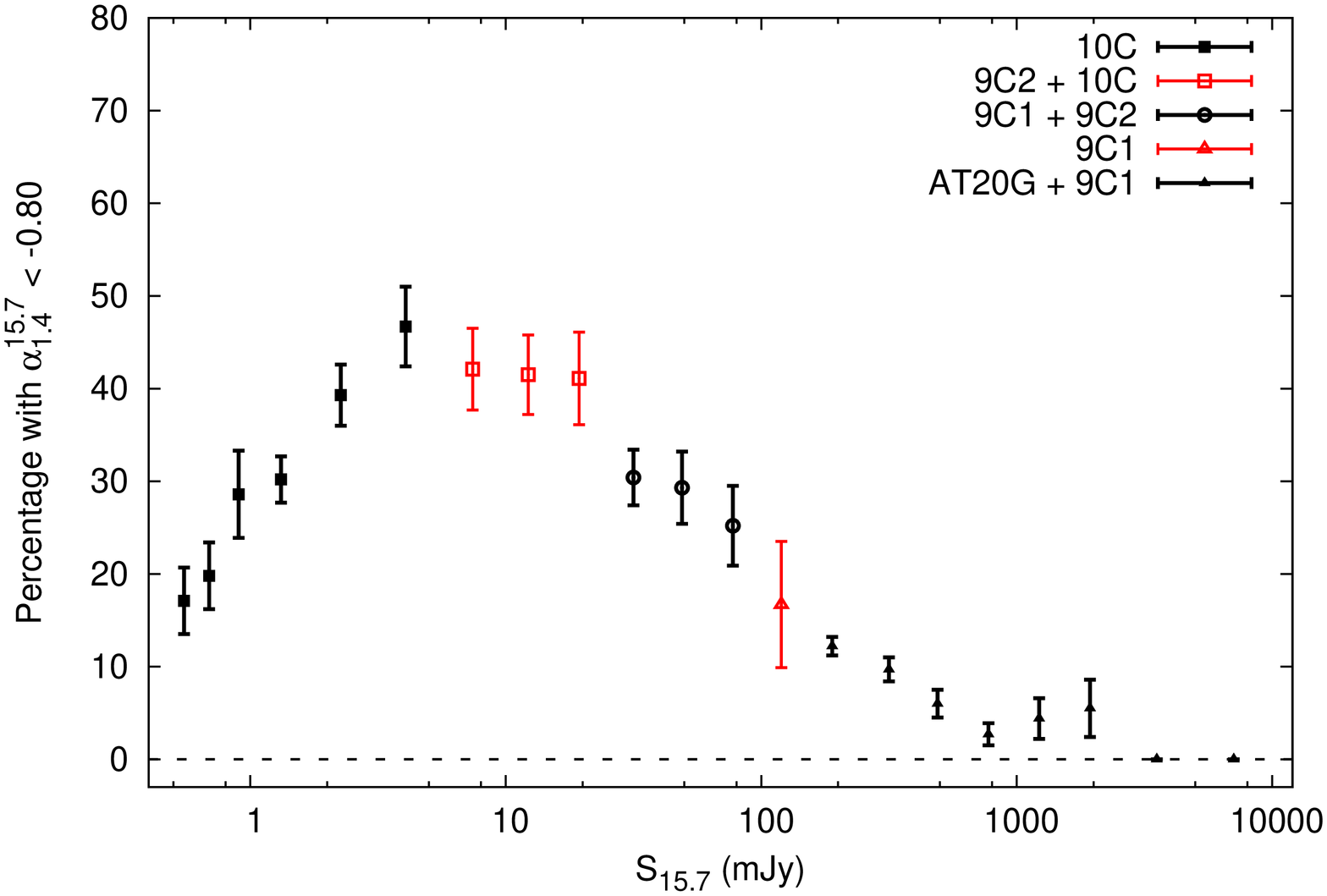}
 \includegraphics[width=0.35\textwidth,angle=270]{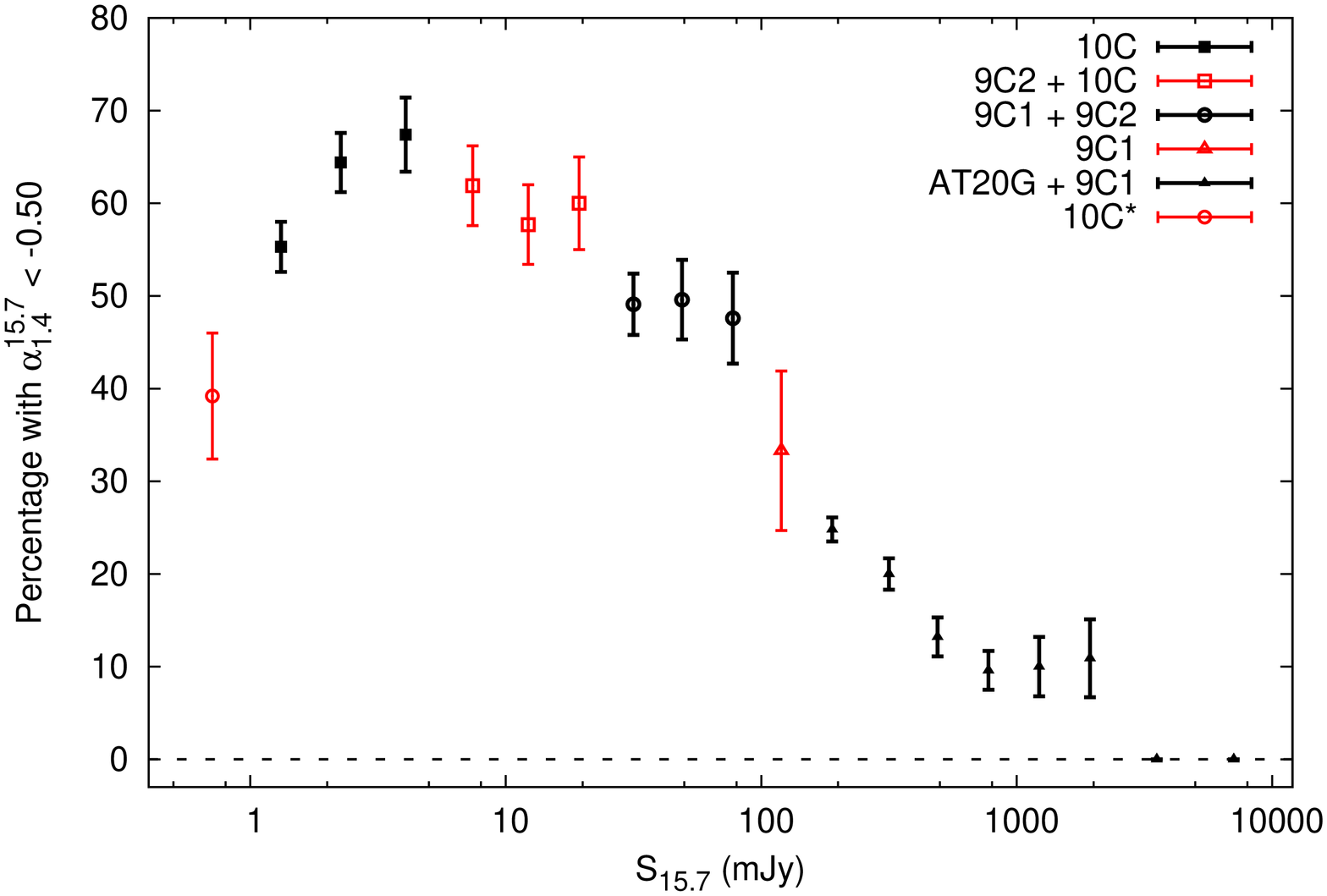}
  \caption{Top: The percentage of sources with $\alpha^{15.7}_{1.4} < -0.8$ as a function of $S_{15.7}$. The different symbols distinguish between the survey regions that were used to produce the plot for the various flux-density bins: `AT20G' designates areas presented in \citet{massardi2011}, `9C1' in \citet{waldram2003}, `9C2' in \citet{waldram2010} and `10C' in \citet{davies2011}. Bottom: Same as above except that a spectral index cut-off of -0.5 is applied. `$10\mathrm{C}^{\star}$' designates a region of the 10C survey in the Lockman Hole in which spectral index information from \citet{whittam2013} was used.}
 \label{fig:alpha}
 \end{center}
\end{figure}

The top panel of Fig.~\ref{fig:alpha} shows the fraction of sources with $\alpha_{1.4}^{15.7} < -0.8$ as a function of flux density. Sources in the 9C and 10C surveys that were too faint to be detected in NVSS and that as a result have no spectral-index measurements were taken to have $\alpha_{1.4}^{15.7} \geq -0.8$. We can be certain that this is the case in all flux-density bins given the completeness limit of NVSS. In any bin, the error on the fraction of sources with $\alpha_{1.4}^{15.7} < -0.8$ was derived assuming Poisson errors on the numbers of sources with $\alpha_{1.4}^{15.7} < -0.8$ and $\alpha_{1.4}^{15.7} \geq -0.8$. It can be seen that in the flux density range between $\sim 1$~Jy and $\sim 5$~mJy, the fraction of sources with $\alpha_{1.4}^{15.7} < -0.8$ increases with decreasing flux density. At fainter flux densities, this trend is reversed, with a move back towards a flatter-spectrum population.

Traditionally, a spectral index cut-off of --0.5 is used to separate flat- and steep-spectrum sources. For spectral indices between 1 and 5~GHz, \cite{chhetri2012} found that the optimum cut-off to separate extragalactic radio sources into compact ($\lesssim 0.15$~arcsec) and extended ($\gtrsim 0.15$~arcsec) sources is --0.46, which is close to the value that is traditionally used. We therefore also looked at the fraction of sources with $\alpha_{1.4}^{15.7} < -0.5$ (see bottom panel of Fig.~\ref{fig:alpha}). 9C and 10C sources too faint to appear in NVSS can be taken to have $\alpha_{1.4}^{15.7} \geq -0.5$ provided that they have $S_{15.7} > 1.0$~mJy. Deeper 1.4-GHz data were therefore required in the deep regions of the 10C survey (where the completeness level goes below 1.0~mJy) to include sources between 0.5 and 1.0~mJy.

We used data from \cite{whittam2013} to include points between 0.5 and 1.0~mJy when using a spectral index cut-off of --0.5. We used their published measurements of $\alpha_{1.4}^{15.7}$ for 51 of the 296 sources which form a sample complete to 0.5~mJy and are between 0.5 and 1.0~mJy. These were derived from 1.4-GHz flux densities in the following order of preference: NVSS, FIRST, the WSRT catalogue \citep{guglielmino2012}, \cite{owen2008} and \cite{biggs2006}. Upper limits on the 1.4-GHz flux densities are quoted for sources with no matches in any of these 1.4-GHz catalogues; such sources can be assumed to have spectra flatter than --0.5. 

A similar trend is observed when using a spectral index cut-off of --0.5. The fraction of sources with $\alpha_{1.4}^{15.7} < -0.5$ peaks in the 3.0-5.5~mJy bin, where it is $(67 \pm 4)$\%. It drops to $(39 \pm 7)$\% in the 0.5--1.0~mJy bin. 

Fig.~\ref{fig:counts_eucl} shows the 15.7-GHz differential counts of sources with $\alpha_{1.4}^{15.7} < -0.8$, $\alpha_{1.4}^{15.7} \geq -0.8$, $\alpha_{1.4}^{15.7} < -0.5$ and $\alpha_{1.4}^{15.7} \geq -0.5$ normalized to $S^{-2.5}$, i.e. divided by the counts expected in a Euclidean universe. The binned differential source count data are provided in Appendix~\ref{Source count data}. The counts of sources with $\alpha_{1.4}^{15.7} < -0.8$ show strong curvature and lie below the counts of sources with $\alpha_{1.4}^{15.7} \geq -0.8$ over the full range of flux densities. The difference between the counts of the sources with $\alpha_{1.4}^{15.7} < -0.5$ and $\alpha_{1.4}^{15.7} \geq -0.5$ is less marked. Flat-spectrum sources dominate the counts above $\sim 25$~mJy, and steep-spectrum sources between $\sim 1.0$ and $\sim 25$~mJy. Below $\sim 1.0$~mJy, flat-spectrum sources dominate the counts again. We note that if extended low brightness emission from steep-spectrum sources was being resolved out in the 10C survey, this would steepen the counts of the steep-spectrum sources at the faint end. We think that this is unlikely to be a significant problem given that only $\approx 5$\% of sources in the 10C survey were found to be extended relative to the beam of $\approx 30$~arcsec.

\subsection{Comparison with the 5-GHz source-count model by Jackson \& Wall}\label{Comparison with the 5-GHz source-count model by Jackson and Wall}

\cite{jackson1999} have produced a source-count model at 5~GHz including contributions from seven different source types: 
\begin{enumerate}
\item[(1)] the unbeamed products of high-excitation FRII sources, with strong optical/UV emission lines
\item[(2)] the beamed products of high-excitation FRII sources, with strong optical/UV emission lines
\item[(3)] the unbeamed products of low-excitation FRII sources, with weak-to-no emission lines
\item[(4)] the beamed products of low-excitation FRII sources, with weak-to-no emission lines
\item[(5)] the unbeamed products of FRI sources
\item[(6)] the beamed products of FRI sources
\item[(7)] star-forming galaxies
\end{enumerate}
Source types 1, 3, 5 and 7 are taken to have steep spectra ($\alpha_{2.7}^{5} < -0.5$) and source types 2, 4 and 6 flat ($\alpha_{2.7}^{5} \geq -0.5$) spectra. They modeled the cosmic evolution of FRI and FRII sources using low-frequency data at 151~MHz to avoid bias due to the effects of Doppler beaming. The evolution was described using luminosity-dependent density evolution. The FRI and FRII populations were found to have very different evolutions, the FRII population undergoing strong cosmic evolution and the FRI population showing little or no evolution. They then accounted for the beamed products of FRI and FRII sources by randomly aligning them with respect to our line of sight. They used higher-frequency data (in particular 5-GHz source count data) to constrain the parameters used to model the beaming. 

\begin{figure}
 \begin{center}
 \includegraphics[width=0.35\textwidth,angle=270]{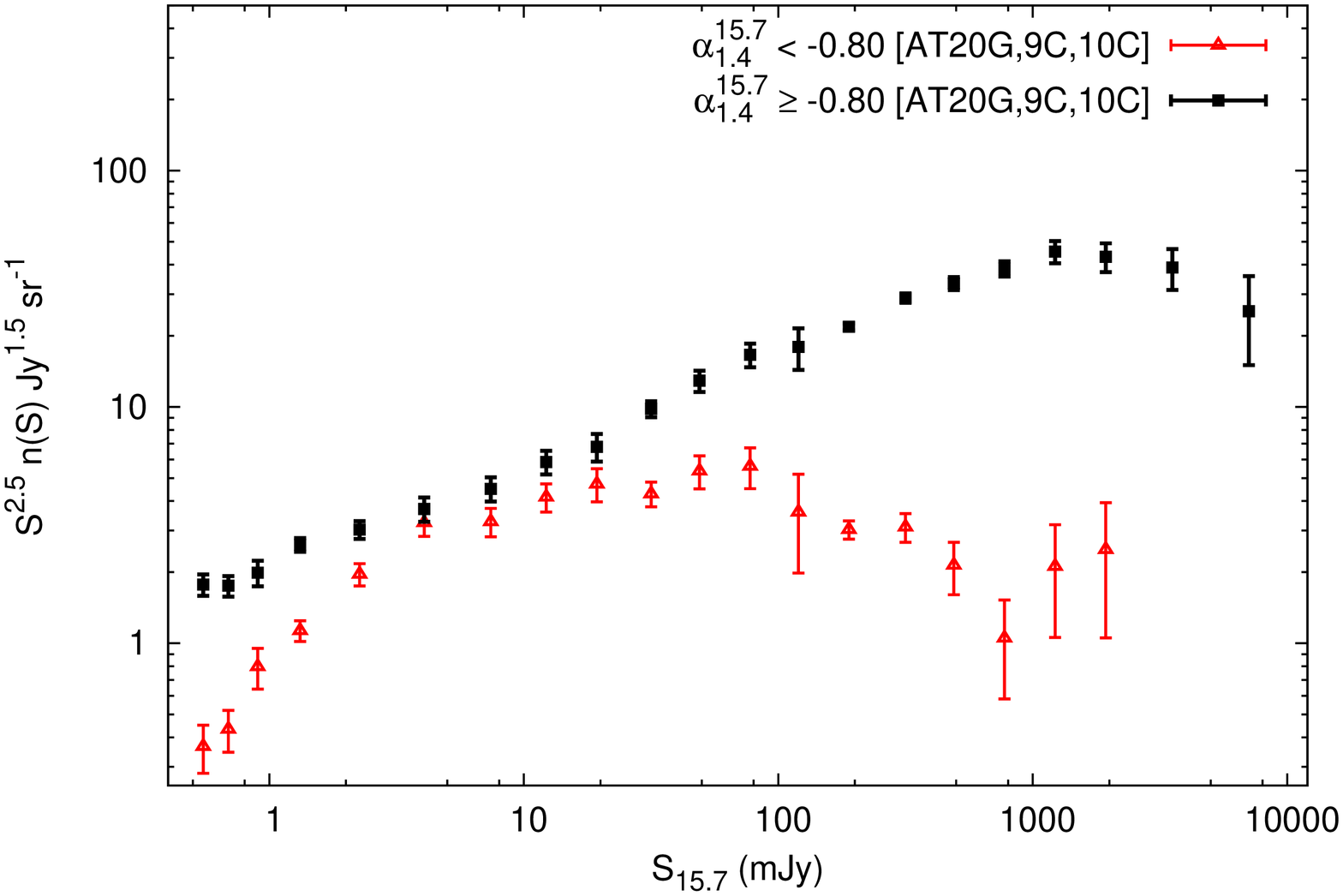}
 \includegraphics[width=0.35\textwidth,angle=270]{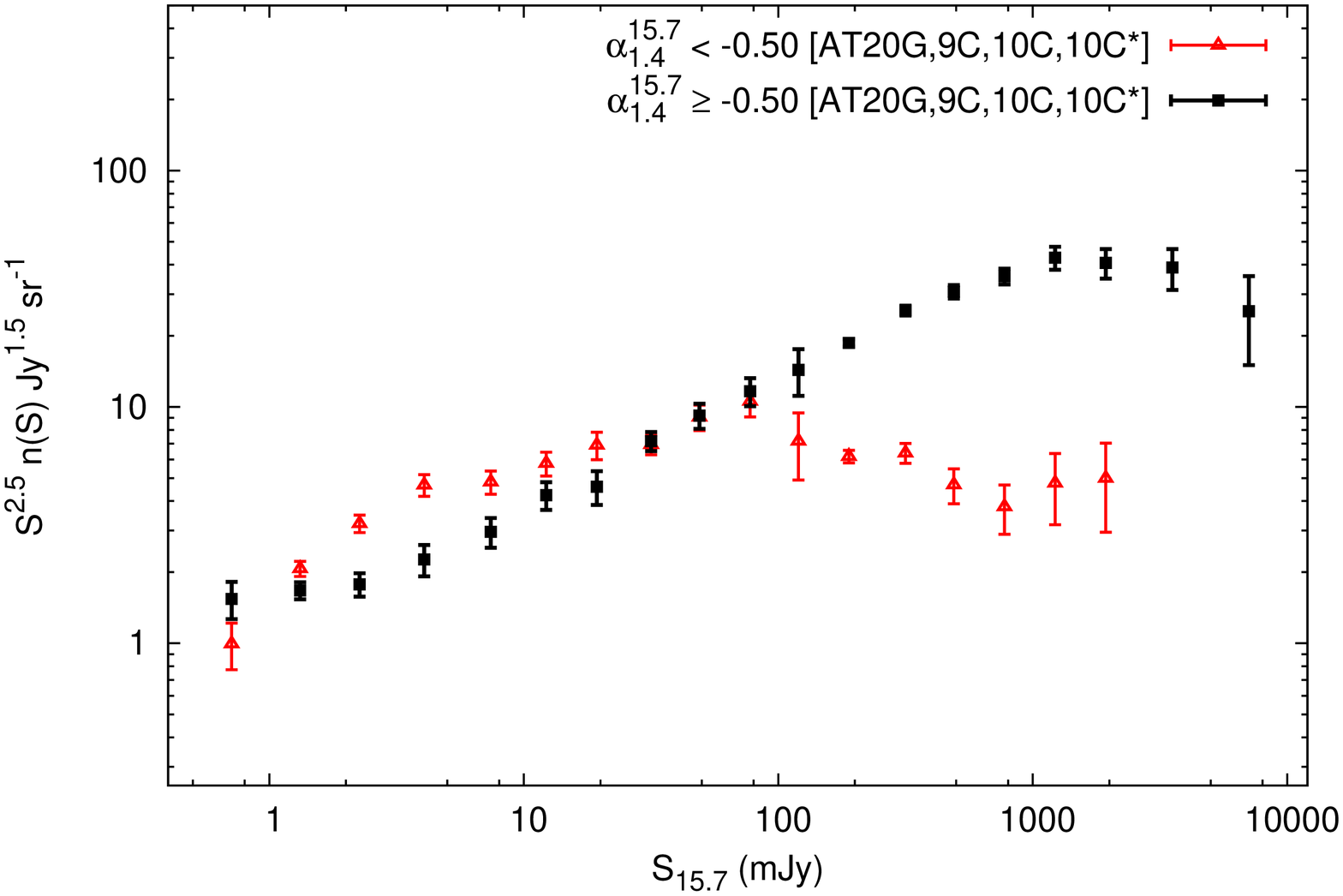}
  \caption{Top: The normalized ($S^{2.5} n(S)$) differential counts for sources with $\alpha_{1.4}^{15.7} < -0.8$ and $\alpha_{1.4}^{15.7} \geq -0.8$. The same flux-density binning scheme is used as in Fig.~\ref{fig:alpha}. Bottom: Same as above except that a spectral index cut-off of -0.5 is applied.}
 \label{fig:counts_eucl}
 \end{center}
\end{figure}

\begin{figure}
\includegraphics[scale=0.45,trim=1.5cm 8cm 0cm 7cm]{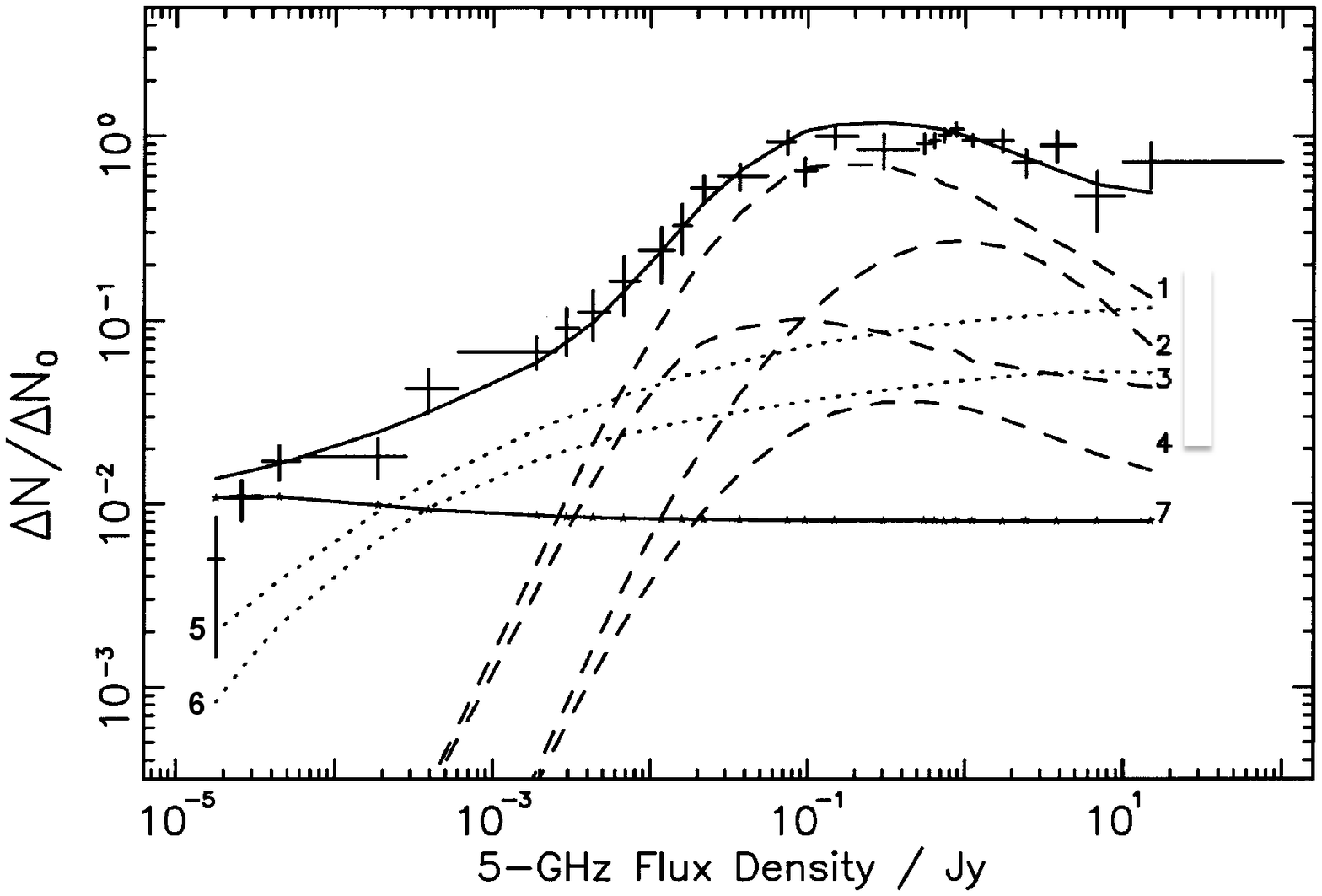}
\caption{Plot from \citet{jackson1999} showing their source count fit at 5~GHz. Data points represent the differential source counts at 5~GHz. The model produces contributions from seven source types as listed above. All counts are shown in relative differential form with $N_{0} = 60(S_{5 \mathrm{GHz}})^{-1.5} \mathrm{sr}^{-1}$.}
\label{fig:jackson_plot}
\end{figure}

Fig.~\ref{fig:jackson_plot} shows their source count fit at 5~GHz, including the contributions from the source types listed above. We compare this with the observed counts at 15.7~GHz, separated into flat- and steep-spectrum sources, shown in the bottom panel of Fig.~\ref{fig:counts_eucl}. Although there is a significant frequency difference, the comparison is still useful as the beamed products of FRI and FRII sources already make a highly significant contribution to the source count at 5~GHz. We note that the counts of the steep-spectrum sources in Fig.~\ref{fig:jackson_plot} will be somewhat shifted to the left at 15.7~GHz.

In the model, the counts of the FRII sources show broad bulges followed by dramatic declines. This is a consequence of the strong evolution attributed to the FRII population. The counts of the beamed sources peak at $\sim 1$~Jy and those of the unbeamed sources at $\sim 100$~mJy  Thus unbeamed FRII sources appear to undergo stronger cosmological evolution than do beamed FRII sources of the same apparent power. \citeauthor{jackson1999} interpret this as being a result of the Doppler enhancement of the radio emission from the beamed FRII sources. The FRI sources have monotonically decreasing counts throughout the entire flux density range since little or no evolution is attributed to the FRI population. They start to become more dominant than FRII sources below a few mJy. The proportion of beamed to unbeamed FRI sources is not predicted to change substantially with flux density. Star-forming galaxies become the dominant source population below $\sim 100 \mu \mathrm{Jy}$ .

Both the flat- and steep-spectrum counts at 15.7~GHz show bulges, presumably due to the influence of the FRII sources. The flat-spectrum counts peak at $\sim 1~\mathrm{Jy}$; the steep-spectrum counts have a broader peak at $\sim$10--100~mJy. This offset between the peaks of the steep- and flat-spectrum counts causes the observed shift towards a steeper-spectrum population when going from $\sim 1$~Jy to $\sim 5$~mJy. The steep-spectrum counts start to fall sharply below $\sim 5$~mJy. Thus the decline in the population of FRII radio galaxies appears to be in large part responsible for the observed decrease in the fraction of steep-spectrum sources below $\sim 5$~mJy. 

On the other hand, the flat-spectrum counts flatten significantly below $\sim 5$~mJy. The counts of BL-Lac objects (source type 6) are not predicted to flatten below 5~mJy. The model predicts that star-forming galaxies play an increasingly important role at fainter flux densities. They have steep spectra between 2.7 and 5~GHz, synchrotron radiation being the dominant emission mechanism in this frequency range. Given that free-free radiation accounts for an increasingly high fraction of the total intensity at higher frequency, we have estimated the spectral index that a typical star-forming galaxy would have between 1.4 and 15.7~GHz. \cite{ysard2012} modeled the radio spectra of star-forming galaxies to estimate their contribution to the cosmic radio background. They considered three emission components: synchrotron, free-free and spinning dust. Based on the work by \cite{peel2011}, \cite{murphy2010b} and \cite{tabatabaei2007} who studied the radio-to-infrared spectra of nearby star-forming galaxies, \citeauthor{ysard2012} modeled the synchrotron spectrum as a power-law with spectral index equal to --1.0, the free-free emission as a power-law with spectral index equal to --0.1, and normalised the intensity of the free-free emission to be equal to 50\% of the synchrotron emission at 10~GHz; the spinning dust component was fully dominated by the synchrotron and free-free components at 10~GHz. This yields a value of $\alpha_{1.4}^{15.7}$ equal to --0.8. We therefore do not expect star-forming galaxies to contribute to the flat-spectrum counts despite the larger contribution of the free-free component at 15.7~GHz. 

Therefore, the \citeauthor{jackson1999} model does not seem to reproduce the flattening of the flat-spectrum counts seen at 15.7~GHz. It is possible that a new population of faint, flat-spectrum sources, not included in the model, is contributing to this effect. \cite{sadler2013} studied the local radio source population at 20~GHz by matching radio sources from the AT20G catalogue with nearby galaxies from the Third Data Release of the 6dF Galaxy Survey \citep{jones2009}. A remarkably high fraction (48\%) of the 201 sources in their sample were found to have flat spectra ($\alpha_{1}^{20} > -0.5$) despite the fact that the sample is not thought to contain any relativistically-beamed sources (the objects were selected from within a sample of normal galaxies from which QSOs were excluded). Many of the flat-spectrum sources showed the characteristics of GPS sources, with 20-GHz sizes less than 1~kpc and radio spectra peaking above 1~GHz. In addition, some of the steep-spectrum ($\alpha_{1}^{20} \leq -0.5$) sources were found to have 20-GHz sizes less 10--20~kpc, making them candidate compact steep-spectrum (CSS) sources. 

It is possible that this same population of flat-spectrum sources, seen at higher redshifts, is starting to contribute significantly to the 15.7-GHz counts at the faint end. GPS and CSS sources are thought to be young versions of large-scale radio galaxies \citep[e.g.][]{odea1998}, but given the observed overabundance of such sources \citep[e.g.][]{odea1997}, it is not clear whether all of these can evolve into FRI or FRII galaxies. Instead, the activity in these sources may only be intermittent and not sustained enough to drive large-scale jets and lobes \citep[e.g.][]{reynolds1997}; or they may not be young but the radio emitting plasma is currently unable to escape from the nuclear regions \citep[e.g.][]{baum1990}.

Another possible explanation is that a large number of FRI radio galaxies have 15.7-GHz core-to-extended flux ratios sufficiently high for them to appear as flat-spectrum sources. FRI sources are predicted to dominate the 5-GHz counts at $\sim 1$~mJy. In the model, it is assumed that the unbeamed products of FRI sources have steep spectra between 2.7 and 5~GHz. At 15.7~GHz, a higher fraction of the total flux density will of course originate from the core. It is also not clear how accurately the 15.7-GHz emission from FRI cores is being modeled given the available data.

It would be very useful to obtain information about the morphology of sources in the AT20G-deep pilot survey, particularly those that have spectra peaking above 1~GHz, to see whether or not they can be identified as FRI or FRII sources; we know that about a quarter of the sources are likely to have radio spectra peaking above 1~GHz (considering the number of inverted-spectrum sources with $\alpha_{1.4}^{20} > 0$ and/or with clearly defined peaks in their spectra between 1.4 and 18~GHz). This may be achieved, for example, using follow-up observations with the Very Large Array at 1.4~GHz, with both high resolution and high surface brightness sensitivity.

Additional information about the source compactness at high radio frequency could be obtained from our ATCA follow-up observations. Using 6-km antenna data from ATCA at 20~GHz, \cite{chhetri2013} were able to identify sources smaller than 0.15~arcsec, or less than 1~kpc at all redshifts, in the AT20G survey, which has a flux density limit of 40~mJy. The larger bandwidth and increased continuum sensitivity of the new CABB correlator on ATCA will allow the technique to be applied to fainter sources in our sample; the results of this analysis are presented in Paper II.

\subsection{Comparison with the de Zotti source-count model at 15~GH\lowercase{z}}\label{Comparison with the de Zotti source-count model at 15 GHz}

\begin{figure}
 \begin{center}
 \includegraphics[width=0.35\textwidth,angle=270]{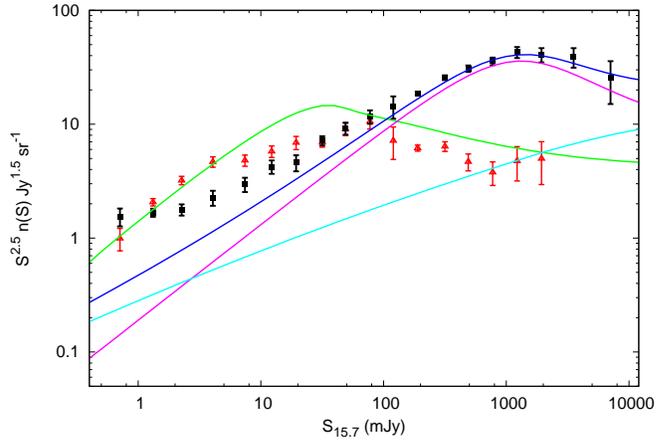}
  \caption{Comparison of the 15.7-GHz counts, separated into flat- and steep-spectrum sources, with the latest version of the \citet{dezotti2005} source-count model at 15~GHz. The red triangles and black squares show the normalized ($S^{2.5} n(S)$) differential counts for sources with $\alpha_{1.4}^{15.7} < -0.5$ and $\alpha_{1.4}^{15.7} \geq -0.5$, respectively. The solid lines indicate the prediction of the de Zotti model. The contribution from FSRQs is shown in purple, BL Lacs in cyan, flat-spectrum sources (sum of FSRQs and BL Lacs) in blue and steep-spectrum sources in green. No attempt was made to correct for the small frequency difference between the measured and model source counts, but this is unlikely to significantly affect any conclusions drawn from the comparison.}
 \label{fig:dezotti_comparison}
 \end{center}
\end{figure}

In Fig.~\ref{fig:dezotti_comparison}, the 15.7-GHz counts of flat- and steep-spectrum sources are compared with the latest version of the 15-GHz source-count model by \cite{dezotti2005}, extracted from their website\footnote{http://web.oapd.inaf.it/rstools/srccnt/srccnt\_tables} on 2013 October 11. The model includes contributions from FSRQs, BL Lacs and steep-spectrum sources. The steep-spectrum sources only include sources powered by nuclear activity; they do not include star-forming galaxies. The data used to constrain the model include source-count measurements from the 9C survey \citep{waldram2003,waldram2010} at 15.2~GHz, which probes flux densities down to 5.5~mJy.  

For the steep-spectrum sources, the \citeauthor{dezotti2005} model is broadly in agreement with the data, although it somewhat overpredicts the measured counts above $\sim 5$~mJy. For the flat-spectrum sources, the model demonstrates good agreement with the data above $\sim 100$~mJy. However, below this flux-density level, the model increasingly underpredicts the measured counts with decreasing flux density. The flat-spectrum counts are predicted to flatten with decreasing flux density due to the contribution from the BL Lacs, but they do not flatten sufficiently to resolve the apparent deficiency of flat-spectrum sources below $\sim 100$~mJy; indeed, BL Lacs are only predicted to overtake FSRQs below $\sim 5$~mJy. The model underpredicts the number of flat-spectrum sources by about a factor of four in the lowest flux-density bin (0.5--1.0~mJy). Therefore, it seems likely that an additional population of unbeamed, flat-spectrum sources is emerging at the low-flux-density end of the measured counts.

\section{Conclusions}\label{Conclusions}

As part of a pilot study for the AT20G-deep survey, we have surveyed an area of $5~\mathrm{deg}^{2}$ to an rms noise level of $\approx 0.3~\mathrm{mJy}$ at 20~GHz with ATCA. The survey covers two regions, the Chandra Deep Field South (03~hr; $-27$~deg) and part of SDSS Stripe 82 (21~hr; $-0.5$~deg). The sources detected in the survey were followed up with ATCA at 5.5, 9 and 18~GHz. 

The catalogue from the AT20G-deep pilot survey contains 85 sources above $5 \sigma$. The reliability of the catalogue is 100\%. The catalogue is estimated to be 90\% complete above 2.5~mJy. The density of sources above 2.5~mJy is $(13.4 \pm 1.7)~\mathrm{deg}^{-2}$.

We have matched our source catalogue with the NVSS, FIRST and ATLAS DR1 catalogues at 1.4~GHz. 55\% of the sources have steep spectra ($\alpha_{1.4}^{20} < -0.5$), 24\% have flat spectra ($-0.5 \leq \alpha_{1.4}^{20} < 0.0$) and 21\% have inverted spectra ($\alpha_{1.4}^{20} \geq 0.0$). Only one source is not detected at 1.4~GHz. It is also not detected in the ATLAS DR3 image, implying that it has $\alpha_{1.4}^{18} > 1.38$.

On the whole, the steep-spectrum sources show little spectral curvature between 1.4 and 18~GHz. There is a strong tendency for the spectral indices of the flat- and inverted-spectrum sources to steepen with frequency. Among the 18 inverted-spectrum sources, 10 have clearly defined peaks in their spectra with $\alpha_{1.4}^{5.5} > 0.15$ and $\alpha_{9}^{18} < -0.15$. 

On a 3-yr timescale, at least 10 sources varied by more than 15\% at 20~GHz; 9 of these have flat or inverted spectra. Sources displaying blazar activity (which the variability would imply) are therefore still relatively common at the flux densities probed by the AT20G-deep pilot survey.

We have combined data from the AT20G survey at 20~GHz, the 9C survey at 15.2~GHz and the 10C survey at 15.7~GHz to investigate the dependence of spectral index on flux density at 15.7~GHz. We find a shift towards a steeper-spectrum population when going from $\sim 1$~Jy to $\sim 5$~mJy. This trend is reversed below $\sim 5$~mJy, with a move back towards a flatter-spectrum population. The counts at 15.7~GHz are dominated by flat-spectrum ($\alpha_{1.4}^{15.7} \geq -0.5$) sources above $\sim 25$~mJy, steep-spectrum ($\alpha_{1.4}^{15.7} < -0.5$) sources between $\sim 1.0$ and $\sim 25$~mJy, and flat-spectrum sources below $\sim 1.0$~mJy.

The Euclidean normalized differential counts of both flat- and steep-spectrum sources show bulges, the flat-spectrum counts peaking at $\sim 1$~Jy and the steep-spectrum counts displaying a broader peak at $\sim$10--100~mJy. Comparison with the 5-GHz source-count model by \cite{jackson1999} suggests that these bulges are caused by the contribution from the rapidly evolving FRII sources. The counts of the beamed, flat-spectrum FRII sources would peak at a higher flux density as a result of the Doppler enhancement of their radio emission. The offset between the peaks of the flat- and steep-spectrum counts results in the observed shift towards a steeper-spectrum population when going from $\sim 1$~Jy to $\sim 5$~mJy. Below $\sim 5$~mJy, the steep-spectrum counts start to steepen strongly, while the flat-spectrum counts start to flatten, causing the trend to reverse. The flattening of the flat-spectrum counts is not predicted by the model, which only includes contributions from FRI and FRII sources, and star-forming galaxies. Given the composition of the 20-GHz population in the local universe, it seems likely that an additional population of compact, flat-spectrum sources is contributing to this effect, although more radio data are required to confirm this. 

Direct comparison of the 15.7-GHz counts with the 15-GHz source-count model by \cite{dezotti2005} for classical radio sources also suggests that an additional population of flat-spectrum sources is emerging at the low-flux-density end of the measured counts: for flat-spectrum sources comprising FSRQs and BL Lacs, there is good agreement with the data above $\sim 100$~mJy, but below this flux-density level, the model increasingly underpredicts the measured counts with decreasing flux density. The model also somewhat overpredicts the measured counts of steep-spectrum sources above $\sim 5$~mJy.

The sources detected in the AT20G-deep pilot survey will be studied further in an accompanying paper where we will study their angular sizes and polarisation properties. We will also combine the radio data with optical and infrared data. 

\section*{Acknowledgments}\label{Acknowledgements}

TMOF acknowledges support from an ARC Super Science Fellowship. IHW acknowledges an STFC studentship. The ATCA is part of the Australia Telescope National Facility which is funded by the Commonwealth of Australia for operation as a National Facility managed by CSIRO. We thank the anonymous referee for helpful comments.

\setlength{\labelwidth}{0pt}

\appendix\section{Source count data}\label{Source count data}

Table~\ref{tab:source_counts_data_-0.8} shows the data for the 15.7-GHz counts of sources with $\alpha_{1.4}^{15.7} < -0.8$ and $\alpha_{1.4}^{15.7} \geq -0.8$. Table~\ref{tab:source_counts_data_-0.5} shows the data for the 15.7-GHz counts of sources with $\alpha_{1.4}^{15.7} < -0.5$ and $\alpha_{1.4}^{15.7} \geq -0.5$ with which the 5-GHz model source counts by \citet{jackson1999} and the 20-GHz model source counts by \citet{dezotti2005} are compared in Sections~\ref{Comparison with the 5-GHz source-count model by Jackson and Wall} and~\ref{Comparison with the de Zotti source-count model at 15 GHz}.

\newpage

\begin{table}
\caption{Differential counts at 15.7~GHz of sources with $\alpha_{1.4}^{15.7} < -0.8$ and $\alpha_{1.4}^{15.7} \geq -0.8$. `AT20G' designates areas presented in \citet{massardi2011}, `9C1' in \citet{waldram2003}, `9C2' in \citet{waldram2010} and `10C' in \citet{davies2011}.}
\label{tab:source_counts_data_-0.8}
\begin{tabular}{@{} c c c c c c } 
\hline
\multicolumn{1}{c}{Bin start}
&\multicolumn{1}{c}{Bin end}
&$N_{\alpha < -0.8}$
&$N_{\alpha \geq -0.8}$
&\multicolumn{1}{c}{Area}
&Survey\\
\multicolumn{1}{c}{$S$ (mJy)}
&\multicolumn{1}{c}{$S$ (mJy)}
&&&\multicolumn{1}{c}{($\mathrm{deg}^{2}$)}&\\
\hline
0.50 &         0.60 &         19 &           92 &           11.96 &        10C \\         
0.60 &         0.80 &         25 &           101 &          11.96 &        10C \\         
0.80 &         1.02 &         26 &           65 &           11.96 &        10C \\         
1.02 &         1.70 &         100 &          231 &          26.86 &        10C \\         
1.70 &         3.00 &         86 &           133 &          26.86 &        10C \\         
3.00 &         5.50 &         63 &           72 &           26.86 &        10C \\         
5.50 &         10.00 &        53 &           73 &           55.96 &        9C2 + 10C \\   
10.00 &        15.00 &        54 &           76 &           141.56 &       9C2 + 10C \\   
15.00 &        25.00 &        39 &           56 &           141.56 &       9C2 + 10C \\   
25.00 &        40.00 &        70 &           160 &          634.70 &       9C1 + 9C2 \\   
40.00 &        60.00 &        39 &           94 &           634.70 &       9C1 + 9C2 \\   
60.00 &        100.00 &       26 &           77 &           634.70 &       9C1 + 9C2 \\   
100.00 &       143.80 &       5 &            25 &           520.00 &       9C1 \\         
143.80 &       250.00 &       129 &          930 &          20606.00 &     AT20G + 9C1 \\ 
250.00 &       400.00 &       52 &           484 &          20606.00 &     AT20G + 9C1 \\ 
400.00 &       600.00 &       16 &           249 &          20606.00 &     AT20G + 9C1 \\ 
600.00 &       1000.00 &      5 &            183 &          20606.00 &     AT20G + 9C1 \\ 
1000.00 &      1500.00 &      4 &            86 &           20606.00 &     AT20G + 9C1 \\ 
1500.00 &      2500.00 &      3 &            52 &           20606.00 &     AT20G + 9C1 \\ 
2500.00 &      5000.00 &      0 &            26 &           20606.00 &     AT20G + 9C1 \\ 
5000.00 &      10000.00 &     0 &            6 &            20606.00 &     AT20G + 9C1 \\ 
\hline
\end{tabular}
\end{table}

\begin{table}
\caption{Differential counts at 15.7~GHz of sources with $\alpha_{1.4}^{15.7} < -0.5$ and $\alpha_{1.4}^{15.7} \geq -0.5$. `$10\mathrm{C}^{\star}$' designates a region of the 10C survey in the Lockman Hole in which spectral index information from \citet{whittam2013} was used.}
\label{tab:source_counts_data_-0.5}
\begin{tabular}{@{} c c c c c c } 
\hline
\multicolumn{1}{c}{Bin start}
&\multicolumn{1}{c}{Bin end}
&$N_{\alpha < -0.5}$
&$N_{\alpha \geq -0.5}$
&\multicolumn{1}{c}{Area}
&Survey\\
\multicolumn{1}{c}{$S$ (mJy)}
&\multicolumn{1}{c}{$S$ (mJy)}
&&&\multicolumn{1}{c}{($\mathrm{deg}^{2}$)}&\\
\hline
0.50 &         1.02 &         20 &           31 &           1.73 &         10C* \\        
1.02 &         1.70 &         183 &          148 &          26.86 &        10C \\         
1.70 &         3.00 &         141 &          78 &           26.86 &        10C \\         
3.00 &         5.50 &         91 &           44 &           26.86 &        10C \\         
5.50 &         10.00 &        78 &           48 &           55.96 &        9C2 + 10C \\   
10.00 &        15.00 &        75 &           55 &           141.56 &       9C2 + 10C \\   
15.00 &        25.00 &        57 &           38 &           141.56 &       9C2 + 10C \\   
25.00 &        40.00 &        113 &          117 &          634.70 &       9C1 + 9C2 \\   
40.00 &        60.00 &        66 &           67 &           634.70 &       9C1 + 9C2 \\   
60.00 &        100.00 &       49 &           54 &           634.70 &       9C1 + 9C2 \\   
100.00 &       143.80 &       10 &           20 &           520.00 &       9C1 \\         
143.80 &       250.00 &       263 &          796 &          20606.00 &     AT20G + 9C1 \\ 
250.00 &       400.00 &       107 &          429 &          20606.00 &     AT20G + 9C1 \\ 
400.00 &       600.00 &       35 &           230 &          20606.00 &     AT20G + 9C1 \\ 
600.00 &       1000.00 &      18 &           170 &          20606.00 &     AT20G + 9C1 \\ 
1000.00 &      1500.00 &      9 &            81 &           20606.00 &     AT20G + 9C1 \\ 
1500.00 &      2500.00 &      6 &            49 &           20606.00 &     AT20G + 9C1 \\ 
2500.00 &      5000.00 &      0 &            26 &           20606.00 &     AT20G + 9C1 \\ 
5000.00 &      10000.00 &     0 &            6 &            20606.00 &     AT20G + 9C1 \\ 
\hline
\end{tabular}
\end{table}

\label{lastpage}
\end{document}